\documentclass{ws-rv9x6}

\usepackage[square]{ws-rv-van}
\usepackage{subfigure}   
\usepackage{ws-rv-thm}   
\usepackage{ws-rv-van}   
\makeindex

\usepackage[hidelinks]{hyperref}
\usepackage{amsmath}
\usepackage{amssymb}
\usepackage{lipsum}

\def\ba{\begin{eqnarray}}
\def\ea{\end{eqnarray}}

\def\be{\begin{equation}}
\def\ee{\end{equation}}

\newcommand{\p}{\vec{p}}

\newcommand{\wbar}{{\overline \tau}}
\newcommand{\Mbar}{{\overline{\mathcal M}}\hspace{0.5mm}}

\DeclareMathOperator{\sign}{sign}

\DeclareFontFamily{OT1}{pzc}{}
\DeclareFontShape{OT1}{pzc}{m}{it}{ <-> s*[1.15] pzcmi7t }{}
\DeclareMathAlphabet{\mathpzc}{OT1}{pzc}{m}{it}

\begin{document}

\chapter[Hydrodynamization and resummed viscous hydrodynamics]{Hydrodynamization and resummed viscous hydrodynamics}

\author[M. Strickland]{Michael Strickland}
\address{Department of Physics\\Kent State University\\Kent, OH 44240 USA}

\begin{abstract}
In this contributed chapter, I review our current understanding of the applicability of hydrodynamics to modeling the quark-gluon plasma (QGP), focusing on the question of hydrodynamization/thermalization of the QGP and the anisotropic hydrodynamics (aHydro) far-from-equilibrium hydrodynamic framework.  I discuss the existence of far-from-equilibrium hydrodynamic attractors and methods for determining attractors within different hydrodynamical frameworks. I also discuss the determination of attractors from exact solutions to the Boltzmann equation in relaxation time approximation and effective kinetic field theory applied to quantum chromodynamics.  I then present comparisons of the kinetic attractors with the attractors obtained in standard second-viscous hydrodynamics frameworks and anisotropic hydrodynamics.  I demonstrate that, due to the resummation of terms to all orders in the inverse Reynolds number, the anisotropic hydrodynamics framework can describe both the weak- and strong-interaction limits.  I then review the phenomenological application of anisotropic hydrodynamics to relativistic heavy-ion collisions using both quasiparticle aHydro and second-order viscous aHydro.  The phenomenological results indicate that aHydro provides a controlled extension of dissipative relativistic hydrodynamics to the early-time far-from-equilibrium stage of heavy-ion collisions.  This allows one to better describe the data and to extract the temperature dependence of transport coefficients at much higher temperatures than linearized second-order viscous hydrodynamics.
\end{abstract}


\body

\newpage 

\tableofcontents

\section{Introduction}

Relativistic viscous hydrodynamics stands as the primary theoretical framework for describing the spatiotemporal evolution of the rapidly expanding quark-gluon plasma (QGP) generated in ultrarelativistic heavy ion collisions \cite{Heinz:2013th}. Despite its notable success, understanding how hydrodynamics can provide a reliable description of the rapidly expanding system created in these collisions poses a formidable challenge. Traditionally, hydrodynamics has been understood as a truncation of a gradient expansion \cite{Chapman_Enskog}. Thus, its region of validity was assumed to be limited to small gradients relative to the inverse microscopic scales of the problem. The gradient expanded theory was historically regarded as a universal macroscopic limit inherent to microscopic theories, achievable at suitably late times.  Recent investigations, however, have revealed that the gradient expansion may possess a zero radius of convergence for flow configurations pertinent to the QGP, both in strong coupling scenarios and within kinetic theory~\cite{Heller:2013fn,Buchel:2016cbj,Heller:2016rtz,Denicol:2016bjh}. Consequently, constructing and refining a hydrodynamic theory by systematically incorporating higher-order terms in this series is fraught. As a result, the conventional notion that relativistic hydrodynamics is only applicable under conditions where gradients of macroscopic quantities are small, as derived from the gradient expansion, appears no longer well-justified, if not altogether unnecessary. Ultimately, these revelations prompt a reevaluation of the very definition of viscous hydrodynamics to assess its range of applicability in the context of heavy ion collisions.

While the phenomenological success of fluid-dynamical models was initially interpreted as an indication of the rapid isotropization and thermalization of the quark-gluon plasma (QGP) \cite{Heinz:2004qz}, subsequent model calculations have suggested that such an interpretation may have been premature~\cite{Chesler:2008hg,Beuf:2009cx,Chesler:2009cy,Heller:2011ju,Heller:2012je,Heller:2012km,vanderSchee:2012qj,Casalderrey-Solana:2013aba,vanderSchee:2013pia,Heller:2013oxa,Keegan:2015avk,Chesler:2015bba,Kurkela:2015qoa,Chesler:2016ceu,Attems:2016ugt,Attems:2016tby,Attems:2017zam,Florkowski:2017olj}. This is due to the fact that systems that significantly depart from equilibrium may already exhibit hydrodynamic behavior through a phenomenon known as ``hydrodynamization''. This novel feature is particularly relevant in rapidly expanding fluids like the QGP.  In practice, the applicability of linearized viscous hydrodynamics is not boundless; it will eventually fail to be accurate as viscosity values become sufficiently large or when applied at very early times when there are large gradients. Nevertheless, even in such extreme scenarios, effective theories capable of describing the quark-gluon plasma exist, with anisotropic hydrodynamics (aHydro) being the most notable among them \cite{Florkowski:2010cf,Martinez:2010sc,Ryblewski:2010ch,Martinez:2012tu,Ryblewski:2012rr,Bazow:2013ifa,Tinti:2013vba,Nopoush:2014pfa,Tinti:2015xwa,Bazow:2015cha,Strickland:2015utc,Alqahtani:2015qja,Molnar:2016vvu,Molnar:2016gwq,Alqahtani:2016rth,Bluhm:2015raa,Bluhm:2015bzi,Alqahtani:2017jwl,Alqahtani:2017tnq}.
In general, hydrodynamization is now expected to occur at a time scale $\tau_{\rm hydro}$, which is shorter than the corresponding time scale for isotropization, being driven by a novel \emph{dynamical attractor} whose details vary according to the theory under consideration, e.g., kinetic theory, hydrodynamics, holography, etc. \cite{Heller:2015dha,Keegan:2015avk,Romatschke:2017vte,Bemfica:2017wps,Spalinski:2017mel}. Such attractor solutions show that hydrodynamics displays a new degree of universality in far-from-equilibrium scenarios regardless of the details of the initial state of the system. In fact, the approach to the dynamical attractor effectively wipes out a subset of information about the specific initial condition used for the evolution, before the true equilibrium state and consequently, full thermalization, is reached. 

In the realm of kinetic theory and conventional statistical mechanics, thermalization is characterized by the emergence of isotropic thermal one-particle distribution functions for the partons constituting the QGP. In the context of high-energy heavy-ion collisions, the substantial longitudinal expansion rate causes the QGP fireball's center to gradually relax toward an approximately isotropic state, on a time scale of $\tau_{\rm iso} \gtrsim 4$ fm/c \cite{Strickland:2013uga}. Notably, the hydrodynamization of the fireball seems to occur on a shorter timescale, as reviewed in Ref.~\cite{Florkowski:2017olj}.\footnote{It is worth mentioning that studies utilizing either the 2PI formalism or holography suggest that, at the highest temperatures explored in heavy-ion collisions, an equation of state may be established well before pressure isotropization \cite{Berges:2004ce,Attems:2016ugt,Attems:2017zam}.} For conformal systems, a crucial factor for determining proximity to universal attractor behavior is the dimensionless variable $w \equiv \tau T$ \cite{Heller:2015dha}. For conformal fluids undergoing Bjorken expansion \cite{Bjorken:1982qr}, this variable is proportional to the inverse of the Knudsen number Kn, with $1/T$ representing the microscopic relaxation time scale. For small gradients where $w \gg 1$, the system exhibits dynamics consistent with the universal hydrodynamical attractor. However, in the large gradient regime where $w \ll 1$, the system's dynamics are dominated by non-hydrodynamic modes (i.e., modes in the linearized dynamics with nonzero frequency even for a spatially homogeneous system \cite{Kovtun:2005ev}), whose evolution depends on the precise initial condition assumed. Considering a fixed proper time after the collision, this implies that, near the edge of the QGP, the system is more sensitive to the genuinely non-equilibrium dynamics linked to non-hydrodynamic modes. Consequently, certain non-universal aspects of the underlying theory, whether rooted in kinetic theory or holographic duality, start to influence the spatiotemporal evolution of the QGP. In such scenarios, a decision must be made regarding as to which underlying microscopic theory best captures the relevant physics. Given the system's dilute nature close to the QGP's edge (large mean free path), a kinetic theory approach appears preferable in this spatial region (large $|x|$, $|y|$, and/or $|\varsigma|$, with the $\varsigma$ being the spatial rapidity).  In addition, at very early times after the nuclear pass through, the energy densities are sufficiently high to justify a weak coupling approach that takes into account perturbative scattering processes, with modified power counting stemming from large amplitude gluon fields.  For this reason, approaches such a anisotropic hydrodynamics, which are explicitly based on a kinetic theory approach are perhaps better suited for understanding the QGP at early times and in dilute regions near the edges of the plasma.

\begin{figure}[t]
    \includegraphics[width=0.95\linewidth]{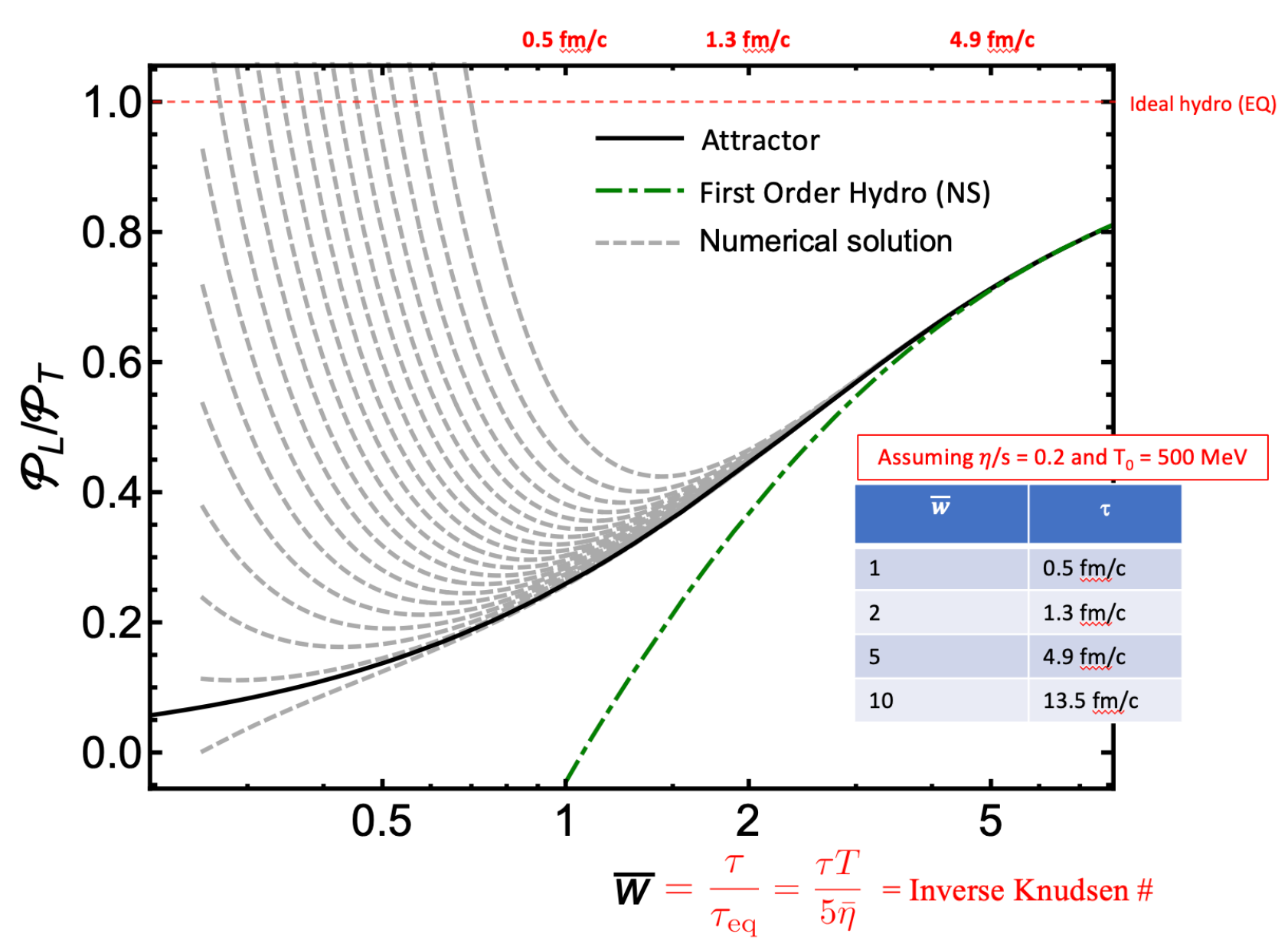}
    \caption{An illustration of the evolution of the pressure anisotropy (${\cal P}_L/{\cal P}_T$) as a function of the rescaled time variable $\bar{w} = \tau/\tau_{\rm eq}$.}
    \label{fig:attractor1}
\end{figure}

In Figure \ref{fig:attractor1}, I show the relaxation time approximation evolution of the typical pressure anisotropy (${\cal P}_L/{\cal P}_T$) as a function of the rescaled time variable $\bar{w} = \tau/\tau_{\rm eq}$.  The black solid line indicates the attractor solution.  The green dot-dashed lines represents the first-order (Navier-Stokes) limit of the hydrodynamical evolution.  The gray dashed lines represent the evolution of the system with different assumed initial pressure anisotropies.  The inset table provides a conversion of $\overline{w}$ to typical times in fm/c, assuming a shear-viscosity to entropy density ratio, $\eta/s = 0.2$, and an initial temperature of $T_0 = $ 500 MeV at $\tau_0 = 0.25$ fm/c.  As can be seen from this figure, irrespective of the initial condition, the particular solutions (gray dashed lines) approach the dynamical attractor rather quickly.  I note, importantly, that the attractor solution is independent of the assumed initial condtions and shear viscosity to entropy density ratio when plotted as a function of $\overline{w}$.  Secondly, I note that this figure demonstrates that at early times after the nuclear pass through, large pressure anisotropies are generated, independent of the assumed initial conditions.  This fact was the historical basis for the introduction of anisotropic hydrodynamics, which is the focus of the second half of this chapter.

Summarizing, in this chapter, I will review recent progress in understanding the hydrodynamization of the QGP, including attractors in hydrodynamics theories themselves and different kinetic theory models including, for example, the relaxation time approximation and effective kinetic theories based on QCD.  Following this, I will discuss the formalism of anisotropic hydrodynamics and present a subset of the phenomenological tests of this framework.

\section{Hydrodynamic attractors}

As mentioned above, in recent years, remarkable progress been made in understanding the evolution of systems subject to the far-from-equilibrium conditions pertinent to heavy-ion collisions, see e.g., Refs.~\cite{Florkowski:2017olj,Romatschke:2017ejr,Berges:2020fwq} for reviews. Substantial progress has occurred in elucidating the onset of hydrodynamic behavior in relativistic systems, both in strong coupling scenarios \cite{Chesler:2010bi,Heller:2011ju,Heller:2012km,vanderSchee:2013pia,Heller:2013oxa,Casalderrey-Solana:2013aba,Chesler:2013lia,Chesler:2015wra,Keegan:2015avk,Spalinski:2017mel} and in weak coupling/kinetic theory \cite{Denicol:2014xca,Denicol:2014tha,Kurkela:2015qoa,Keegan:2015avk,Bazow:2015dha,Denicol:2016bjh,Heller:2016rtz,Romatschke:2017vte,Strickland:2017kux,Blaizot:2017ucy,Strickland:2018ayk,Kurkela:2018wud,Kurkela:2018vqr,Kurkela:2019set,Strickland:2019hff,Blaizot:2019scw,Denicol:2019lio,Brewer:2019oha,Almaalol:2020rnu,Ambrus:2021fej,Blaizot:2021cdv,Jaiswal:2022udf,Alalawi:2022pmg,Mullins:2022fbx,Alalawi:2022pmg,Ambrus:2022qya,Ambrus:2022koq,Rocha:2022ind,Du:2022bel}.  The introduction of hydrodynamic attractors \cite{Heller:2015dha} has led to an emerging framework for estimating typical values of $\tau_{\rm hydro}$, which is the time scale at which some form of hydrodynamics becomes applicable as a function of charged particle multiplicity. Notably, references such as \cite{Keegan:2015avk,Strickland:2018ayk} have attempted to make estimates of this time scale.

To understand the main arguments underpinning these advancements, we first consider a simplified model of the QGP based on conformal kinetic theory \cite{Baier:2007ix,Denicol:2014xca,Denicol:2014tha} within the relaxation time approximation \cite{Baym:1984np,Florkowski:2013lza,Florkowski:2013lya,Florkowski:2014sfa}. This approximation has proven to be a powerful tool for gaining insights into the onset of hydrodynamic behavior in rapidly expanding systems \cite{Florkowski:2013lza,Florkowski:2013lya,Florkowski:2014sfa,Denicol:2014xca,Denicol:2014tha,Strickland:2015utc,Alalawi:2022pmg}.  Despite the seemingly rudimentary nature of the relaxation time approach and the non-conformal nature of the QGP, the conclusions drawn are expected to possess a semi-universal character \cite{Du:2023bwi}. Fundamentally, the behavior seen reflects the interplay between free streaming and dissipative dynamics in an approximately conformal system undergoing rapid longitudinal expansion. As discussed below, simple conformal kinetics captures the scaling behavior anticipated from QCD at weak coupling \cite{Baier:2000sb} and the scaling properties exhibited by strongly coupled conformal theories based on the AdS/CFT correspondence \cite{Heller:2011ju}. While the initial focus here is on conformal systems for simplicity, it is worth noting recent evidence indicating the existence of hydrodynamic attractors for certain moments of the one-particle distribution function in non-conformal systems \cite{Chattopadhyay:2021ive,Jaiswal:2021uvv,Alalawi:2022pmg,Jaiswal:2022udf}.  I will discuss these most recent advances in detail later in this chapter.
To start with, I will review our understanding of attractors within dissipative hydrodynamics, focusing first on the conformal dynamics of systems subject to 0+1-dimensional boost-invariant Bjorken flow.

\subsection{Mueller-Israel-Stewart type theories}
\label{sect:mis}

We start by assuming a 0+1-dimensional system (0+1D), which is transversally homogeneous and boost-invariant, following the framework presented in Ref.~\cite{Bjorken:1982qr}. Consequently, all variables depend solely on the longitudinal proper time, denoted as \( \tau = \sqrt{t^2-z^2} \). The metric is considered ``mostly minus'' with spacetime coordinates \( x^\mu = (t, x, y, z) \), and the line element is given by \( ds^2=g_{\mu\nu} dx^\mu dx^\nu=dt^2-dx^2-dy^2-dz^2 \), where $g^{\mu\nu}$ is the metric tensor in Minkowski space.  The longitudinal spacetime rapidity, denoted as \( \varsigma \), is given by \( \varsigma = \tanh^{-1} \left(\frac{z}{t}\right) \). Assuming conformality, the system has an equation of state stemming from \( N_{\rm dof} \) massless degrees of freedom, which is Landau-matched to the general non-equilibrium energy density. Within this framework, it follows that \( \varepsilon = \varepsilon_0(T) = 3 P_0(T) \), and the temperature \( T \) is expressed as \( \gamma \varepsilon^{1/4} \), with \( \gamma \) being proportional to \( N_{\rm dof}^{-1/4} \). Additionally, in the context of a longitudinally boost-invariant system, the flow velocity is represented as \( u^\mu = (\cosh\varsigma,0,0,\sinh\varsigma) \) (Bjorken flow).

In this section, we will employ kinetic theory to derive the dynamical attractors for anisotropic hydrodynamics (aHydro) and second-order viscous hydrodynamics (vHydro). To achieve this, we begin with the Boltzmann equation within the relaxation time approximation (RTA) \cite{anderson1974relativistic}
\be
p^\mu\partial_\mu f = - \frac{p_\mu u^\mu}{\tau_{\rm eq}} \left( f - f_{\rm eq}\right) \, .
\label{BoltzmannRTA}
\ee
The momentum- and energy-independent relaxation time \(\tau_{\rm eq}\) is defined as \(\tau_{\rm eq} = 5\eta/sT \) \cite{Denicol:2010xn,Denicol:2011fa}, where \(\eta\) represents shear viscosity, \(T\) is the local effective temperature obtained through Landau matching, and \(s\) is the entropy density.  For simplicity, in this section I will assume a classical Boltzmann distribution for the underlying thermal distribution function.

In kinetic theory the covariantly conserved energy-momentum tensor is given by
\be
T^{\mu\nu} = N_{\textrm{dof}}\int dP \,p^{\mu} p^\nu \,f \, ,
\ee
with $\int dP = \int d^3{\vec p}/(2\pi)^3E$ being the appropriate Lorentz invariant integration measure \cite{anderson1974relativistic}. The local energy density is obtained via $\varepsilon = u_\mu u_\nu T^{\mu\nu}$ and the shear stress tensor is obtained by projecting with a transverse projector
\be
\pi^{\mu\nu} = \Delta^{\mu\nu}_{\alpha\beta}T^{\alpha\beta},
\ee
where $\Delta^{\mu\nu}_{\alpha\beta} = \left(\Delta^{\mu}_\alpha\Delta^{\nu}_\beta + \Delta^{\mu}_\beta\Delta^{\nu}_\alpha\right)/2 - \Delta^{\mu\nu}\Delta_{\alpha\beta}/3$ is the tensor projector orthogonal to the flow constructed using $\Delta^{\mu\nu} = g^{\mu\nu}-u^\mu u^\nu$.

Bjorken symmetry and conformal invariance allow for the reduction of the energy-momentum conservation laws derived from the first moment of the Boltzmann equation to a single dynamical equation
\be
\tau \frac{d\log \varepsilon}{d\tau}  = -\frac{4}{3} + \frac{\pi}{\varepsilon} \, ,
\label{eq:firstmom}
\ee 
involving the energy density and $\pi \equiv \pi^\varsigma_\varsigma$.  In second-order hydrodynamic frameworks such as the Mueller-Israel-Stewart (MIS) \cite{Muller:1967zza,Israel:1976tn,Israel:1979wp} and Denicol-Niemi-Molnar-Rischke (DNMR) \cite{Denicol:2012cn,Denicol:2014loa} approaches, the 14-moment approximation for the single-particle distribution function is employed to derive a differential equation for the evolution of \(\pi\). For a 0+1D system undergoing Bjorken expansion, the dynamical for \(\pi\) equation takes the following form
\be
\dot\pi = \frac{4\eta}{3\tau\tau_\pi} - \beta_{\pi\pi}\frac{\pi}{\tau} - \frac{\pi}{\tau_\pi} \, ,
\label{2ndorderhydro}
\ee
where $\,\dot{}\,=d/d\tau$.  In RTA, $\beta_{\pi\pi} = 38/21$ and $\tau_\pi = \tau_{\rm eq}$ in the complete second order calculation of DNMR \cite{Denicol:2010xn,Denicol:2012cn,Denicol:2014loa,Jaiswal:2013vta,Jaiswal:2013npa}, while in MIS $\beta_{\pi\pi} = 4/3$ and $\tau_\pi = 6\tau_{\rm eq}/5$ \cite{Muronga:2003ta}.\footnote{I emphasize that the DNMR results are the correct values resulting from a complete treatment, while the MIS equations discard certain terms.  Despite this, MIS is frequently used as a standard reference point for discussions.}  Solving Eqs.~\eqref{eq:firstmom} and \eqref{2ndorderhydro} allows one to determine the dynamical evolution of a viscous fluid described by second-order hydrodynamics. From these simple equations we can understand the emergence of hydrodynamic attractor behavior, as demonstrated in Ref.~\cite{Heller:2015dha}.

\subsection{0+1D conformal anisotropic hydrodynamics} 
\label{sect:aHydrosection}   

Later in this chapter we will discuss the general 3+1D non-conformal aHydro formalism.  Here we only present the formalism for a 0+1D conformal system.   In the 0+1D case, aHydro requires only one anisotropy direction and parameter, $\hat{\vec n}$ and $\xi$, respectively.  This leads to a distribution function of the form \cite{Romatschke:2003ms,Romatschke:2004jh}
\be
f(\tau,{\vec x},{\vec p}) = f_{\rm eq} \!\left( \frac{1}{\Lambda(\tau,{\vec x})} \sqrt{ p_T^2 + [1+\xi(\tau,{\vec x})] p_L^2 } \right) ,
\ee
where $\Lambda$ can be interpreted as the local ``transverse temperature'' field and $-1 < \xi < \infty$ is the anisotropy field.  For a 0+1D system boost-invariant system, both of these fields only depend on the longitudinal proper time.
Using this form, the integrals defining the conformal energy density, transverse pressure, and longitudinal pressure factorize, resulting in
\ba
\varepsilon &=& {\cal R}(\xi) \varepsilon_0(\Lambda) \, , \nonumber \\
{\cal P}_T &=& {\cal R}_T(\xi) P_0(\Lambda) \, , \nonumber \\
{\cal P}_L &=& {\cal R}_L(\xi) P_0(\Lambda) \, , \nonumber
\ea
with \cite{Rebhan:2008uj,Martinez:2010sc}
\ba
{\cal R}(\xi) &=& \frac{1}{2}\left[\frac{1}{1+\xi}
+\frac{\arctan\sqrt{\xi}}{\sqrt{\xi}} \right] ,
\label{eq:rfunc}
\\
{\cal R}_{T}(\xi) &=& \frac{3}{2 \xi} 
\left[ \frac{1+(\xi^2-1){\cal R}(\xi)}{\xi + 1}\right] ,
\\
{\cal R}_{L}(\xi) &=& \frac{3}{\xi} 
\left[ \frac{(\xi+1){\cal R}(\xi)-1}{\xi+1}\right] ,
\ea
which satisfy $3{\cal R} = 2 {\cal R}_T + {\cal R}_L$ (the conformal isotropic pressure is $P_0 = \varepsilon/3$).  In these expressions, \(L\) and \(T\) represent the directions parallel and perpendicular to \(\hat{\vec n}\), respectively. Conventionally, the anisotropy direction is aligned along the beam line direction in heavy-ion applications (\(\hat{\vec n} = \hat{\vec z}\)). Employing Landau matching, one obtains \(\varepsilon = \varepsilon_0(T)\).  For a conformal system with a momentum- and energy-independent relaxation time, this leads to
\be
T = {\cal R}^{1/4}(\xi) \Lambda \, .
\label{defineLambda}
\ee
To dynamically evolve the system, we need an equation of motion for $\xi$, since $\Lambda$ is already connected to the temperature (energy density) using Eq.~\eqref{defineLambda}.

To do this, we will employ the following moment of the Boltzmann distribution \cite{Nopoush:2014pfa}
\be
I^{\mu\nu\lambda} = N_{\textrm{dof}}\int dP\, p^{\mu}p^\nu p^\lambda\,f \, ,
\ee
which will be important for the aHydro approach. Using the Boltzmann equation in RTA \eqref{BoltzmannRTA}, the equation of motion for this moment is
\be
\partial_\alpha I^{\alpha\mu\nu} = \frac{1}{\tau_{\rm eq}} ( u_\alpha I^{\alpha\mu\nu}_{\rm eq} - u_\alpha I^{\alpha\mu\nu} ) \, .
\label{eq:secondmom1}
\ee
We note that $I^{\mu\nu\lambda}$ is symmetric with respect to interchanges of $\mu$, $\nu$, and $\lambda$ and traceless in any pair of indices (massless particles/conformal invariance).  In an isotropic system, one finds $I_{xxx} = I _{yyy} = I_{zzz} = I_0$ with
\be
I_0(\Lambda) = \frac{4 N_{\rm dof}}{\pi^2} \Lambda^5 \, .
\ee
Using the aHydro form for the one-particle distribution function, one obtains
\ba
I_{uuu} &=& \frac{3+2\xi}{(1+\xi)^{3/2}} I_0(\Lambda) \, , \nonumber \\
I_{xxx} = I_{yyy} &=& \frac{1}{\sqrt{1+\xi}} I_0(\Lambda) \, , \nonumber \\
I_{zzz} &=& \frac{1}{(1+\xi)^{3/2}} I_0(\Lambda) \, ,
\ea
with, e.g. $I_{uuu} \equiv u_\mu u_\nu u_\lambda I^{\mu\nu\lambda}$, etc.  Taking the $zz$ projection of Eq.~(\ref{eq:secondmom1}) minus one-third of the sum of its $xx$, $yy$, and $zz$ projections gives our second equation of motion~\cite{Tinti:2013vba} 
\be
\frac{1}{1+\xi} \dot\xi - \frac{2}{\tau} + \frac{{\cal R}^{5/4}(\xi)}{\tau_{\rm eq}} \xi \sqrt{1+\xi} = 0\, ,
\label{eq:2ndmomf}
\ee
which can be used to obtain the proper-time evolution of the anisotropy parameter. 

\subsubsection{Connection with shear stress tensor and the inverse Reynolds number}

\begin{figure}[t!]
\centerline{
\includegraphics[width=.45\linewidth]{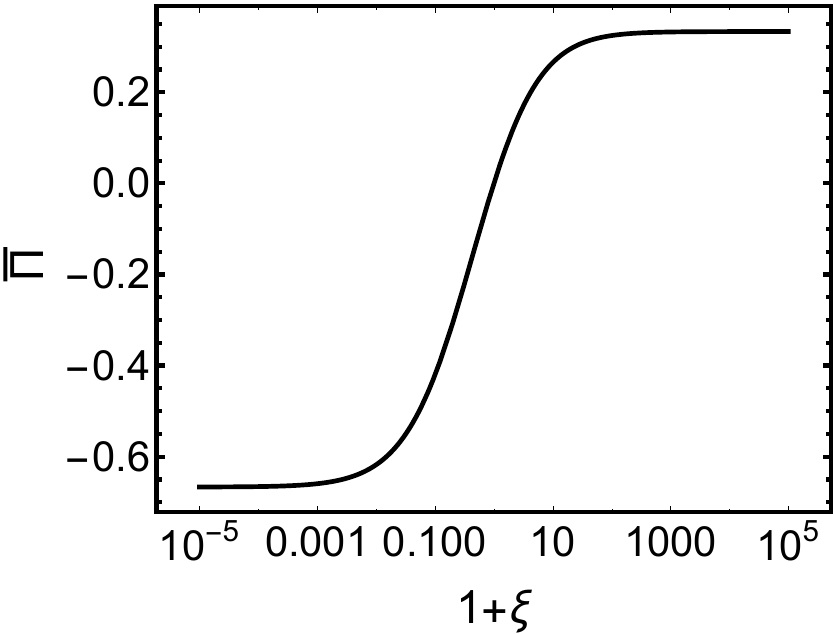}
\hspace{5mm}
\includegraphics[width=.45\linewidth]{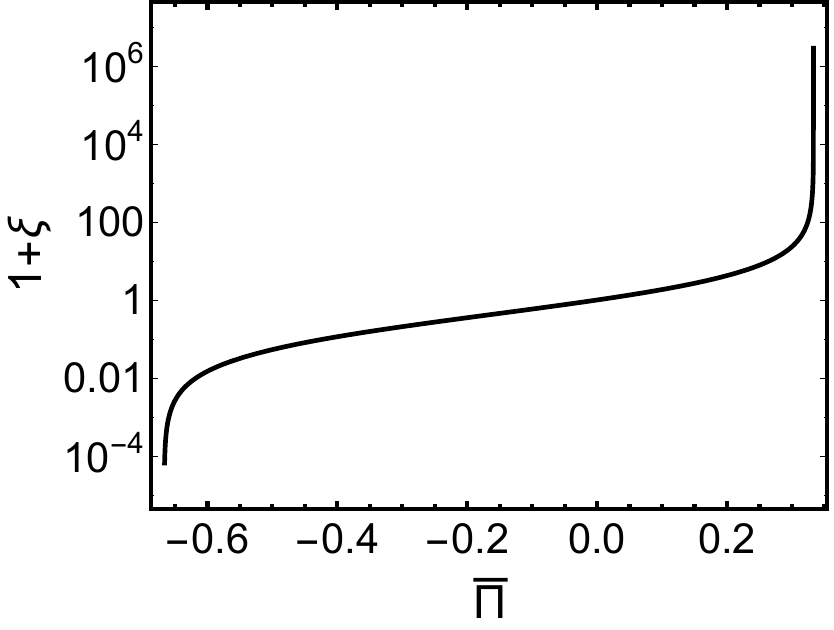}
}
\caption{The left panel shows $\overline{\pi}$ as a function of $\xi$ determined via Eq.~(\ref{eq:pixirel}).  The right panel shows $\xi$ as a function of $\overline{\pi}$ determined via numerical inversion of Eq.~(\ref{eq:pixirel}).}
\label{fig:pibar}
\end{figure}

To enable a more transparent comparison between the equations of motion for aHydro and those of standard viscous hydrodynamics, one can make a change of variables in Eq.~(\ref{eq:2ndmomf}) to express it in terms of the shear stress tensor component \(\pi\). This can be achieved by using \(\pi = P_0 - {\cal P}_L\), yielding
\be
\overline\pi(\xi) \equiv \frac{\pi}{\varepsilon} = \frac{1}{3} \left[ 1 - \frac{{\cal R}_L(\xi)}{\cal R(\xi)} \right]  .
\label{eq:pixirel}
\ee 
In the left panel of Fig.\ \ref{fig:pibar}, we present the dependence of $\overline{\pi}$ on $\xi$, determined by Eq.~(\ref{eq:pixirel}). In the right panel, we present the dependence of $\xi$ on $\overline{\pi}$ as obtained through the numerical inversion of Eq.~(\ref{eq:pixirel}). It is important to note that in aHydro, $\overline\pi$ is constrained within $-2/3 < \overline{\pi} < 1/3$, which is a restriction associated with the naturally occuring positivity of longitudinal and transverse pressures in aHydro. Next we note that, for a 0+1D boost-invariant system, the magnitude of $\overline\pi$ is proportional to the inverse Reynolds number
\be
R_\pi^{-1} = \frac{\sqrt{\pi^{\mu\nu} \pi_{\mu\nu}}}{P_0} = 3 \sqrt{\frac{3}{2}} |\overline\pi| \, .
\label{eq:reynoldsnumber}
\ee
As a consequence, a series in $\overline\pi$ can be understood as an expansion in $R_\pi^{-1}$.

We will also need the relation between the time derivatives of $\pi$ and $\xi$.  This relation can be obtained from Eq.~(\ref{eq:pixirel})
\be
\frac{\dot\pi}{\varepsilon} = \overline\pi^\prime \dot\xi + \overline\pi \partial_\tau \!\log\varepsilon \, ,
\ee
which upon using Eqs.~(\ref{eq:pixirel}) and (\ref{eq:firstmom}) gives
\be
\dot\xi = \frac{1}{\overline\pi'} \left[ \frac{\dot\pi}{\varepsilon} + \frac{\pi}{\varepsilon\tau} \left( \frac{4}{3} - \frac{\pi}{\varepsilon}  \right)  \right] ,
\label{eq:xidot2}
\ee
where $\overline\pi' \equiv d\overline\pi/d\xi$.

Plugging (\ref{eq:xidot2}) into (\ref{eq:2ndmomf}), one obtains
\be
\frac{\dot\pi}{\varepsilon} + \frac{\pi}{\varepsilon\tau} \left( \frac{4}{3} - \frac{\pi}{\varepsilon}  \right) - \left[ \frac{2(1+\xi)}{\tau} - \frac{{\cal H}(\xi)}{\tau_{\rm eq}} \right]\overline\pi'(\xi) = 0\, ,
\label{eq:2ndmomf3}
\ee
with
\be
{\cal H}(\xi) \equiv \xi (1+\xi)^{3/2}{\cal R}^{5/4}(\xi) \, ,
\ee
and the understanding that $\xi = \xi(\overline\pi)$ with $\xi(\overline\pi)$ being the inverse function of $\overline\pi(\xi)$ (shown in the right panel of figure \ref{fig:pibar}).  Written in this form, we can see explicitly that the aHydro second-moment equation sums an infinite number of terms in the expansion in the inverse Reynolds number (\ref{eq:reynoldsnumber}).  In the next section we will expand this equation in powers of the inverse Reynolds number through second order in order to compare it to standard viscous hydrodynamics.

\subsubsection{Small \texorpdfstring{$\xi$}{xi} expansion}

To establish the final connection to standard viscous hydrodynamics, one can expand Eq.~(\ref{eq:2ndmomf3}) in terms of \(\xi\) around \(\xi=0\). To accomplish this, we need the \(\xi\) expansions of the various functions in order to construct an explicit inversion and rewrite the equations exclusively in terms of \(\overline\pi\). At the second order in \(\xi\), the expression is
\ba
\overline\pi &=& \frac{8}{45} \xi  \left[1 - \frac{13}{21} \xi + {\cal O}(\xi^2) \right] , \nonumber \\
{\cal H} &=& \xi + \frac{2}{3} \xi^2 + {\cal O}(\xi^3) \, .
\ea
Inverting the relationship between $\overline\pi$ and $\xi$ to second-order in $\overline\pi$ gives
\be
\xi = \frac{45}{8} \overline\pi \left[ 1 + \frac{195}{56} \overline\pi + {\cal O}(\pi^2)  \right] , 
\ee
which results in
\ba
\overline\pi^\prime  &=& \frac{8}{45} - \frac{26}{21} \overline\pi + \frac{1061}{392} \overline\pi^2 + {\cal O}(\overline\pi^3) \, , \nonumber \\
{\cal H} &=& \frac{45}{8} \overline\pi \left[ 1 + \frac{405}{56} \overline\pi + {\cal O}(\overline\pi^3) \right] , 
\ea

Applying this to the equation of motion (\ref{eq:2ndmomf3}) and keeping only terms through $\pi^2$ gives
\be
\dot\pi - \frac{4 \eta}{3 \tau_\pi \tau} + \frac{38}{21} \frac{\pi}{\tau} - \frac{36\tau_\pi}{245\eta} \frac{\pi^2}{\tau}
= - \frac{\pi}{\tau_\pi} - \frac{15}{56} \frac{\pi^2}{\tau_\pi \varepsilon} \, , %
\label{eq:2ndmomf4}
\ee
where, on the left-hand side, we employed the fact that the energy density can be eliminated by expressing it in terms of the transport coefficients as
\be
\varepsilon = \frac{15}{4} \frac{\eta}{\tau_{\rm eq}} \, ,
\ee
and we have relabeled $\tau_{\rm eq} \rightarrow \tau_\pi$ in order cast the equations in second order viscous hydrodynamics form.  Note that, to linear order in $\pi$, Eq.~(\ref{eq:2ndmomf4}) agrees with the previously obtained RTA second-order viscous hydrodynamics results \cite{Denicol:2010xn,Denicol:2012cn,Denicol:2014loa,Jaiswal:2013vta,Jaiswal:2013npa}.  However, one can go beyond this truncation by extending the power series expansions used above or simply solving the non-linear equation of motion \eqref{eq:2ndmomf3} numerically without Taylor expansion \cite{Strickland:2017kux}.

\subsection{Comparison of attractors in viscous and anisotropic hydrodynamics models}
\label{sec:attractorvars}

In this subsection, we will explore the hydrodynamic attractor behavior of aHydro and contrast it with the corresponding outcomes in MIS and DNMR theories. In each of these cases, the dynamics of the system is governed by solving the differential equations for \(\varepsilon\) and \(\pi\). However, to establish a connection with the literature, we adopt the methodology introduced in \cite{Heller:2015dha} and introduce the dimensionless ``time'' variable.
\be
w \equiv \tau T(\tau) \, ,
\ee
with which one may define the {\em amplitude} 
\be
\varphi(w) \equiv \tau \frac{\dot w}{w} = 1 + \frac{\tau}{4} \partial_\tau\!\log \varepsilon \, ,
\label{wdot1}
\ee
which is related to the shear correction $\pi$ as follows
\be
\frac{\pi}{\varepsilon} = 4\left(\varphi -\frac{2}{3}\right). 
\ee 
As a consequence, a solution for the proper-time evolution of the energy density uniquely determines the dependence of the amplitude \(\varphi\) on \(w\). Additionally, it is important to note that the positive energy condition imposes a bound on \(\varphi\) within the region \(0 \leq \varphi \leq 1\)  \cite{Janik:2005zt}.

\subsubsection{The hydrodynamic attractor in the DNMR framework}

The change of variables from $\{\varepsilon,\pi\} \to \{w,\varphi\}$ is convenient because it allows one to express the coupled set of first-order ODEs for $\{\varepsilon,\pi\} $ in terms of a single first-order ODE for $\varphi(w)$ \cite{Heller:2015dha}. In the case of MIS and DNMR, this procedure gives
\be
c_\pi w \varphi \varphi' + 4 c_\pi \varphi^2 + \left[ w + \left( \beta_{\pi\pi} - \frac{20}{3} \right) c_\pi \right] \varphi - \frac{4 c_\eta}{9} -\frac{2c_\pi}{3} ( \beta_{\pi\pi} - 4) - \frac{2w}{3}  = 0 \, ,
\label{attractor2ndorder} 
\ee
where $\varphi' = d\varphi(w)/dw$, $c_\pi \equiv \tau_\pi T$, and $c_\eta = \eta/s$ (with $c_\pi = 5 c_\eta$ in the cases considered here). After defining the rescaled variable $\overline{w} = w/c_\pi$ one obtains
\be
\overline{w} \varphi \varphi' + 4 \varphi^2 + \left[ \overline{w} + \left( \beta_{\pi\pi} - \frac{20}{3} \right)  \right] \varphi - \frac{4 c_{\eta/\pi}}{9} -\frac{2}{3} ( \beta_{\pi\pi} - 4) - \frac{2\overline{w}}{3}  = 0\,.
\label{attractor2ndordernew} 
\ee
The form above makes it clear that the solution only depends on the ratio $c_{\eta/\pi} \equiv c_\eta/c_\pi = (\eta/s)/(\tau_\pi T)$ and the value chosen for $\beta_{\pi\pi}$. To connect these equations with the RTA Boltzmann one must set $c_{\eta/\pi}=1/5$. 

Using the MIS value $\beta_{\pi\pi}=4/3$ one obtains
\be
\overline{w} \varphi \varphi' + 4 \varphi^2 + \left( \overline{w} - \frac{16}{3} \right) \varphi - \frac{4 c_{\eta/\pi}}{9} + \frac{16}{9} - \frac{2\overline{w}}{3}  = 0 \, ,
\label{eq:MISeq}
\ee
which agrees with Eq.~(9) of Ref.\ \cite{Heller:2015dha}; however, for RTA this value for $\beta_{\pi\pi}$ is incorrect. Using the correct value for $\beta_{\pi\pi}=38/21$ obtained using the complete second order formalism of DNMR and again neglecting quadratic terms in $\pi$ one obtains\footnote{As demonstrated in Eq.~(\ref{eq:2ndmomf4}), aHydro naturally reproduces this equation when truncated at leading order in $\xi$ (linear order in the inverse Reynolds number).}
\be
\overline{w} \varphi \varphi' + 4 \varphi^2 + \left(\overline{w} - \frac{34}{7} \right) \varphi - \frac{4 c_{\eta/\pi}}{9} + \frac{92}{63} - \frac{2\overline{w}}{3}  = 0 \, .
\label{eq:DNMReq}
\ee

Following Ref.~\cite{Heller:2015dha}, the underlying dynamical attractor can be deduced from Eq.\ \eqref{attractor2ndorder} through a procedure similar to the slow-roll expansion in cosmology \cite{Liddle:1994dx}. This process can be described as follows: a small parameter \(\delta\) is formally introduced as a pre-factor in the term \( \overline{w} \varphi \varphi'\) in \eqref{attractor2ndordernew}. It is assumed that the solution of the differential equation \(\varphi(\overline{w};\delta)\) can be expressed as a power series expansion \(\varphi(\overline{w};\delta) = \varphi_0(\overline{w})+\varphi_1(\overline{w})\,\delta + \mathcal{O}(\delta^2)\). After considering all orders, one may then take the limit \(\delta \to 1\). The $0^{\rm th}$-order truncation is obtained by solving the simple quadratic equation
\be
4  \varphi_0^2 + \left[ \overline{w} + \left( \beta_{\pi\pi} - \frac{20}{3} \right) \right] \varphi_0 - \frac{4 c_{\eta/\pi}}{9} -\frac{2}{3} ( \beta_{\pi\pi} - 4) - \frac{2\overline{w}}{3}  = 0 \, .
\label{attractor2ndorder0order} 
\ee
Out of the two possible solutions, only one is stable and remains finite in the large $\overline{w}$ limit and is
\be
\varphi_0(\overline{w}) = \frac{1}{24} \left(-3 \beta_{\pi\pi} +\sqrt{64 c_{\eta/\pi}+(3
   \beta_{\pi\pi} +3 \overline{w}-4)^2}-3 \overline{w}+20\right).
   \label{attractor2ndordersol}
\ee
It is possible to compute higher order corrections in the slow-roll expansion, however, it is better to define a numerically determined attractor solution using the boundary condition  \mbox{$\lim_{\overline{w} \rightarrow 0} \overline{w} \varphi \varphi' = 0$}, which then implies that \cite{Heller:2015dha}
\be
\lim_{\overline{w}\to 0}\varphi(\overline{w})=\frac{1}{24} \left(-3 \beta_{\pi\pi} +\sqrt{64 c_{\eta/\pi}+(3
   \beta_{\pi\pi} -4)^2}+20\right).
\ee 
Numerical solution of Eq.~\eqref{attractor2ndordernew} subject to this boundary condition allows one to construct the attractor without having to resort to the slow-roll expansion.   In the next section, we generalize this analysis to determine the dynamical attractor in aHydro. 

\subsection{The hydrodynamic attractor in the aHydro framework}

I now discuss how to obtain the hydrodynamic attractor in aHydro. This can be achieved by combining the two first-order aHydro differential equations into a single second-order differential equation written in terms of \(\varphi\) and \(w\). To obtain the necessary aHydro dynamical equation, one can use the following identity
\be
w \varphi \varphi' = -\frac{8}{3} + \frac{20}{3}\varphi - 4\varphi^2 + \frac{\tau}{4}\frac{\dot\pi}{\varepsilon}
\label{eq:finalfirstmom}
\ee
and \eqref{eq:2ndmomf3}.  We first express  Eq.~(\ref{eq:2ndmomf3}) in terms of $\varphi$ and $w$ and then using $\tau\partial_\tau\!\log \varepsilon = 4(\varphi - 1)=-4/3+\pi/\varepsilon$, one finds 
\be
\frac{\tau}{4} \frac{\dot\pi}{\varepsilon} =  \frac{8}{3} - \frac{20}{3} \varphi + 4 \varphi^2 + \left[ \frac{1}{2} (1+\xi)  - \frac{w}{4 c_\pi} {\cal H} \right] \overline\pi' \,.
\ee
Inserting this into Eq.~(\ref{eq:finalfirstmom}) and converting to $\overline{w}$ gives our final result for the aHydro attractor equation
\be
\overline{w} { \varphi} \frac{\partial \varphi}{\partial \overline{w}}  = \left[ \frac{1}{2} (1+\xi) - \frac{\overline{w}}{4} {\cal H} \right] \overline\pi' \, .
\label{eq:ahydroattractoreq2}
\ee
Note that above $\xi = \xi(\overline\pi) = \xi(4\varphi - 8/3)$ and likewise for $\overline\pi'$.  It is worth noting that the aHydro equation above does not contain $c_{\eta/\pi}$ since it cancels explicitly. The aHydro attractor solution remains universal when plotted as a function of $\overline{w}$. This universality holds true for second-order viscous hydrodynamic approximations as well.

Eq.~\eqref{eq:ahydroattractoreq2}, when expanded in gradients (powers of $1/\overline{w}$ here), has zero radius of convergence \cite{Florkowski:2016zsi}. Thus, the numerical solution of the differential equation \eqref{eq:ahydroattractoreq2} may be considered an all-orders resummation of the gradient series, as in MIS theory \cite{Heller:2015dha}.  However, it is important to highlight that the right-hand side of Eq.~(\ref{eq:ahydroattractoreq2}) involves a sum of an infinite number of terms in the inverse Reynolds number. This conceptual difference sets it apart from DNMR, which derived their equations of motion assuming a perturbative series in $R_\pi^{-1}$.

In the case of aHydro, solving even the $0^{\rm th}$-order approximation in the slow-roll expansion requires a numerical approach. Hence, we directly proceed to solve the differential equation numerically. Once again, for this purpose, the attractor solution is determined by imposing the same boundary condition as before at $\overline{w}=0$. Utilizing the numerical solution of the slow-roll equation, one finds that in aHydro the attractor boundary condition is
\be
\lim_{\overline{w} \rightarrow 0} \varphi(\overline{w}) =  \frac{3}{4} \, .
\label{eq:ahydrobc}
\ee
With this boundary condition, we simply numerically solve Eq.~(\ref{eq:ahydroattractoreq2}).  Note that the boundary condition above guarantees the positivity of the longitudinal pressure of the attractor solution as $\overline{w} \rightarrow 0$.

\begin{figure}[t!]
\centerline{
\includegraphics[width=0.48\linewidth]{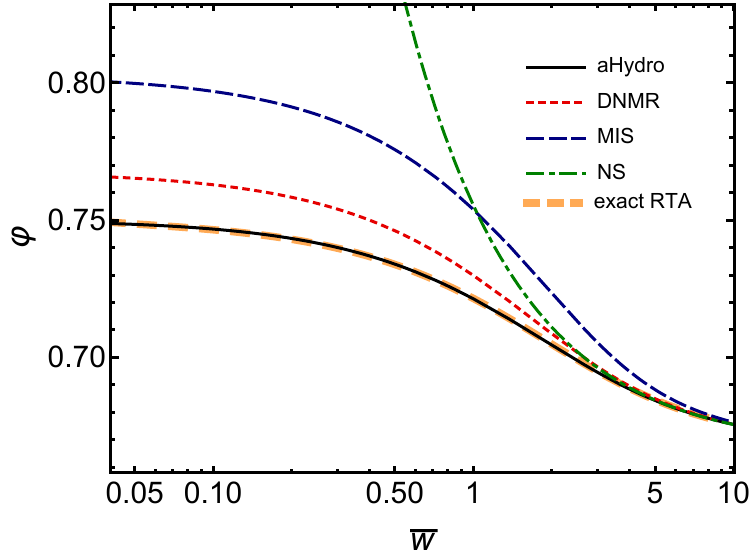} \,
\includegraphics[width=0.48\linewidth]{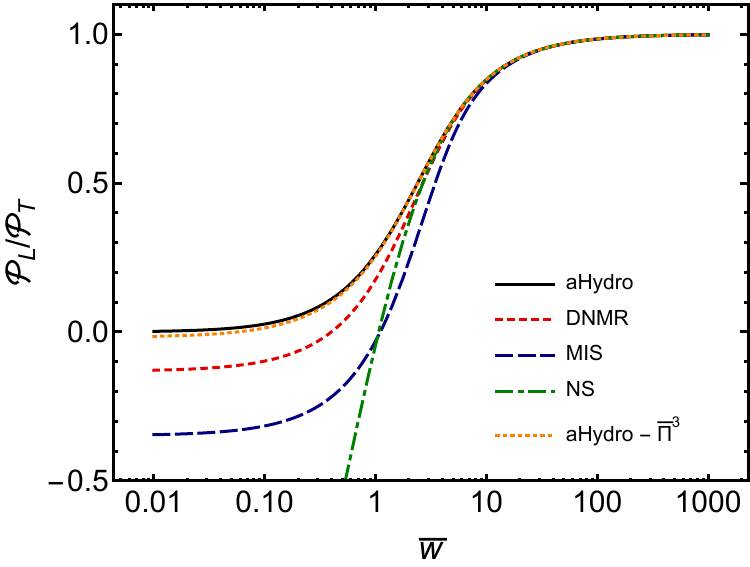}
}
\caption{aHydro, MIS, and DNMR attractors compared to the attractor obtained from exact solution to the RTA Boltzmann equation.  The left panel shows the variable $\varphi$ while the right panel shows the pressure ratio.}
\label{fig:attractor_2_compare}
\end{figure}

In the left panel of Fig.\ \ref{fig:attractor_2_compare} we compare the aHydro, MIS, and DNMR attractors to the corresponding quantity obtained from the exact solution to the 0+1d RTA Boltzmann equation~\cite{Florkowski:2013lza,Florkowski:2013lya}.  We also include a curve showing the Navier-Stokes (NS) result~\cite{Heller:2015dha}
\be
\varphi_{\rm NS} = \frac{2}{3} + \frac{4}{9} \frac{c_{\eta/\pi}}{\overline{w}} \, .
\ee
As the left panel of Fig.~\ref{fig:attractor_2_compare} demonstrates, the aHydro attractor solution closely resembles the exact RTA attractor \cite{Florkowski:2013lya,Florkowski:2013lza,Romatschke:2017vte,Strickland:2018ayk}. Given that aHydro involves resummation not only in the Knudsen number but also in the inverse Reynolds number, the remarkable agreement with the exact kinetic theory result suggests that the inverse Reynolds number resummation is critical. This observation could serve as a guide for developing alternative approaches to far-from-equilibrium hydrodynamics that do not depend on perturbative expansions in both the Knudsen and the inverse Reynolds number.  Finally, in the right panel of Fig.~\ref{fig:attractor_2_compare}, I  present the attractor solution for the pressure ratio which can be obtained using
\be
\frac{{\cal P}_L}{{\cal P}_T} = \frac{3 - 4 \varphi}{2 \varphi - 1} \, .
\ee
In this figure we have also included a result from Ref.~\cite{Strickland:2017kux} for the aHydro equations truncated at third order in the inverse Reynolds number.  As can be seen from this figure, only the resummed aHydro attractor gives positive longitudinal pressure for all $\overline w$.

\begin{figure}[t!]
\centerline{
\includegraphics[width=0.48\linewidth]{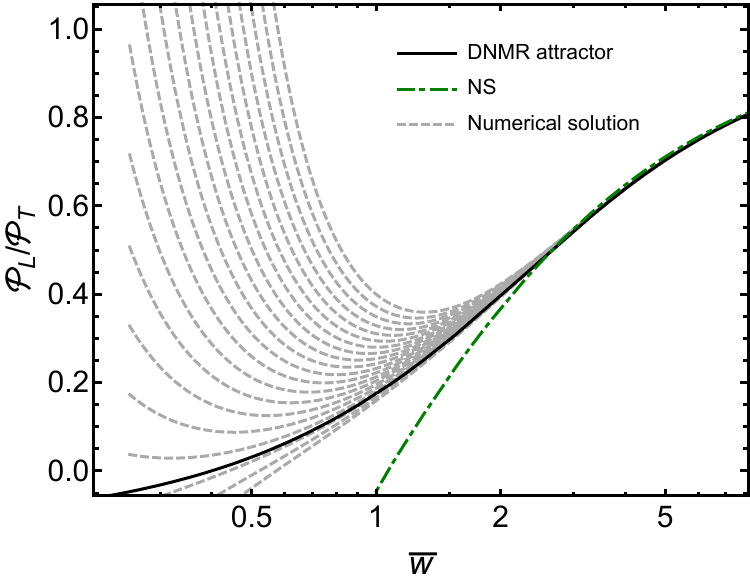} \,
\includegraphics[width=0.48\linewidth]{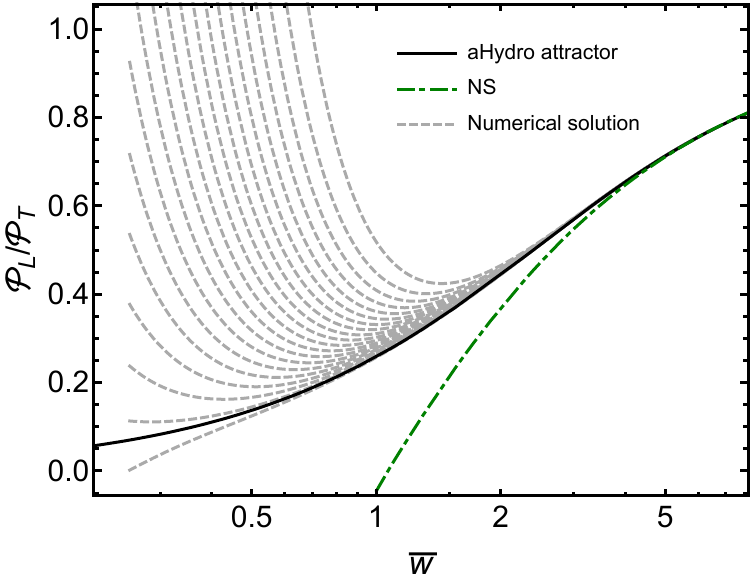}
}
\caption{(Left) The DNMR ${\cal P}_L/{\cal P}_T$ attractor (solid black line) and numerical solutions (grey dashed lines) corresponding to a variety of initial conditions for ${\cal P}_L/{\cal P}_T$. (Right) The aHydro ${\cal P}_L/{\cal P}_T$ attractor (solid black line) and numerical solutions (grey dashed lines) corresponding to a variety of initial conditions for ${\cal P}_L/{\cal P}_T$. }
\label{fig:attractor_3_compare}
\end{figure}

In Fig.~\ref{fig:attractor_3_compare}, I show the DNMR and aHydro attractors expressed in terms of ${\cal P}_L/{\cal P}_T$ along with specific numerical solutions corresponding to different assumed values of the initial plasma anisotropy ratio.  As can be seen from this figure, all numerical solutions collapse to their respective attractors for $\overline w \gtrsim 2$.  This indicates that information about the precise initial conditions for ${\cal P}_L/{\cal P}_T$ quickly becomes ignorable.  In this figure, we also see the feature that truncations in the inverse Reynolds number lead to attractors with negative longitudinal pressure at early times.

\section{Attractors in  non-equilibrium kinetic theory}
\label{sect:exactsolutions}

Having established the existence of hydrodynamic attractors in conformal 0+1D dissipative hydrodynamics, I now turn to a discussion of the existence of dynamical attractors in non-equilibrium kinetic theory.  I review the solutions obtained for both conformal (massless) \cite{Strickland:2018ayk} and non-conformal (massive) systems \cite{Chattopadhyay:2021ive,Jaiswal:2021uvv,Alalawi:2022pmg}.  In kinetic theory, one finds that attractors exist at very early times for a large set of moments of the Boltzmann equation implying the existence of an attractor for the entire one-particle distribution function~\cite{Strickland:2018ayk}.\footnote{For a recent analytic understanding of this, see Ref.~\cite{Aniceto:2024pyc}.}

\subsection{Non-conformal Boltzmann equation in relaxation time approximation}

We continue with the Boltzmann equation in relaxation time approximation (RTA)
\begin{equation}
 p^\mu \partial_\mu  f(x,p) =  C[ f(x,p)] \, , 
\label{kineq}
\end{equation}
where $f$ is the one-particle distribution function, $p^\mu$ is the particle four-momentum, and $C$ is the collision kernel 
\begin{eqnarray}
C[f] = \frac{p \cdot u}{\tau_{\rm eq}} \left( f_{\rm eq}-f \right) ,
\label{col-term}
\end{eqnarray}
with $u^\mu$ being the four-velocity of the local rest frame and $a \cdot b \equiv a^\mu b_\mu$.  The quantity $\tau_{\rm eq}$ appearing above is the relaxation time, which will be precisely specified below.  For the equilibrium distribution, we will follow Ref.~\cite{Florkowski:2014sfa} and assume a Boltzmann distribution\footnote{It is possible to investigate the emergence of attractors with underlying Fermi-Dirac or Bose statistics using the exact solution presented in Ref.~\cite{Florkowski:2014sda}.}
\begin{eqnarray}
f_{\rm eq} = \exp\left(- \frac{p \cdot u}{T} \right) .
\label{Boltzmann}
\end{eqnarray}
Herein, we will assume Bjorken flow, in which case in Milne coordinates one has $u^\tau=1$ and $u^{x,y,\varsigma}=0$, where $\tau $ is the longitudinal proper-time, $\tau = \sqrt{t^2-z^2}$, and $\varsigma$ is the spatial rapidity, $\varsigma = \tanh^{-1}(z/t)$.

\subsubsection{Thermodynamic variables}
\label{sect:thermo}

For a single-component massive gas obeying Boltzmann statistics, the equilibrium thermodynamic quantities are
\ba
{\mathpzc n} &=& \frac{T^3}{2 \pi^2} \, \hat{m}^2 K_2\left( \hat{m}\right) , \label{eq:neq} \nonumber \\
s &=& \frac{T^3}{2 \pi^2} \, \hat{m}^2 \Big[4K_2\left( \hat{m}\right)+\hat{m}K_1\left( \hat{m}\right)\Big] ,
\label{eq:Seq} \nonumber \\
\varepsilon &=& \frac{T^4}{2 \pi^2} \, \hat{m}^2
 \Big[ 3 K_{2}\left( \hat{m} \right) + \hat{m} K_{1} \left( \hat{m} \right) \Big]  \, , 
\label{eq:Eeq} \nonumber \\
 P &=& {\mathpzc n} T = \frac{T^4}{2 \pi^2} \, \hat{m}^2 K_2\left( \hat{m}\right)  \, ,
\label{eq:Peq}
\ea
with $\hat{m} \equiv m/T$ and $K_n$ being modified Bessel functions of the second kind.  Above $\mathpzc n$ is the number density, $s$ is the entropy density, $\varepsilon$ is the energy density, and $P$ is the pressure.  These satisfy $\varepsilon + P = Ts$ and, from the above relations, one can determine the speed of sound squared
\be
c_s^2 = \frac{dP}{d\varepsilon} = \frac{\varepsilon+P}{3\varepsilon+(3+\hat{m}^2)P} \, .
\ee

\subsubsection{Relaxation time for a massive gas}
\label{sect:relaxationtime}

\begin{figure}[t!]
\centerline{\includegraphics[width=0.5\linewidth]{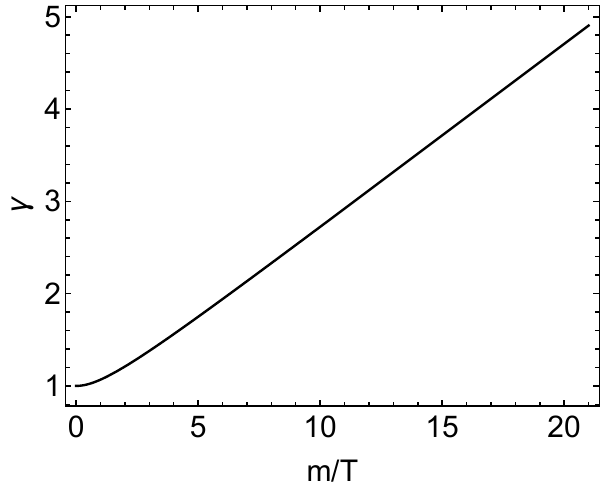}}

\caption{The non-conformal relaxation time modification factor $\gamma$ \eqref{eq:gammadef} as a function of $m/T$. }
\label{fig:teqfac}
\end{figure}

For a massive system, the shear viscosity $\eta$ can be expressed as \cite{anderson1974relativistic,Czyz:1986mr,Alqahtani:2015qja}
\be
\eta = \frac{\tau_{\rm eq} P}{15}\kappa(\hat{m})\, ,
\ee
with
\be 
\kappa(x)\equiv x^3 \bigg[\frac{3}{x^2}\frac{K_3(x)}{K_2(x)}-\frac{1}{x}+\frac{K_1(x)}{K_2(x)}-\frac{\pi}{2}\frac{1-xK_0(x)L_{-1}(x)-xK_1(x)L_0(x)}{K_2(x)}\bigg] \, ,
\ee
and $L_n(x)$ being modified Struve functions. 
For fixed specific shear viscosity, $\bar\eta \equiv \eta/s$, using $\varepsilon+P=Ts$ one obtains
\be 
\tau_{\rm eq}(T,m)=\frac{5 \bar{\eta}}{T} \gamma(\hat{m}) \, ,
\label{eq:teq2}
\ee
with 
\be
\gamma(\hat{m}) \equiv \frac{3}{\kappa(\hat{m})} \bigg(1+\frac{\varepsilon}{P}\bigg) \, .
\label{eq:gammadef}
\ee
Note that, in the massless limit, $m\rightarrow 0$, one has $\kappa(\hat{m})\rightarrow12$, $\varepsilon \rightarrow 3P$, and $\gamma \rightarrow 1$, giving the usual conformal RTA relaxation time 
\be
\tau_{\rm eq}(T,0)= \frac{5 \bar{\eta}}{T} \, .
\label{eq:teq0}
\ee
For small $\hat{m}$, one has
\be 
\gamma(\hat{m}) = 1+\frac{\hat{m}^2}{12}-\frac{13 \hat{m}^4}{288}+{\cal O}\!\left(\hat{m}^5\right),
\ee
and in the large $\hat{m}$ limit, one has
\be
\gamma(\hat{m}) = \frac{\hat{m}}{5} + \frac{7}{10} + {\cal O}\!\left(\frac{1}{\hat{m}}\right).
\ee
In Fig.~\ref{fig:teqfac}, I plot $\gamma(\hat{m})$.  As can be seen from this figure, $\gamma(\hat{m})$ goes to unity in the massless limit and grows linearly at large $m/T$, which corresponds either to fixed temperature and large mass or fixed mass and small temperature.  The fact that $\gamma(\hat{m}) \geq 1$ implies that a massive gas always relaxes more slowly to equilibrium than a massless one in physical units, however, it is unclear a priori how things will change as a function of the rescaled time $\wbar \equiv \tau/\tau_{\rm eq}$.  We note that the strong enhancement of the relaxation time at low temperatures modifies the asymptotic approach to equilibrium.

\subsubsection{Exact solution for the distribution function and its solution}
\label{sect:exactsolf}

Here I review the derivation of the exact solution presented in Ref.~\cite{Florkowski:2014sfa} and derive the integral equation obeyed by all moments of the distribution function.  I will also generalize the results contained in that reference to the full set of integral moments.  For details concerning the derivation of the exact solution in the non-conformal case, I refer the reader to Ref.~\cite{Alalawi:2022pmg}.

In the case of one-dimensional boost-invariant expansion (0+1D), all scalar quantities depend only on the longitudinal proper time $\tau$.  To describe boost-invariant 0+1D dynamics, one can introduce a spacelike vector $z^\mu$, which is orthogonal to the fluid four-velocity $u^\mu$ in all frames and corresponds to the z-direction in the local rest frame of the matter \cite{Ryblewski:2010ch,Martinez:2012tu}.  The requirement of boost invariance implies that $f(x,p)$ may depend only on three variables, $\tau$, $w$, and $\vec{p}_T$ \cite{Bialas:1984wv,Bialas:1987en}, with the boost-invariant variable $w$ defined by\footnote{In Eq.~\eqref{eq:w}, $z$ is the spatial coordinate, which is not to be confused with the basis vector $z^\mu$.}
\be
w =  t p_L - z E \, .
\label{eq:w}
\ee
Using $w$ and $\vec{p}_T$ one can define
\be
v = Et-p_L z = 
\sqrt{w^2+\left( m^2+\vec{p}_T^{\,\,2}\right) \tau^2} \, .
\label{eq:v}
\ee
Using these variables, one can write the energy and the longitudinal momentum as
\ba
E &=& \frac{vt+wz}{\tau^2} \, , \\
p_L &=& \frac{wt+vz}{\tau^2} \, ,
\label{eq:p0p3}
\ea
and the Lorentz-invariant momentum integration measure becomes
\begin{equation}
dP =   \frac{d^4p}{(2\pi)^4} \, 2\pi \delta \left( p^2-m^2\right) 2 \theta (p^0)
=\frac{dp_L}{(2\pi)^3p^0}d^2p_T =\frac{dw \, d^2p_T }{(2\pi)^3v}\, .  
\label{dP}
\end{equation}

When written in terms of these variables, the 0+1D RTA Boltzmann equation takes a particularly simple form~\cite{Florkowski:2013lza,Florkowski:2013lya,Florkowski:2014sfa}
\begin{eqnarray}
\frac{\partial f}{\partial \tau}  &=& 
\frac{f_{\rm eq}-f}{\tau_{\rm eq}} \, ,
\label{eq:simpleformrta}
\end{eqnarray} 
with $\tau_{\rm eq}$ specified in Eq.~\eqref{eq:teq2} and the equilibrium distribution function given by
\begin{eqnarray}
f_{\rm eq}(\tau,w,p_T) =
\exp\!\left[
- \frac{\sqrt{w^2+ (p_T^2+m^2)\tau^2}}{T \tau}  \right] .
\label{eqdistform}
\end{eqnarray}
The exact solution to Eq.~\eqref{eq:simpleformrta} is~\cite{Florkowski:2013lza,Florkowski:2013lya,
Baym:1984np,Baym:1985tna,Heiselberg:1995sh,Wong:1996va,Florkowski:2014sfa}
\begin{equation}
f(\tau,w,p_T) = D(\tau,\tau_0) f_0(w,p_T)  + \int_{\tau_0}^\tau \frac{d\tau^\prime}{\tau_{\rm eq}(\tau^\prime)} \, D(\tau,\tau^\prime) \, 
f_{\rm eq}(\tau^\prime,w,p_T) \, ,  \label{eq:solf}
\end{equation}
where $f_0(w,p_T)$ is the initial distribution function specified at $\tau = \tau_0$ and the damping function $D$ is defined as
\begin{equation}
D(\tau_2,\tau_1) = \exp\left[-\int_{\tau_1}^{\tau_2}
\frac{d\tau^{\prime\prime}}{\tau_{\rm eq}(\tau^{\prime\prime})} \right].
\end{equation}

Here, I will assume that the initial distribution function $f_0$ can be expressed in spheroidally-deformed form~\cite{Romatschke:2003ms,Romatschke:2004jh}
\begin{eqnarray}
f_0(w,p_T) &=& 
\exp\left[
-\frac{\sqrt{(p\cdot u)^2 + \xi_0 (p\cdot z)^2}}{\Lambda_0} \, \right] \nonumber \\
&=& 
\exp\left[
-\frac{\sqrt{(1+\xi_0) w^2 + (m^2+p_T^2) \tau_0^2}}{\Lambda_0 \tau_0}\, \right] ,
\label{G0}
\end{eqnarray}
where $\xi_0$ is the initial anisotropy parameter and $\Lambda_0$ is the initial transverse momentum scale.
For $-1 < \xi_0 <0$, this corresponds to an initially prolate distribution in the local rest frame and, conversely, for $\xi_0 > 0$ this corresponds to an initially oblate distribution function. For $\xi_0 = 0$, one obtains an isotropic Boltzmann distribution function as the initial condition.

\subsubsection{The integral equation obeyed by all moments}
\label{sect:moments}

To understand the emergence of the kinetic attractor, I introduce the following moments of the one-particle distribution function~\cite{Strickland:2018ayk,Strickland:2019hff}
\be
{\cal M}^{nl}[f] \equiv \int dP \,(p \cdot u)^n \, (p \cdot z)^{2l} \, f(\tau,w,p_T) \, .
\label{eq:genmom1}
\ee
In principle, powers of $p_T^2$ could also appear in a general moment, however, such moments can be expressed as a linear combination of the two-index moment appearing above using $p^2=m^2$ to write $p_T^2 = (p \cdot u)^2 - (p\cdot z)^2 - m^2$.

Some specific cases of ${\cal M}^{nl}$ map to familiar quantities, e.g., $n=1$ and $l=0$ maps to the number density ${\mathpzc n} = {\cal M}^{10}$, $n=2$ and $l=0$ maps to the energy density, and $n=0$ and $l=1$ maps to the longitudinal pressure, $P_L$.  The transverse pressure, $P_T$, can be obtained by using $p_T^2 = (p \cdot u)^2 - (p\cdot z)^2 - m^2$ to obtain $P_T = {\cal M}^{20} - {\cal M}^{01} - m^2 {\cal M}^{00}$.

For a Boltzmann equilibrium distribution function, these moments reduce to 
\ba
{\cal M}^{nl}_{\rm eq}(T,m) &\equiv& {\cal M}^{nl}[f_{\rm eq}] \nonumber \\
&=& \frac{2 T^{n+2l+2}}{(2\pi)^2(2l+1)} \int_0^\infty d\hat{p} \, \hat{p}^{n+2l+1} \left( 1 + \frac{\hat{m}^2}{\hat{p}^2} \right)^{(n-1)/2}  e^{-\sqrt{\hat{p}^2+\hat{m}^2}} \, . \nonumber \\
\ea

I will present results for these general moments scaled by their equilibrium values, i.e., 
\be
{\Mbar}^{nl} \equiv \frac{{\cal M}^{nl}}{{\cal M}^{nl}_{\rm eq}} \, .
\ee
In the late-time limit ($\tau \rightarrow \infty$), if the system approaches equilibrium, then ${\Mbar}^{nl} \rightarrow 1$.

In the general case, using the boost-invariant variables introduced earlier, one finds that the general moments can be expressed as
\ba
{\cal M}^{nl}[f] &=& \int  \frac{dw \, d^2p_T }{(2\pi)^3v} \left( \frac{v}{\tau} \right)^n \left( \frac{w}{\tau} \right)^{2l} \, f(\tau,w,p_T) \, , \nonumber \\
&=& \frac{1}{(2\pi)^3 \, \tau^{n+2l}} \int  dw \, d^2p_T  \, v^{n-1} w^{2l} \, f(\tau,w,p_T) \, ,
\ea
Taking a general moment of Eq.~(\ref{eq:solf}), one obtains
\be
{\cal M}^{nl}(\tau) = D(\tau,\tau_0) {\cal M}^{nl}_0(\tau)  + \int_{\tau_0}^\tau \frac{d\tau^\prime}{\tau_{\rm eq}(\tau^\prime)} \, D(\tau,\tau^\prime) \, 
{\cal M}^{nl}_{\rm eq}(\tau') \, .  \nonumber
\ee
Evaluating the integrals necessary results in
\ba
{\cal M}^{nl} &=&  \frac{D(\tau,\tau_0) \Lambda_0^{n+2l+2}}{(2\pi)^2}\tilde{H}^{nl}\left(\frac{\tau_0}{\tau \sqrt{1+\xi_0}}, \frac{m}{\Lambda_0} \right) \nonumber \\
&& + \frac{1}{(2\pi)^2} \int_{\tau_0}^\tau \frac{d\tau^\prime}{\tau_{\rm eq}(\tau^\prime)} \, D(\tau,\tau^\prime) \, 
T^{n+2l+2}(\tau^\prime) \, \tilde{H}^{nl}\left(\frac{\tau'}{\tau}, \frac{m}{T(\tau^\prime)} \right)  , \;\;
 \label{eq:mnleq}
\ea
where
\be
\tilde{H}^{nl}(y,z) =
\int_{0}^{\infty} du \, u^{n+2l+1} e^{-\sqrt{u^2 + z^2}} \,  H^{nl}\!\left(y,\frac{z}{u}\right) ,
\ee
with
\be
H^{nl}(y,x) = \frac{2\,y^{2l+1}
   (1+x^2)^{\frac{n-1}{2}}}{2l+1}
   \,_2F_1\!\left(l+\frac{1}{2},\frac{1-n}{2};l
   +\frac{3}{2};\frac{1-y^2}{1+x^2}\right),
\ee
where $_2F_1$ is a hypergeometric function.

\begin{figure}[t]
    \includegraphics[width=1\linewidth]{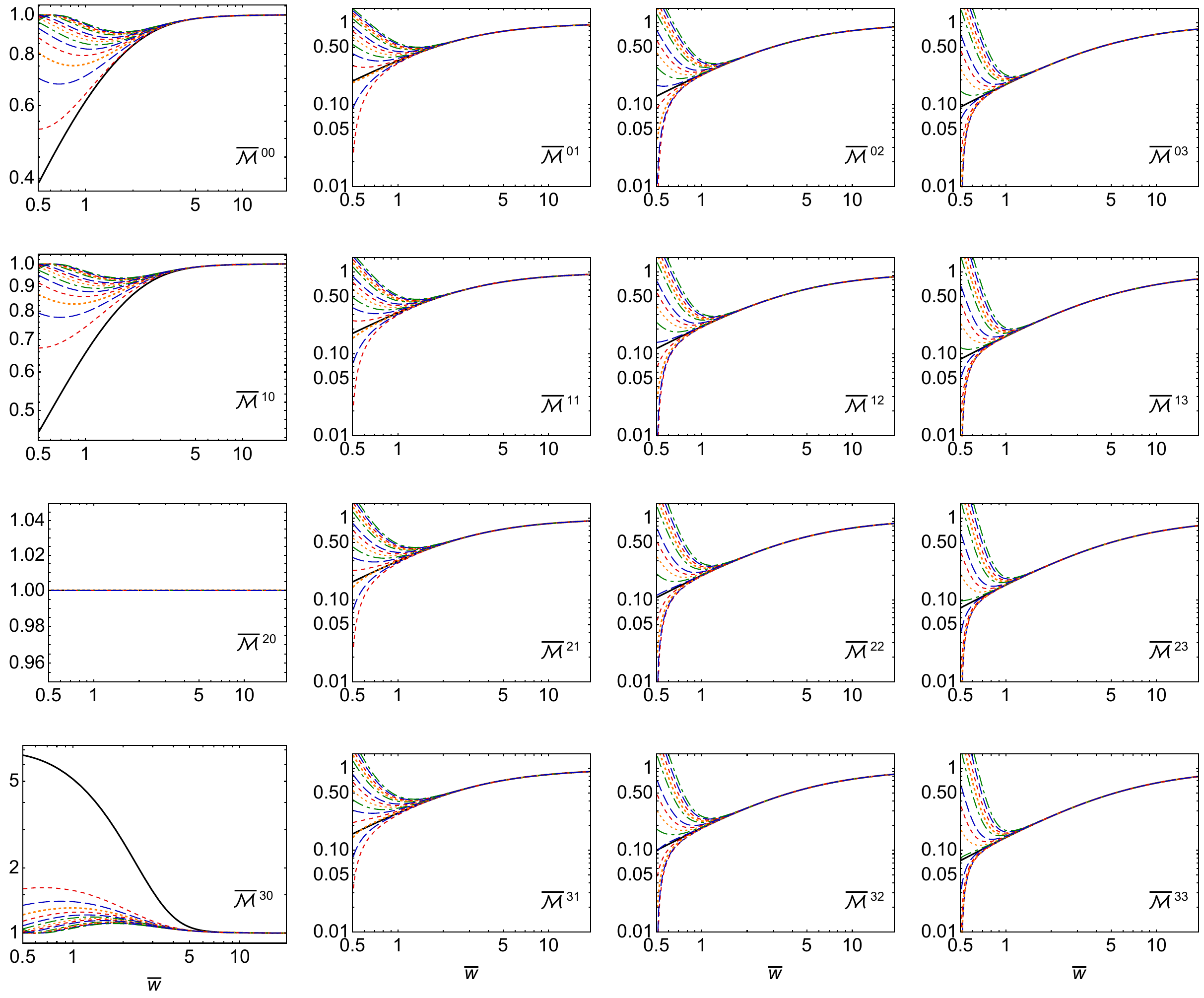}
    \caption{Scaled moments $\overline M_{nl}$ as a function of rescaled time for a conformal system ($m=0$).  The black solid line indicates the attractor solution for each moment and the non-solid lines indicate particular solutions with a range of anisotropic initial conditions.}
    \label{fig:attractorGridSols}
\end{figure}

Finally, for a momentum- and energy-independent relaxation time, specializing to the case $n=2$ and $l=0$ and requiring conservation of energy $\varepsilon(\tau) =\varepsilon_{\rm eq}(T)$, also known as Landau matching, we obtain the following integral equation
\ba
&& \hspace{-1cm} 2 T^4(\tau) \, \hat{m}^2
 \left[ 3 K_2\!\left( \frac{m}{T(\tau)} \right) + \hat{m} K_1\!\left( \frac{m}{T(\tau)} \right) \right]
 \nonumber \\ 
 && = D(\tau,\tau_0) \Lambda_0^4 \tilde{H}^{20}\!\left(\frac{\tau_0}{\tau\sqrt{1{+}\xi_0}},\frac{m}{\Lambda_0}\right) \nonumber \\
 && \hspace{2cm} +  \int_{\tau_0}^\tau \frac{d\tau^\prime}{\tau_{\rm eq}(\tau^\prime)} \, D(\tau,\tau^\prime) \, 
T^4(\tau^\prime) \tilde{H}^{20}\!\left(\frac{\tau'}{\tau}, \frac{m}{T(\tau^\prime)} \right) . 
\label{eq:20final}
\ea
This is the integral equation obtained originally in Ref.~\cite{Florkowski:2014sfa} with the understanding that \mbox{$\tilde{H}^{20} = \tilde{\cal H}_2$} defined therein.  This equation can be solved iteratively for $T(\tau)$ and, once converged to the desired numerical accuracy, the solution can be used in Eq.~\eqref{eq:mnleq} to compute all moments. 

In Fig.~\ref{fig:attractorGridSols}, I present the resulting solutions for the scaled moments $\overline M_{nl}$ as a function of rescaled time in the conformal limit ($m=0$).  The black solid line indicates the attractor solution for each moment and the non-solid lines indicate particular solutions with a range of anisotropic initial conditions.  As can be seen from this figure, there exists an attractor for all moments shown.  This implies that there is an attractor for the entire one-particle distribution function.  I remind the reader that $M^{20} = \varepsilon$, $M^{10} = {\mathpzc n}$, and $M^{01} = P_L$.  All other moments represent different modes of the distribution function not associated with basic quantities.  I note that in the case of $\overline M^{20}$, it should be equal to 1 at all times due to energy conservation.  

\begin{figure}[t]
    \includegraphics[width=1\linewidth]{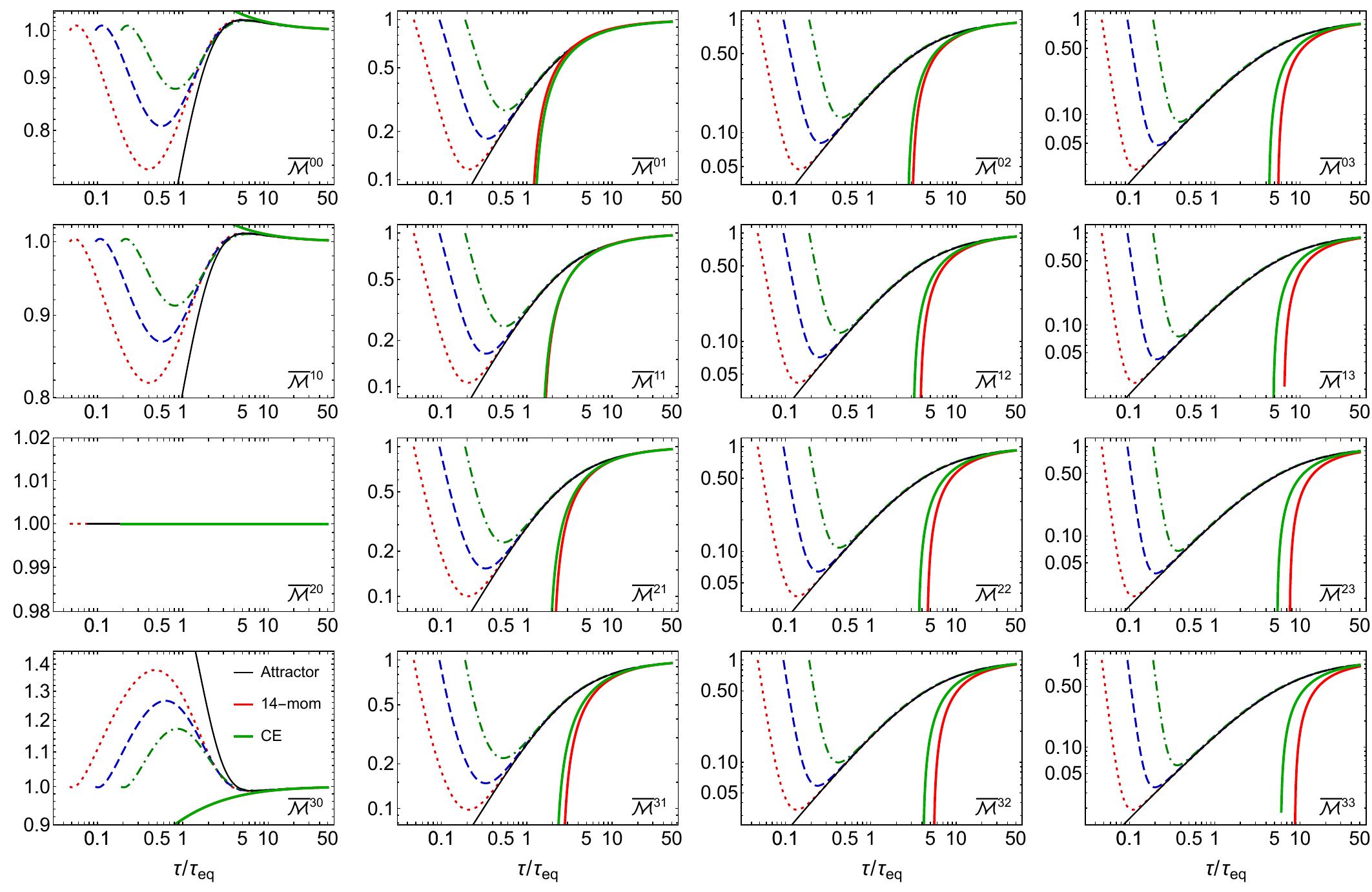}
       \caption{Scaled moments $\overline M_{nl}$ as a function of rescaled time for a non-conformal system ($m=1$ GeV).  The black solid line indicates the attractor solution for each moment and the non-solid lines indicate particular solutions with different initialization times.}
    \label{fig:t0scan-m1}
\end{figure}

In Fig.~\ref{fig:t0scan-m1}, I present the resulting solutions for the scaled moments $\overline M_{nl}$ as a function of rescaled time for a non-conformal system ($m=1$ GeV).  In this figure, I varied the initialization time while holding the initial momentum-space anisotropy and energy density fixed.  This is done in order to search for the existence of an early-time attractor or {\em pull-back attractor}.  As before, the black solid line indicates the attractor solution for each moment and the non-solid lines indicate particular solutions with different initialization times.  As can be seen from this figure, there exists an attractor for all moments with $l>0$.  For moments with $l=0$ there are indications that the early time behavior is only semi-universal.  This semi-universality does not detract from the usefulness of the phenomenon, however, it does result in a small inherent uncertainties if one wants to use the attractor in order to extrapolate to early times.    

\subsection{Attractors in QCD kinetic theory}

Having discussed results obtained within the relaxation time approximation, I present some results obtained in QCD using effective kinetic theory with an emphasis on what this may imply for freeze-out prescriptions in heavy-ion phenomenology.  This section is based on the findings reported in Ref.~\cite{Almaalol:2020rnu}.  The use of fluid dynamics is instrumental in understanding ultra-relativistic nuclear collisions \cite{Averbeck:2015jja,Jeon:2016uym,Romatschke:2017ejr}. Within this framework, only select degrees of freedom evolve dynamically, namely those stemming from the energy-momentum tensor $T^{\mu\nu}$. These encompass local temperatures, velocities, and, within viscous hydrodynamics, the shear and bulk viscous tensors. Despite this, experimental measurements do not directly capture fluid-dynamic variables; instead, they measure distributions of particles that have undergone``freeze-out'' and stream unimpeded to the detectors. The angular and momentum distributions of these particles provide crucial details about the fluid's characteristics \cite{Teaney:2003kp}. 

Extensive research has focused on assessing the ability of different formulations of viscous fluid dynamics to accurately portray the time evolution of energy-momentum tensor components. Notably, it has been found that hydrodynamic constitutive equations, which correlate the stress tensor with flow field gradients, are well fulfilled in systems that are significantly far from equilibrium, particularly in cases characterized by highly symmetrical flow.  The freeze-out procedure's validity far from equilibrium has received considerably less scrutiny. One challenge lies in the fact that models with simplistic momentum dependence, like kinetic theory in relaxation time approximation, offer limited insights into freeze-out procedure validity. Conversely, strongly coupled models do not possess quasiparticle structure and hence lack underlying particle distributions altogether. 

In this section I discuss the possibility of reconstructing particle distributions from the energy-momentum tensor within Effective Kinetic Theory (EKT) for weak-coupling quantum chromodynamics~\cite{Arnold:2002zm}. The detailed momentum-dependent structure of the EKT collision kernel allows for a meaningful examination of the freeze-out procedure in this theoretically clean case. Focusing again on 0+1D far-from-equilibrium Bjorken flow, I compare various moments of the distribution function with predictions from hydrodynamic freeze-out prescriptions. Interestingly, one finds qualitative similarities between EKT and RTA kinetic theory, as well as Israel-Stewart-type hydrodynamics \cite{Heller:2015dha, Romatschke:2017vte, Kurkela:2019set}, both in early (early-time or pullback attractor) and late times (late-time or hydrodynamic attractor). 

As discussed in the previous section, this attractor behavior extends beyond the energy-momentum tensor components to other integral moments of the one-particle distribution function. While commonly used freeze-out prescriptions well reproduce low-order moments of the distribution at late times, they may not be accurate at early times or when addressing moments that are sensitive to significantly higher momenta than the temperature scale.
To study this one can use numerical implementations of the EKT introduced in Refs.~\cite{Arnold:2002zm,York:2014wja,Kurkela:2015qoa}. In parametrically isotropic systems,  EKT gives a leading order accurate description (in $\alpha_s$) of the QCD time evolution of the one-particle distribution function and allows for a numerical realization of the so-called bottom-up thermalization scenario \cite{Baier:2000sb}.  In practice, we solved the EKT Boltzmann equation for a gluonic plasma undergoing one-dimensional Bjorken expansion with transverse translational symmetry such that the effective Boltzmann equation reads \cite{Mueller:1999pi}
\be
\frac{d f(\p)}{d\tau} - \frac{p_z}{\tau}\partial_{p_z} f = \mathcal{C}_{1\leftrightarrow 2}[f(\p)] + \mathcal{C}_{2\leftrightarrow 2}[f(\p)] \, ,
\label{eq:ekt1}
\ee
where $f(\p)$ is the gluonic one-particle distribution function (per degree of freedom).  The elastic scattering term $\mathcal{C}_{2\leftrightarrow 2}$ and the effective inelastic term $\mathcal{C}_{1 \leftrightarrow 2}$ include physics of dynamical screening and Landau-Pomeranchuck-Migdal suppression and, in order to find the form of the collision kernels, self-energy and ladder resummations are required~\cite{Arnold:2002zm,York:2014wja,Kurkela:2015qoa}. 

To numerically solve Eq.~\eqref{eq:ekt1}, one can discretize $n(\p) = p^2 f(\p)$ on an optimized momentum-space grid and employ Monte Carlo sampling to compute the integrals within the elastic and inelastic collisional kernels. The methodology used is based on the methods outlined in Refs.~\cite{York:2014wja,Kurkela:2015qoa}.  The method ensures exact conservation of energy, while accurately addressing particle number violation originating from inelastic contributions to the collisional kernel. Leveraging the azimuthal symmetry inherent in Bjorken flow, one can reduce the momentum-space discretization to a two-dimensional grid.  In Ref.~\cite{Almaalol:2020rnu} we utilized 250 points in the $p$ direction and 2000 points in the $x=\cos\theta$ direction. The momenta $p$ were distributed logarithmically within the ranges $[0.02,45]\,\Lambda$, where $\Lambda$ denotes the typical energy scale of the initial condition. In the results presented in this section we used a 't Hooft coupling $\lambda = N_c g^2 = 10$, corresponding to a specific shear viscosity of $\bar\eta = \eta/s \approx 0.62$ \cite{Arnold:2003zc,Keegan:2015avk}.

We track the temporal evolution of the same set of integral moments considered in RTA~\cite{Strickland:2018ayk}
\be
{\cal M}^{nl}(\tau) \equiv  \int \frac{d^3 {\vec p}}{(2\pi)^3} \, p^{n-1} \, p_z^{2l} \, f(\tau,\p) \, ,
\label{eq:genmom2}
\ee
where $p = |{\vec p}|$. Notably, the energy density is expressed as $\varepsilon=N_{\rm dof}{\cal M}^{20}$, longitudinal pressure as $P_L =  N_{\rm dof}{\cal M}^{01}$, and number density as $n=N_{\rm dof}{\cal M}^{10}$. Other moments lack interpretation in terms of conventional hydrodynamic moments often discussed in literature.

We scaled these moments by their corresponding equilibrium values with \mbox{$\overline{\cal{M}}^{nl}(\tau) \equiv {\cal M}^{nl}(\tau) / {\cal M}^{nl}_{\rm eq}(\tau)$}.  Using a Bose distribution, one obtains
\be
{\cal M}^{nl}_{\rm eq} =  \frac{ T^{n+2l+2}\Gamma(n+2l+2) \zeta(n+2l+2)}{2 \pi^2 (2l+1)} \, .
\ee
The temperature $T$ here corresponds to the temperature of an equilibrium system with the same energy density, given by 
$T = (30 \varepsilon/N_{\rm dof} \pi^2)^{1/4}$.

\begin{figure}[t]
    \includegraphics[width=1\linewidth]{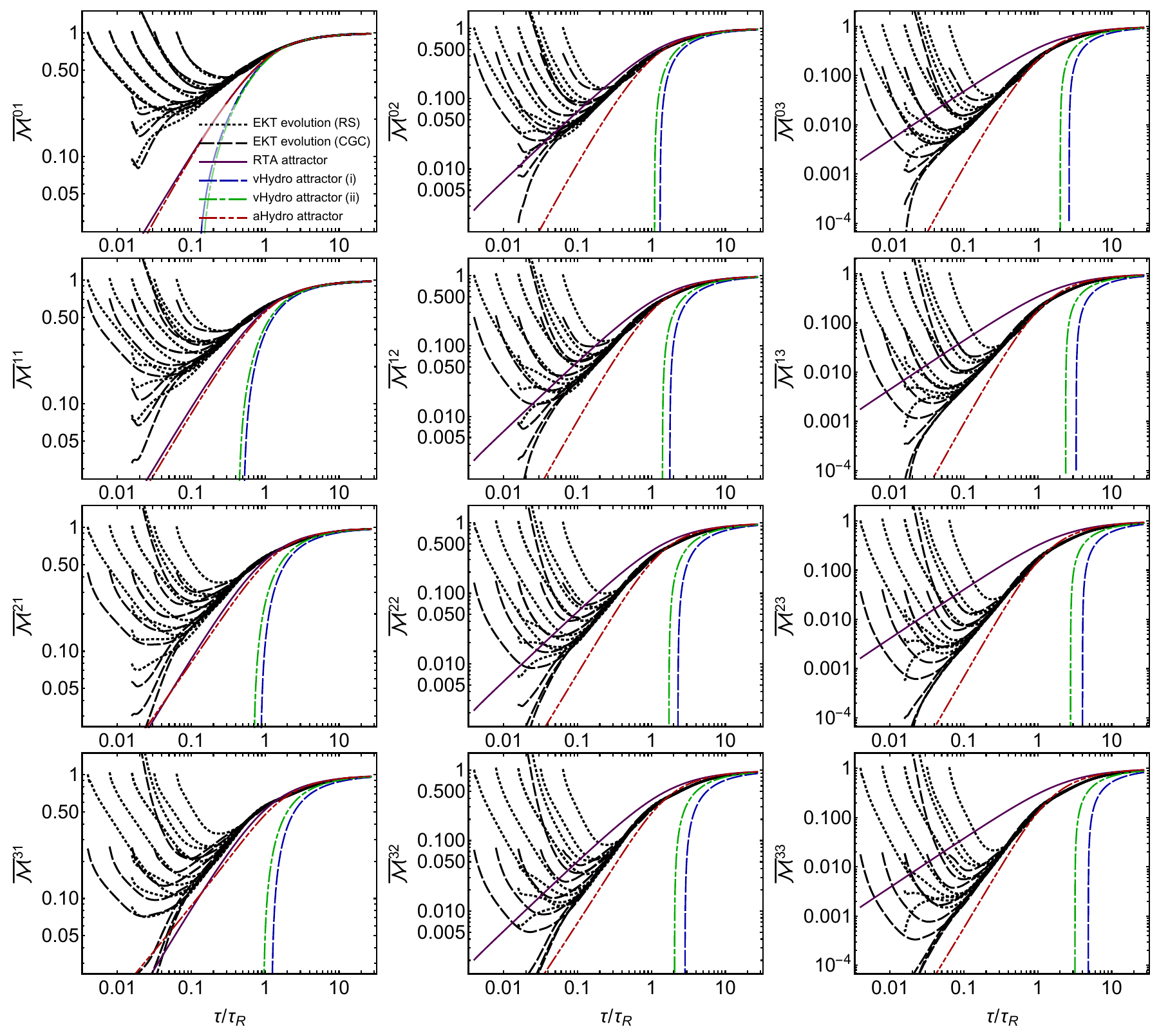}
    \caption{The evolution of scaled moments $\overline{\cal{M}}^{nl}$ with $n \in \{0,1,2\}$ and $m \in \{0,1,2,3\}$. The black dotted and dashed lines represent EKT evolution with RS and CGC initial conditions, respectively. The purple solid line denotes the exact RTA attractor, while the blue long-dashed line corresponds to the DNMR vHydro attractor using $\delta f$ parameterization (i). The green dot-dashed line illustrates the DNMR vHydro attractor using $\delta f$ parameterization (ii), and the red dot-dot-dashed line signifies the aHydro attractor.}
    \label{fig:qcdfig1}
\end{figure}

Our simulations were initialized with two types of initial conditions: either spheroidally-deformed thermal initial conditions, referred to as ``RS'' initial conditions \cite{Romatschke:2003ms}, or non-thermal color-glass-condensate (CGC) inspired initial conditions \cite{Kurkela:2015qoa}. In the former scenario, the initial gluonic one-particle distribution function takes the following form
\be
f_{0,{\rm RS}}(\p) = f_{\rm Bose}\!\left(\sqrt{\p^2 + \xi_0 p_z^2 }/\Lambda_0\right) ,
\ee
where $-1 < \xi_0 < \infty$ encodes the initial momentum-anisotropy and $\Lambda_0$ is a temperature-like scale which sets the magnitude of the initial average transverse momentum.  In the second case, we take for the form of the initial gluonic one-particle distribution
\ba
f_{0,\rm CGC}(\p) &=&  \frac{2A}{\lambda} \frac{ \tilde \Lambda_0}{\sqrt{ {\vec p}^2 +\xi_0  p_z^2}} e^{-\frac{2}{3}\left(\p ^2 +  \xi_0 \hat{p}_z^2 \right)/\tilde \Lambda_0^2} \, .
\ea
This latter form has been utilized in several prior studies (see, for instance, \cite{Kurkela:2015qoa,Keegan:2016cpi,Kurkela:2018vqr,Kurkela:2018vqr,Kurkela:2018oqw,Kurkela:2018xxd}) and is motivated by the saturation framework. Here, the initial average transverse momentum scale $\tilde\Lambda_0$ is connected to the saturation scale, which is $\tilde\Lambda_0 = \langle p_T \rangle_0 \approx 1.8\,Q_s$ \cite{Mueller:1999fp,Kovchegov:2000hz,Lappi:2011ju}. The overall constant $A$ is determined by adjusting the initial energy density to match an expected value $\tau_0 \varepsilon_0 = 0.358\, \nu Q_s^3/\lambda$ from a classical Yang-Mills simulation~\cite{Lappi:2011ju}.

In Fig.~\ref{fig:qcdfig1}, I present the evolution of the scaled moments $\overline{\cal{M}}^{nl}$ with $n \in \{0,1,2\}$ and $m \in \{0,1,2,3\}$. As shown in Fig.~\ref{fig:qcdfig1}, evidence of a non-equilibrium attractor emerges both at early and late times across all moments.   The integral moments are plotted against a rescaled time variable $\bar{w} = \tau/\tau_R(\tau)$, which represents the system's age in units of the instantaneous interaction time $\tau_R(\tau)$. This interaction time scale $\tau_R(\tau) = 4 \pi \bar\eta/T(\tau)$ varies in time. The chosen time scaling ensures that, as long as the system is closely described by hydrodynamics near thermal equilibrium, $\overline{\mathcal{M}}^{01}$ will eventually be described by the first-order gradient expansion, $\overline{\mathcal{M}}^{01} = 1 - (120\zeta(5)/\pi^5) \tau_R/\tau$~\cite{Heller:2016rtz,Strickland:2018ayk}. This behavior remains unaffected by microscopic details or specific values of macroscopic hydrodynamic parameters. A similar convergence was also observed in RTA and it occurs before the system is effectively characterized by the hydrodynamic gradient expansion.

While the late-time attractor behavior for the longitudinal pressure, $\overline{\mathcal{M}}^{01}$, has previously been noted in simplified kinetic theories, e.g. RTA, QCD EKT evolution allows one to examine the extent to which this attractive behavior governs the overall shape of the distribution function. The timescale at which various solutions to Eq.~\eqref{eq:ekt1} converge to the attractor is approximately $\tau/\tau_R\sim 0.5$. While all theories ultimately converge onto a single curve, the timescale for individual solutions to collapse to the attractor depends on specifics of the model. In \cite{Kurkela:2019set}, two distinct patterns were identified. In RTA kinetic theory and IS hydrodynamics, the transition to the attractor occurs via a power law, with the scale determined by the initial time $\tau_0$. Consequently, a unique attractor emerges at arbitrarily early times, discernible through initialization with decreasing $\tau_0$. Conversely, in AdS/CFT, the transition to the attractor occurs solely on the timescale dictated by the decay of the quasi-normal modes.

\section{Two phenomenological implications of attractors}

We now highlight two phenomenological implications of attractors.  The first is based on the ability to use attractors to extrapolate backwards in time and the second stems from the detailed QCD EKT simulations presented in the previous section.

\subsection{Entropy generation constraints}

As mentioned above, for a system undergoing Bjorken expansion, conformality, along with energy-momentum conservation $\partial_{\mu}T^{\mu\nu}=0$, leads to the following evolution equation for the energy density
\be
\frac{d\varepsilon}{d\tau} = \frac{\varepsilon + P_L}{\tau} \, ,
\ee
where $P_{L}$ is the longitudinal pressure of the system, and $\tau$ is the longitudinal proper time.  In first-order Navier-Stokes (NS) viscous hydrodynamics, the longitudinal pressure is given by
\be
\frac{P_L^{\rm NS}}{\varepsilon} = \frac{1}{3} - \frac{16 \bar\eta}{9 \tau T} \, ,
\ee
with second-order viscous corrections being of order $(\bar\eta/T \tau)^2$ \cite{Romatschke:2017ejr}, where $\eta$ is the shear viscosity, $s$ is the entropy density, and $\bar\eta \equiv \eta/s$ is the specific shear viscosity. The hydrodynamic attractor concept \cite{Heller:2015dha} envisions an all-order resummation of these terms into a constitutive relation of the form $P_L/e=f(w)$, where $w(\tau) \propto \tau T_{\rm eff}/\bar\eta$ \cite{Florkowski:2017olj}. The effective temperature $T_{\rm eff}$ is defined through the non-equilibrium energy density and the equation of state, $ s T = \tfrac{4}{3} \varepsilon =  b_{\rm qgp} T^4$, where $b_{\rm qgp} \simeq 17.6$ is estimated from the lattice equation of state at high temperatures \cite{Borsanyi:2010cj,HotQCD:2014kol}.

In the conventional view, hydrodynamics is applicable only when the Knudsen number, defined as the ratio of microscopic scales like $\ell$ (such as the mean free path of a gas) to macroscopic scales $L$ associated with spatial gradients of conserved quantities, becomes significantly smaller than one. This Knudsen number is given by $\mathrm{Kn} = \ell/L$. In this context, we have $w\sim 1/\mathrm{Kn}$, meaning that the hydrodynamic regime becomes valid at later times when $w \gg 1$. Corrections are then organized in powers of $w^{-1}$ (or equivalently, in powers of $\mathrm{Kn}$).  If the appropriate attractor constitutive relation is in play, the energy density can be expressed as
\be
\frac{\varepsilon(\tau) \tau^{4/3}}{ (\varepsilon \tau^{4/3})_{\rm \infty} }  \equiv  \mathcal  E(w) \, ,
\label{eq:attractor}
\ee
where $(\varepsilon\tau^{4/3})_{\infty} \propto (\tau s)_{\infty}^{4/3}$ normalizes the entropy in the system, and $\mathcal E \rightarrow 1$ at late times.

Different types of simulations, encompassing the relaxation time approximation~\cite{Romatschke:2017vte,Strickland:2017kux,Strickland:2018ayk,Heller:2018qvh,Strickland:2019hff,Jaiswal:2022udf,Alalawi:2022pmg}, QCD kinetic theory~\cite{Kurkela:2015qoa,Keegan:2015avk,Kurkela:2018wud,Kurkela:2018vqr,Kurkela:2019set,Almaalol:2020rnu}, and the AdS/CFT correspondence~\cite{Heller:2011ju,Heller:2012km,vanderSchee:2013pia,Heller:2013oxa,Casalderrey-Solana:2013aba,Chesler:2013lia,Chesler:2015wra,Keegan:2015avk,Spalinski:2017mel}, suggest that the simplified interpolating expression in Eq.~\eqref{eq:attractor} effectively captures the overall dynamics. The behavior of $\mathcal E(w)$ is highly constrained by its characteristics at both early and late times. Specifically, at late times, the entropy per rapidity $(\tau s)_{\infty}$ remains constant, whereas at early times when $\tau\rightarrow 0$, the energy per rapidity $(\tau \varepsilon(\tau))_0$ remains constant, dictating the behavior as $\mathcal E \simeq C_{\infty}^{-1} w^{4/9}$. The constant $C_{\infty}$ exhibits a mild dependence on the underlying theory and is approximately unity. Simulations within QCD kinetic theory suggest $C_{\infty} \simeq 0.87$, while the AdS/CFT correspondence yields $C_{\infty}=1.06$ \cite{Heller:2011ju,Heller:2015dha}.

The hydrodynamic attractor, therefore, becomes very useful, allowing one to determine the energy density with an accuracy of 20\% during the initial stages, using the measured charged particle multiplicity, nearly independent of the underlying theoretical framework~\cite{Giacalone:2019ldn}. This precision level serves to significantly constrain models for the initial state, including those based on the color-glass condensate \cite{Gelis:2010nm}. The second noteworthy significance of the attractor solution is its role as a criterion for the initiation of hydrodynamics. This criterion has undergone validation through detailed studies, particularly highlighting that the hydrodynamic gradient expansion becomes applicable at a time $\tau_{\rm hydro}$ when $w\simeq 1$ or larger. To translate this theoretical criterion into an experimental context, it is important to recognize that the total entropy per rapidity in hydrodynamics is directly linked to the hydrodynamic multiplicity~\cite{Kurkela:2018vqr}.
\be
   \frac{dS}{dy} = b_{\rm hrg}  \frac{dN}{d\eta}  \, ,
\ee
where $b_{\rm hrg}\simeq 8.3$ is based on the particle composition of the hadron resonance gas model and the lattice equation of state, with a 15\% uncertainty \cite{Kurkela:2018vqr}. The temperature during later times is ascertained through the constant entropy or multiplicity, where $dN_{\rm ch}/d\eta \propto \tau T^3$, resulting in
\ba
   w  \equiv \frac{\tau T}{4\pi \eta/s} &=& 
   \ \frac{1}{4\pi \eta/s} \left(\frac{1}{N_0}\frac{dN_{\rm ch}}{d\eta} \right)^{1/3}   \left(\frac{\tau}{\tau_R} \right)^{2/3} \\
   &\equiv&  \ \chi \left(\frac{\tau}{\tau_R} \right)^{2/3},
\ea
where we have defined the \emph{opacity}  of the system
\be
    \chi = \frac{1}{4\pi \eta/s} \left(\frac{1}{N_0}\frac{dN_{\rm ch}}{d\eta} \right)^{1/3}\,,   \qquad  N_0 \equiv \frac{\pi b_{\rm hrg} }{b_{\rm qgp}} \simeq 6.68 \, .
\ee
We note that the multiplicity factor $N_0$ is primarily determined by the properties of the equation of state, which is known from  lattice QCD. Consequently the uncertainties in $N_0$ are only of order 20\%.
The system will have a strong hydrodynamic response if  $\tau_{\rm hydro}/\tau_R \ll 1$~\cite{Kurkela:2018wud}
\be
    \frac{\tau_{\rm hydro}}{\tau_R}  = \frac{1}{\chi^{2/3} }  \propto \frac{1}{\sqrt{dN_{\rm ch}/d\eta} } \, ,
\ee
which  amounts to a requirement that opacity is large $\chi \gg 1$.

\subsection{Freeze-out in heavy-ion collisions}

Transforming hydrodynamic fields into particle distributions requires a ``freeze-out'' procedure. While the energy-momentum tensor relies solely on the first momentum-integral moments of the distribution function, particle distributions span all moments. Therefore, translating results from hydrodynamic simulations into particle distributions requires the incorporation of additional information via theoretical assumptions. The standard approach involves assuming a distribution function with a near-equilibrium configuration, wherein deviations from equilibrium stem from formally small corrections. These corrections' structure is dictated by the linearized collision kernel's response, within some assumed kinetic theory, to infinitesimal strain \cite{Teaney:2009qa,Dusling:2009df}.  Since the freeze-out procedure strongly affects the phenomenological analysis and conclusions about the matter created in ultra-relativistic heavy-ion collisions, it is of great interest to assess how well justified the theoretical assumptions about the shape of the non-equilibrium distribution functions are. The need for such an assessment becomes increasingly important in the case of small systems, \emph{e.g.}, peripheral nucleus-nucleus collisions, proton-nucleus, and high-multiplicity proton-proton collisions, where hydrodynamical descriptions are being applied to situations that remain far from equilibrium throughout their dynamical evolution.

Although hydrodynamics does not explicitly describe higher moments of the distribution functions, it is customary to deduce the entire distribution's shape solely from the shear components of the energy-momentum tensor. For a given $T^{\mu\nu}$, the linearized viscous correction to the one-particle distribution function, $\delta f$, can be locally determined based on some assumptions about the collision kernel. Here, I will present results obtained with two forms for $\delta f$. Firstly, the (i) quadratic ansatz,
\be
\frac{\delta f_{(i)}}{f_{\rm eq}(1+f_{\rm eq})} =  \frac{3 \overline\Pi}{16 T^2} ( p^2 - 3 p_z^2 ) \, .
\label{eq:fa}
\ee
This form arises from a broad array of models, including RTA with momentum-independent relaxation time, the momentum diffusion approximation, scalar field theory, and the EKT in the leading-log approximation \cite{Dusling:2009df}. Here, $\overline{\Pi} = \Pi/\varepsilon = 1/3 - T^{zz}/\varepsilon$ represents the shear viscous correction to the longitudinal pressure, normalized by the energy density. However, QCD EKT exhibits additional structure; for large momenta $p \gg T$, it reduces to a power law form of the (ii) LPM ansatz,
\be
\frac{\delta f_{(ii)}}{f_{\rm eq}(1+f_{\rm eq})} = \frac{16 \overline\Pi}{21 \sqrt{\pi}\, T^{3/2}} \! \left( p^{3/2} - \frac{3 p_z^2}{\sqrt{p}} \right) .
\label{eq:fb}
\ee
This $p^{1.5}$ power-law is numerically close to $\propto p^{1.38}$ that is found to describe the full EKT in the relevant momentum range in \cite{Dusling:2009df}.

\begin{figure}[t]
    \includegraphics[width=1\linewidth]{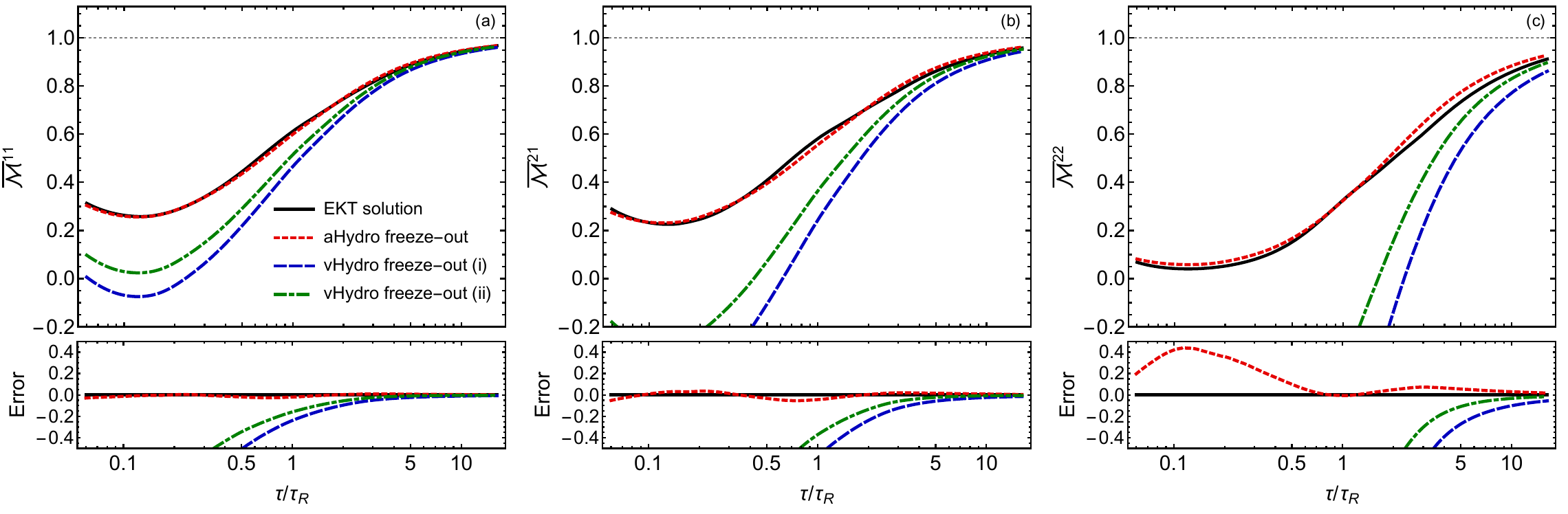}
    \caption{The evolution of scaled moments (a) $\overline{\cal{M}}^{11}$, (b) $\overline{\cal{M}}^{21}$, and (c) $\overline{\cal{M}}^{22}$. The black solid line represents typical EKT evolution, the red dashed line denotes the $P_L$-matched aHydro result for a given moment, and the blue and green dot-dashed lines correspond to second-order viscous hydrodynamics results using Eqs.~\eqref{eq:fa} and \eqref{eq:fb}, respectively. The relative error, depicted in the bottom panels, is calculated as (${\rm approximation}/{\rm EKT} -1$).}
    \label{fig:qcdfig2}
\end{figure}

One can also explore a straightforward (iii) aHydro freeze-out ansatz, which does not rely on linearization around equilibrium. Instead, in this approach, the non-equilibrium distribution function is approximated by a spheroidally-deformed Bose distribution, $f(p) = f_{\rm Bose}(\sqrt{{\vec p}^2 + \xi p_z^2}/\Lambda)$ \cite{Romatschke:2003ms,Florkowski:2010cf,Martinez:2010sc}. To assess this method, we determine $\xi(\tau)$ such that the energy-momentum tensor of the ansatz reproduces the full simulation, enabling predictions for higher-order moments.

In Fig.~\ref{fig:qcdfig2}, I compare the different moments obtained from the prescriptions mentioned above with the QCD EKT attractor solution. At late times ($\tau \gtrsim 5\,\tau_R$), all prescriptions describe the low-order moments within a few percent. However, some discrepancy persists even at $\tau \sim 20\,\tau_R$ between the quadratic ansatz (i) and the QCD EKT results. This disparity gradually worsens at earlier times, particularly around $\tau \sim \tau_R$, where corrections to the longitudinal pressure become substantial ($P_L/P_L^{\rm eq} \sim 65\%$). For instance, ${\cal M}^{11}$ exhibits an approximately $20\%$ discrepancy between EKT and both linearized ansatze. This disagreement increases for higher moments and at earlier times. Conversely, we observe a remarkably good agreement between the aHydro ansatz and the QCD EKT results at all times.

In the phenomenological analysis of nuclear collisions, a critical step is the freeze-out procedure, wherein hydrodynamic fields are transformed into particle distributions. Presently, the quadratic ansatz (i) is the most commonly used form. However, this approach assumes linear deviations from thermal equilibrium, a stark contrast to the far-from-equilibrium conditions prevalent in current phenomenological applications, especially in modeling small systems (see, e.g., Refs.~\cite{Bozek:2013uha,Shen:2016zpp,Alqahtani:2016rth,Weller:2017tsr,Mantysaari:2017cni,Strickland:2018exs}). To investigate whether these linearized procedures remain quantitatively predictive in far-from-equilibrium scenarios, I have compared them with far-from-equilibrium simulations of QCD effective kinetic theory. The findings shown in Fig.~\ref{fig:qcdfig2} indicate that the non-linear aHydro freeze-out ansatz performs better in reconstructing moments of the distribution function compared to linearized ansatze in far-from-equilibrium systems.

\section{3+1D anisotropic hydrodynamics}

The results in the first part of this chapter demonstrated that the aHydro formulation of dissipative hydrodynamics resums an infinite number of terms in the expansion in the inverse Reynolds number.  As a result, aHydro best reproduces the evolution of all moments of the one-particle distribution function and also provides a superior formulation when it comes to freeze-out.  I now describe phenomenological application of 3+1D anisotropic hydrodynamics, allowing for a non-conformal realistic equation of state and full dynamics in both the transverse plane and longitudinal (spatial rapidity) direction.

To model heavy-ion collisions one must obtain the evolution equations for the resulting 3+1D configurations and include the non-conformality of QCD consistent with a realistic lattice-based equation of state. To do this one can use anisotropic hydrodynamics (aHydro) \cite{Martinez:2010sc,Florkowski:2010cf,Alqahtani:2017mhy}, specifically quasiparticle anisotropic hydrodynamics (aHydroQP) in which one assumes that the QGP consists of massive relativistic quasiparticles with temperature-dependent masses $m(T)$~\cite{Alqahtani:2017mhy}. In aHydroQP the system is assumed to obey a relativistic Boltzmann equation with $m(T)$ determined from lattice QCD computations of QCD thermodynamics. Since the masses are temperature dependent, the Boltzmann equation contains an additional force term on the left-hand side related to gradients in the temperature
\be
p^\mu \partial_\mu f + \frac{1}{2}\partial_i m^2 \partial_{(p)}^{i} f =  C[f] \, .
\label{eq:be}
\ee
with
\be
C[f] = - \frac{p \cdot u}{\tau_{\rm eq}(T)} [ f - f_\text{eq}(T) ] \, ,
\ee
being the collisional kernel in relaxation time approximation (RTA). Above, $u^\mu$ is the four-velocity associated with the local rest frame (LRF) of the matter and Latin indices such as $i \in \{x,y,z\}$ are spatial indices.
 
For a gas of massive quasiparticles, the relaxation time is given by \cite{Alqahtani:2017mhy}
\be
\tau_{\rm eq}(T)= \bar{\eta} \, \frac{\varepsilon+P}{I_{3,2}(\hat{m}_{\rm eq})}
\label{eq:teq}
\ee
where $\hat{m}_{\rm eq} = m/T$, $\bar\eta = \eta/s$ is the specfic shear viscosity, $\varepsilon$ is the energy density, $P$ is the pressure, which is fixed by the equation of state.  The definition of the special function $I_{3,2}(\hat{m}_{\rm eq})$ can be found in Refs.~\cite{Alqahtani:2017mhy,Alalawi:2021jwn}.
The effective temperature $T(\tau)$ is determined by requiring energy-momentum conservation.  For a momentum- and energy-independent relaxation time this requires that the non-equilibrium kinetic energy densities calculated from $f$ are equal to the equilibrium kinetic energy density calculated from the equilibrium distribution, $f_\text{eq}(T,m)$.

In all leading-order aHydro phenomenological works to date, one assumes that the distribution is parameterized by a diagonal anisotropy tensor in the LRF\,\footnote{In Ref.~\cite{Nopoush:2019vqc} a method for non-perturbatively including off-diagonal components of the anisotropy tensor was presented, however, to date it has not been applied to phenomenology.}
\ba
f_{\rm LRF}(x,p) &=& f_{\rm eq}\!\left(\frac{1}{\lambda}\sqrt{\sum_i \frac{p_i^2}{\alpha_i^2} + m^2}\right)  \, .
\label{eq:fdef}
\ea
As indicated above, in the LRF, the argument of the distribution function can be expressed in terms of three independent momentum-anisotropy parameters $\alpha_i$ and a scale parameter $\lambda$, which are space-time dependent fields.  Herein we will assume that $f_{\rm eq}$ is given by a Boltzmann distribution.

In order to determine the space-time evolution of the fields $\vec{u}$, $\vec{\alpha}$, and $\lambda$ one must obtain seven dynamical equations. The first aHydroQP equation of motion is obtained from the first moment of the left-hand side of the quasiparticle Boltzmann equation \eqref{eq:be}, which reduces to $\partial_\mu T^{\mu\nu}$. In the relaxation time approximation, however, the first moment of the collisional kernel on the right hand side results in a constraint that must be satisfied in order to conserve energy and momentum, i.e. $\int dP \, p^\mu C[f]=0$. This constraint can be enforced by expressing the effective temperature in terms of the microscopic parameters $\lambda$ and $\vec\alpha$.  As a consequence, computing the first moment of the Boltzmann equation gives the partial differential equation resulting from energy-momentum conservation
\be
\partial_\mu T^{\mu\nu}=0 \, ,
\label{eq:1m}
\ee
where 
\be
T^{\mu \nu}=\int \frac{d^3{\vec p}}{(2\pi)^3} \frac{1}{E} \, p^\mu p^\nu f \equiv \int dP \, p^\mu p^\nu f \, .
\ee
For the second equation of motion, one can perform a similar procedure using the second moment of the quasiparticle Boltzmann equation  
\be
\partial_\alpha  I^{\alpha\nu\lambda} - J^{(\nu} \partial^{\lambda)} m^2 =-\int dP \, p^\nu p^\lambda \, {\cal C}[f]\, \label{eq:I-conservation} ,
\ee
with $I^{\mu\nu\lambda} = \int dP \, p^\mu p^\nu p^\lambda f $ and $J^{\mu} = \int dP \, p^\mu f$. 

\subsection{The equation of state for aHydroQP}
\label{sec:EoS}

For a system of massive particles obeying Boltzmann statistics, the equilibrium energy density, pressure, and entropy density are given by 
\ba
\varepsilon_{\rm eq}(T,m) &=& \hat{N} T^4 \, \hat{m}_{\rm eq}^2
 \Big[ 3 K_{2}\left( \hat{m}_{\rm eq} \right) + \hat{m}_{\rm eq} K_{1} \left( \hat{m}_{\rm eq} \right) \Big] , \nonumber \\
 P_{\rm eq}(T,m) &=& \hat{N} T^4 \, \hat{m}_{\rm eq}^2 K_2\left( \hat{m}_{\rm eq}\right) ,  \nonumber  \\
 s_{\rm eq}(T,m) &=& \hat{N} T^3 \, \hat{m}_{\rm eq}^2 \Big[4K_2\left( \hat{m}_{\rm eq}\right)+\hat{m}_{\rm eq}K_1\left( \hat{m}_{\rm eq}\right)\Big] ,
\label{eq:SeqConstantM}
\ea
where $\hat{m}_{\rm eq} = m/T$, $K_i$ are modified Bessel functions of the second kind and $\hat{N} = N_{\rm dof}/2\pi^2$, with $N_{\rm dof}$ being the number of degrees of freedom present in the theory under consideration.

In the quasiparticle approach, one assumes the mass to be temperature dependent. This results in a change in the bulk variables in Eqs.~\eqref{eq:SeqConstantM}. Due to the temperature dependence of the quasiparticle mass, one can not simply insert $m(T)$ into the bulk variables since this will not be thermodynamically consistent.   This is due to the fact that the entropy density may be obtained in two ways: $s_{\rm eq} = (\varepsilon_{\rm eq} + P_{\rm eq})/T$ and $ s_{\rm eq} =\partial P_{\rm eq}/\partial T$. In order to guarantee that both methods result in the same expression for the entropy density, the energy-momentum tensor definition must include a background field $B_{\rm eq}$, i.e.,
\be
T^{\mu\nu} = T^{\mu\nu}_{\rm kinetic} + g^{\mu\nu} B_{\rm eq}(T)  \, .
\label{eq:Tmunu}
\ee
where $B_{\rm eq}$ is the additional background contribution. 
As a result, in an equilibrium gas of massive quasiparticles, the bulk thermodynamic variables for the gas become
\ba
\varepsilon_{\rm eq}(T,m) &=& \varepsilon_{\rm kinetic} +B_{\rm eq} \, ,  \\
 P_{\rm eq}(T,m) &=& P_{\rm kinetic} -B_{\rm eq}\, ,\\
 s_{\rm eq}(T,m) &=&s_{\rm kinetic} \, .
\ea

To determine the temperature dependence of $B_{\rm eq}$ one requires thermodynamic consistency
\be
 s_{\rm eq} = \varepsilon_{\rm eq} + P_{\rm eq} = T \frac{\partial P_{\rm eq}}{\partial T}
\ee
To fix the equation of state, one can determine $m(T)$ using
\be
\varepsilon_{\rm eq} + P_{\rm eq} = T s_{\rm eq} =  \hat{N} T^4 \, \hat{m}_{\rm eq}^3 K_3\left( \hat{m}_{\rm eq}\right) .
\ee
This can be solved based on the equilibrium energy density and pressure determined from lattice QCD calculations~\cite{Alqahtani:2015qja}.  Once $m(T)$ is fixed, the background field $B(T)$ can be determined using the requirement of thermodynamic consistency.  The resulting effective mass scaled by $T$ extracted from continuum extrapolated Wuppertal-Budapest lattice data \cite{Borsanyi:2010cj} can be found in Refs.~\cite{Alqahtani:2015qja,Ryblewski:2017ybw}.   At high temperatures ($T \sim$ 0.6 GeV) the scaled mass is $\propto T$ in agreement with the expected high-temperature behavior of QCD~\cite{Ryblewski:2017ybw}.  Once $m(T)$ is fixed, one can self-consistently determine all transport coefficients.  

\begin{figure}[t!]
\centering
\includegraphics[width=0.6\linewidth]{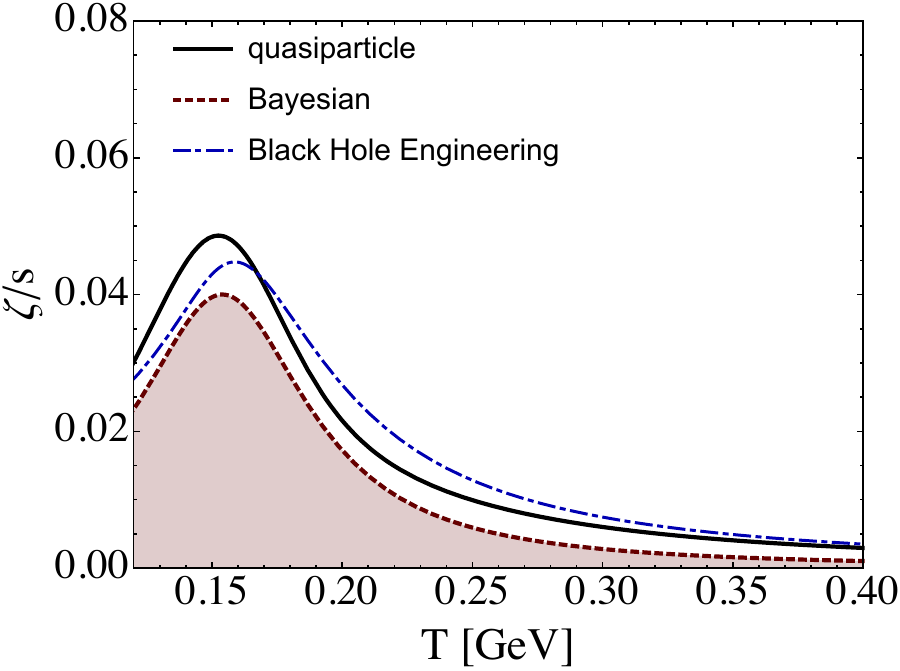}
\caption{The bulk viscosity scaled by the entropy density $\zeta/s$ as a function of $T$.  The solid black line shows the aHydroQP result, the red short-dashed line shows the result of a Bayesian analysis \cite{Bass:2017zyn}, and the blue long dot-dashed line shows the result of an AdS/CFT-inspired fit to lattice data \cite{Rougemont:2017tlu}. }
\label{fig:zeta}
\end{figure}

In Fig.~\ref{fig:zeta}, I show the result for the bulk viscosity to entropy density ratio.  I note that the peak value of $\zeta/s$ is smaller than obtained from some other analyses, but consistent with Bayesian extractions and AdS/CFT-inspired calculations of this ratio.  As can be seen from Fig.~\ref{fig:zeta}, the peak value obtained using aHydroQP is $\sim 0.05$, whereas in some other fits to data~\cite{Ryu:2015vwa}, the peak value can be as large as  $\zeta/s \sim 0.5$.  Finally, I point out that, in practice, one does not have to encode this in the dynamical equations since it is automatically incorporated.  In aHydroQP one does not have this as a independent function to fit to data.  In fact, not only this transport coefficient, but an infinite number of non-conformal transport coefficients are self-consistently included by the aHydroQP resummation, without the need for arbitrary parameterizations. 

\subsection{Leading-order aHydroQP evolution equations}
\label{sec:ahydroeqs}

The evolution equations for $u^\mu$, $\lambda$, and $\alpha_i$ are obtained from moments of the quasiparticle Boltzmann equation.  These can be expressed compactly by introducing a timelike vector $u^\mu$ which is normalized as \mbox{$u^\mu u_\mu = 1$} and three spacelike vectors $X_i^\mu$ which are individually normalized as \mbox{$X^\mu_i X_{\mu,i} = -1$}~\cite{Ryblewski:2010ch,Martinez:2012tu}.  These vectors are mutually orthogonal and obey $u_\mu X^\mu_i = 0$ and $X_{\mu,i} X^\mu_j = 0$ for $i \neq j$.  The four equations resulting from the first moment of the Boltzmann equation are
\ba
&& D_u\varepsilon +\varepsilon \theta_u + \sum_j P_j u_\mu D_j X^\mu_j = 0\, , \label{eq:1stmomOne} \\
&& D_i P_i+P_i\theta_i -\varepsilon X_{\mu,i} D_uu^\mu + P_i X_{\mu,i} D_i X^\mu_i  - \sum_j P_j X_{\mu,i} D_j X^\mu_j  = 0 \, , \;\;\;\;\;\;
\label{eq:1stmomTwo} 
\ea
where $D_u \equiv u^\mu \partial_\mu$ and $D_i \equiv X^\mu_i \partial_\mu$.  The expansion scalars are $\theta_u = \partial_\mu u^\mu$ and $\theta_i = \partial_\mu X^\mu_i$.  Expressions for the basis vectors, derivative operators, and expansion scalars appearing above can be found in Refs.~\cite{Nopoush:2014pfa,Alqahtani:2015qja,Alqahtani:2016rth,Alqahtani:2017tnq}.  The quantities $\varepsilon$ and $P_i$ are the energy density and pressures obtained using the anisotropic hydrodynamics ansatz for the one-particle distribution function including the background contribution $B(T)$ necessary to enforce thermodynamic consistency
\ba
\varepsilon &=& \varepsilon_{\rm kinetic}(\lambda,\vec\alpha,m) + B(\lambda,\vec\alpha) \, ,  \\
P_i &=& P_{i, \rm kinetic}(\lambda,\vec\alpha,m) - B(\lambda,\vec\alpha) \, .
\ea

The three equations resulting from the second moment of the Boltzmann equation are
\be
D_u I_i + I_i (\theta_u + 2 u_\mu D_i X_i^\mu)
= \frac{1}{\tau_{\rm eq}} \Big[ I_{\rm eq}(T,m) - I_i \Big] ,
\label{eq:2ndmoment} 
\ee
with \cite{Nopoush:2014pfa}
\ba
I_i &=& \alpha \, \alpha_i^2 \, I_{\rm eq}(\lambda,m) \, , \nonumber \\ 
I_{\rm eq}(\lambda,m) &=&  \hat{N} \lambda^5 \hat{m}^3 K_3(\hat{m}) \, ,
\ea
where $\hat{m} = m/\lambda$ and $\alpha = \alpha_x \alpha_y \alpha_z$.

Equations~\eqref{eq:1stmomOne}, \eqref{eq:1stmomTwo}, and \eqref{eq:2ndmoment} provide seven partial differential equations for $\vec{u}$, $\vec{\alpha}$, and $\lambda$, which can be solved numerically.  To determine the local effective temperature, one make uses of Landau matching; requiring the equilibrium and non-equilibrium energy densities in the LRF to be equal and solving for $T$.  This system of partial differential equations are evolved until the effective temperature in the entire simulation volume falls below a given freeze-out temperature, $T_{\rm FO}$.  

\subsection*{Initial transverse and longitudinal profiles}

For the results reviewed herein, the initial energy density distribution in the transverse plane is computed from a ``tilted'' profile \cite{Bozek:2010bi}.  The distribution used is a linear combination of smooth Glauber wounded-nucleon and binary-collision density profiles, with a binary-collision mixing factor of $\chi = 0.15$.  In the longitudinal direction, we used a profile with a central plateau and Gaussian ``tails'', resulting in a longitudinal profile function of the form 
\be
\rho(\varsigma) \equiv \exp \left[ - (\varsigma - \Delta \varsigma)^2/(2 \sigma_\varsigma^2) \, \Theta (|\varsigma| - \Delta \varsigma) \right] .
\label{eq:rhofunc}
\ee
We fit $\Delta\varsigma$ and $\sigma_{\varsigma}$ to the pseudorapidity distribution of charged hadron production, with the results being $\Delta\varsigma = 2.3$ and $\sigma_{\varsigma} = 1.6$~\cite{Alqahtani:2017jwl,Alqahtani:2017tnq,Alqahtani:2020paa} at LHC energies and $\Delta\varsigma = 1.4$ and $\sigma_{\varsigma} = 1.4$ in $\sqrt{s_{NN}} = 200$ GeV collisions \cite{Almaalol:2018gjh}.

The resulting initial energy density at a given transverse position ${\vec x}_T$ and spatial rapidity $\varsigma$ was computed using 
\be
{\cal E}({\vec x}_T,\varsigma) \propto (1-\chi) \rho(\varsigma) \Big[ W_A({\vec x}_T) g(\varsigma) + W_B({\vec x}_T) g(-\varsigma)\Big] + \chi \rho(\varsigma) C({\vec x}_T) \, ,
\ee
where $W_{A,B}({\vec x}_T)$ are the wounded nucleon densities for nuclei $A$ and $B$, $C({\vec x}_T)$ is the binary collision density, and $g(\varsigma)$ is 
\ba 
g(\varsigma) =
\left\{ \begin{array}{lcccc}
0  & \, & \mbox{if} & \,\,\,
& \varsigma < -y_N \, ,
 \\ (\varsigma+y_N)/(2y_N) & & \mbox{if} &
& -y_N \leq \varsigma \leq y_N \, , \\
1 & & \mbox{if} & 
& \varsigma > y_N\, ,
\end{array}\right. \,  
\ea 
where $y_N = \log(2\sqrt{s_{NN}}/(m_p + m_n))$ is the nucleon momentum rapidity \cite{Bozek:2010bi}.

\subsection{Freeze-out and hadronic decays}

From the 3+1D aHydroQP solution one extracts a three-dimensional freeze-out hypersurface at a fixed energy density (temperature).  For this purpose, we assume the same fluid anisotropy and scale parameters for all hadronic species.  We also assume that all hadrons are created in chemical equilibrium. Employing an extended Cooper-Frye prescription \cite{Alqahtani:2017mhy},  one translates the underlying hydrodynamic evolution parameters, including flow velocity, anisotropy parameters, and scale, into explicit `primordial' hadronic distribution functions on this hypersurface.  The aHydroQP parameters on the freeze-out hypersurface are then processed by a modified version of \textsc{Therminator 2}, which generates hadronic configurations using Monte Carlo sampling \cite{Chojnacki:2011hb}. Following the sampling of primordial hadrons, hadronic decays are accounted for using standard routines in \textsc{Therminator 2}. The source code for aHydroQP and the custom version of \textsc{Therminator 2} utilized are publicly accessible \cite{kent-code-library}.

\subsection{Phenomenological applications of leading-order aHydroQP}
\label{sec:qppheno}

Next, I discuss the phenomenological use of leading-order aHydroQP with a diagonal anisotropy tensor.  This includes application at \mbox{$\sqrt{s_{NN}} = 2.76$ TeV},  5.02 TeV, and 200 GeV.  I present comparisons with data from both LHC and RHIC.

\begin{figure}[t!]
\includegraphics[width=0.99\linewidth]{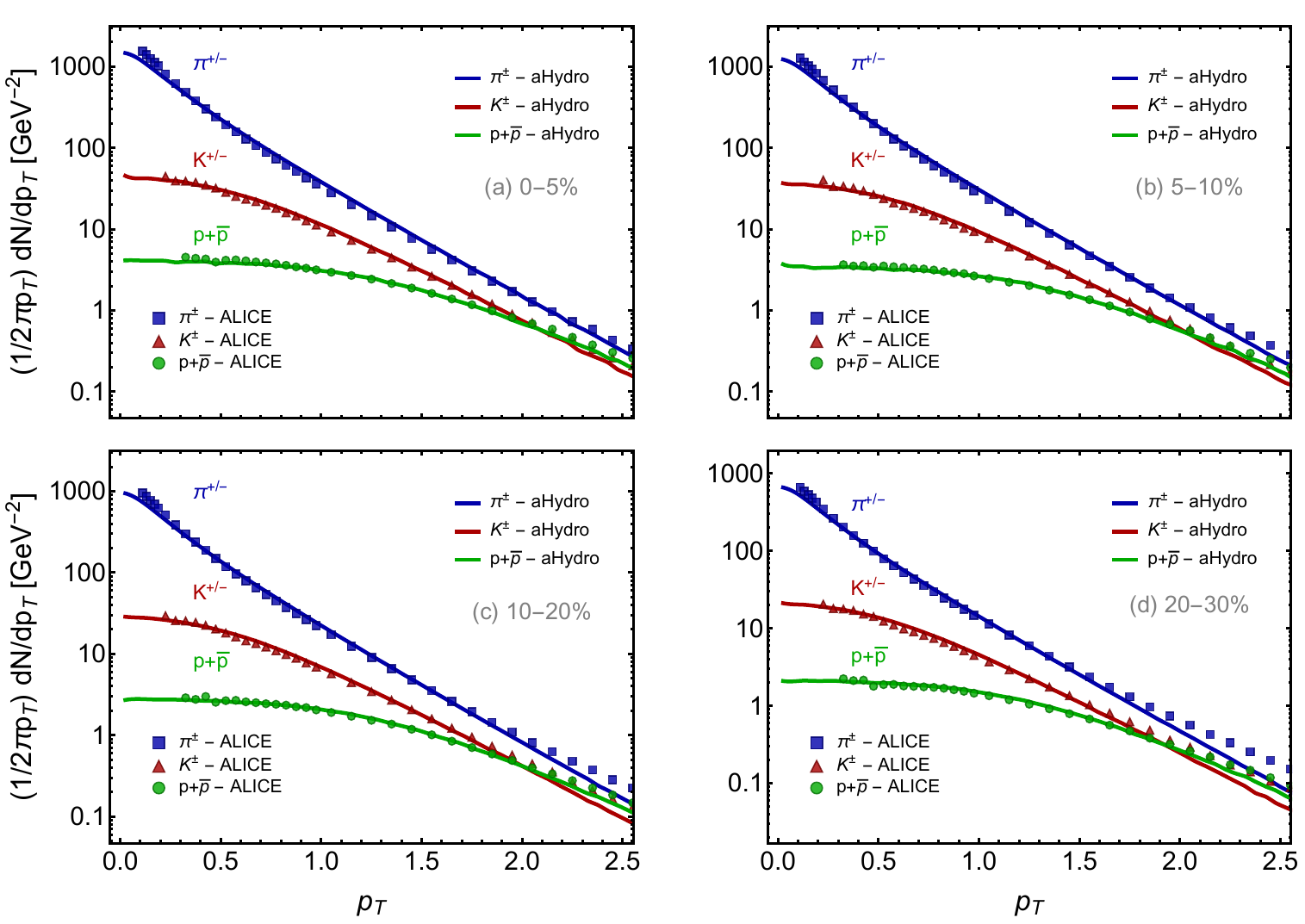}
\caption{The spectra of $\pi^\pm$, $K^\pm$, and $p+\bar{p}$ as a function of $p_T$ in four centrality classes for $\sqrt{s_{NN}} = 2.76$ TeV.  Data shown come from the ALICE Collaboration \cite{Abelev:2013vea}. The initial central temperature was taken to be $T_0= 600$ MeV at $\tau_0=0.25$ fm/c, $\eta/s = 0.159$, and $T_{\rm FO} = 130$ MeV. For full simulation details, I refer the reader to Ref.~\cite{Alqahtani:2017tnq}.}
\label{fig:spectra2.76}
\end{figure}

\begin{figure*}[t!]
\centerline{
\includegraphics[width=0.51\linewidth]{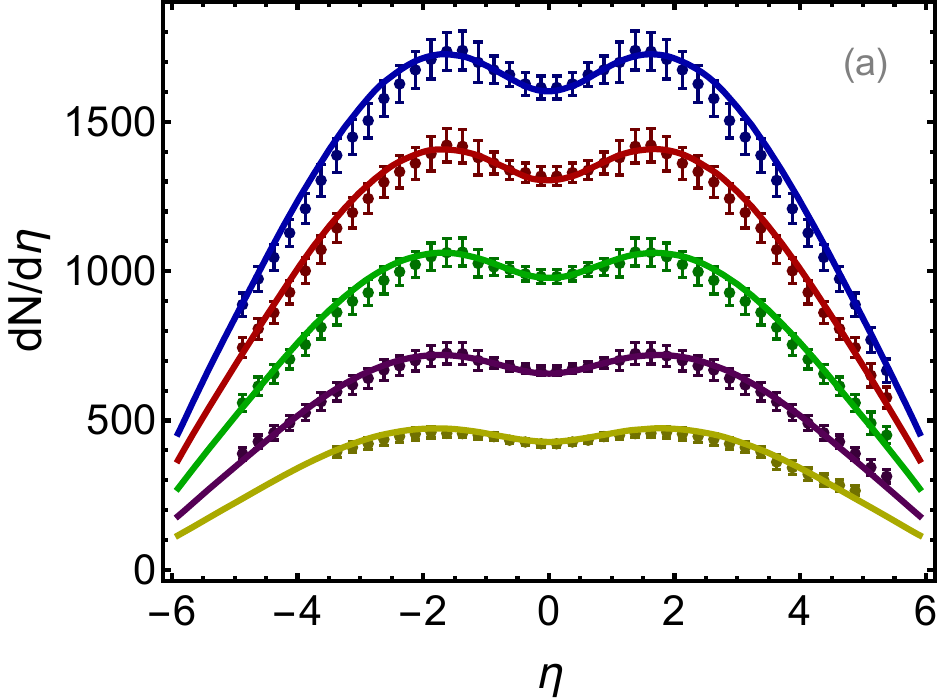}
\includegraphics[width=.49\linewidth]{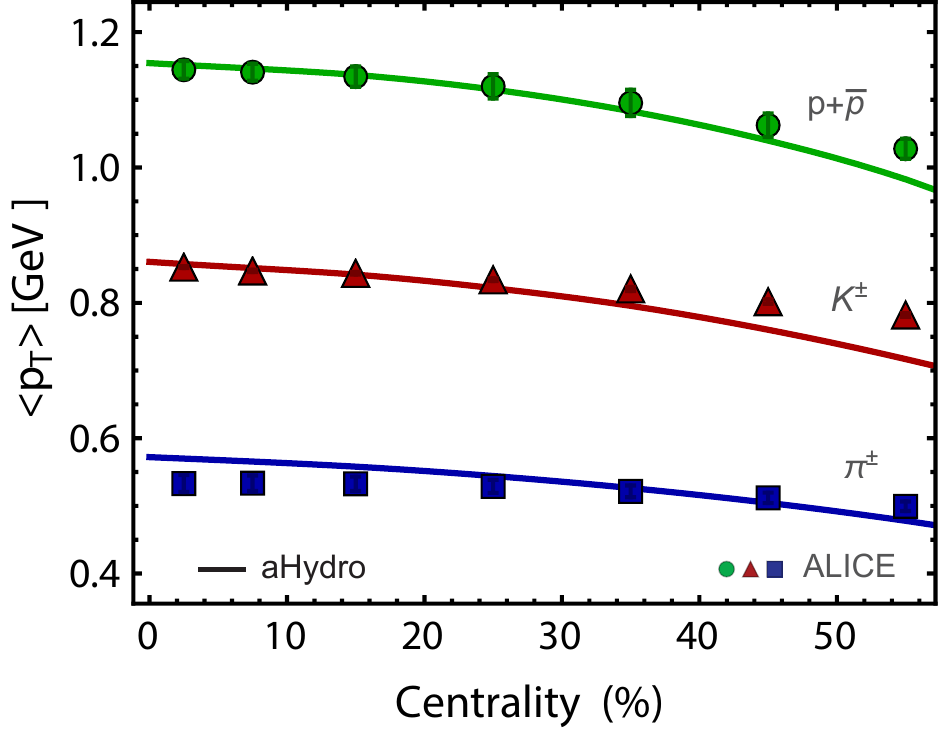}
}
\caption{In panel (a), the charged-hadron multiplicity $dN/d\eta$ as a function of pseudorapidity $\eta$  for $\sqrt{s_{NN}} = 2.76$ TeV is shown for five centrality classes (0-5\%, 5-10\%, 10-20\%, 20-30\%, and 30-40\%, from top to bottom) where  data  are from the ALICE Collaboration Refs.~\cite{Abbas:2013bpa,Adam:2015kda}. In panel (b), we show $\langle p_T \rangle $ of pions, kaons, and protons at $\sqrt{s_{NN}} = 2.76$ TeV as a function of centrality where data are from the ALICE Collaboration Ref.~\cite{Abelev:2013vea}. }
\label{fig:ptavg}
\end{figure*}

\begin{figure}[t!]
\includegraphics[width=1\linewidth]{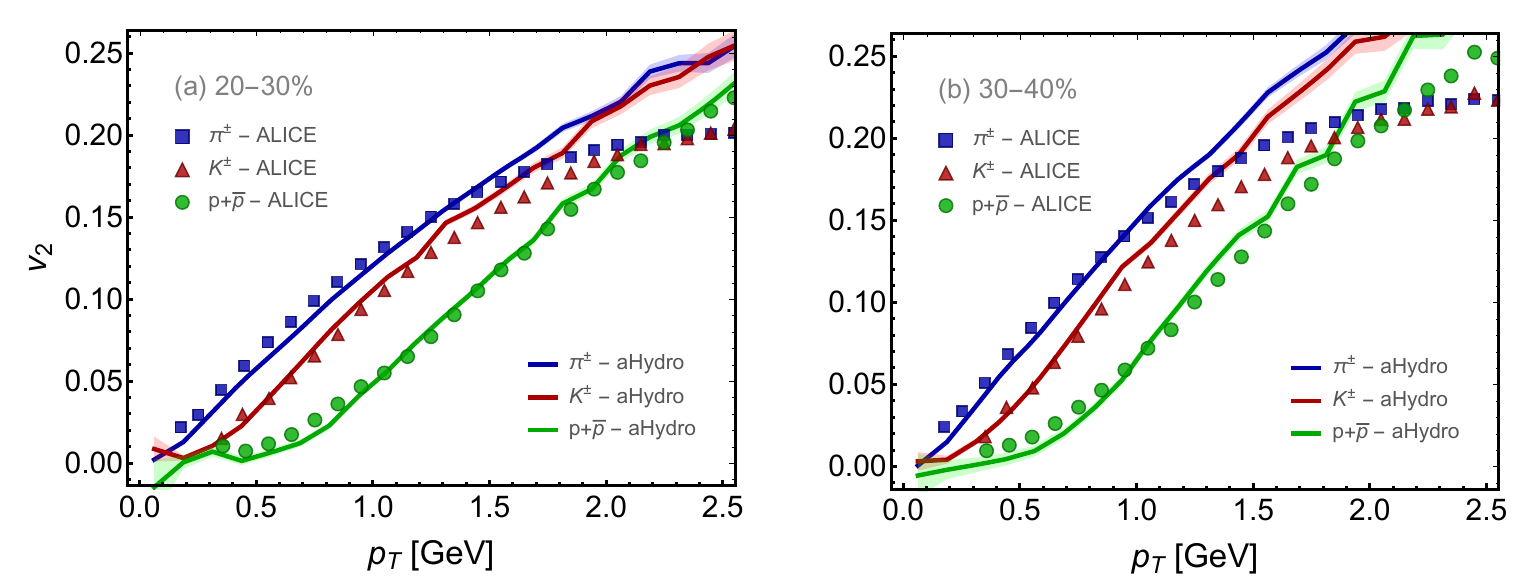}
\caption{The identified elliptic flow coefficient as a function of  $p_T$ is shown for $\pi^\pm$, $K^\pm$, and $p+\bar{p}$ in 20-30\% and 30-40\% centrality classes $\sqrt{s_{NN}} = 2.76$ TeV. The experimental data shown  are from the ALICE Collaboration~\cite{Abelev:2014pua}. }
\label{fig:v2}
\end{figure}

\begin{figure}[t!]
\centerline{
\hspace{-1.5mm}
\includegraphics[width=1\linewidth]{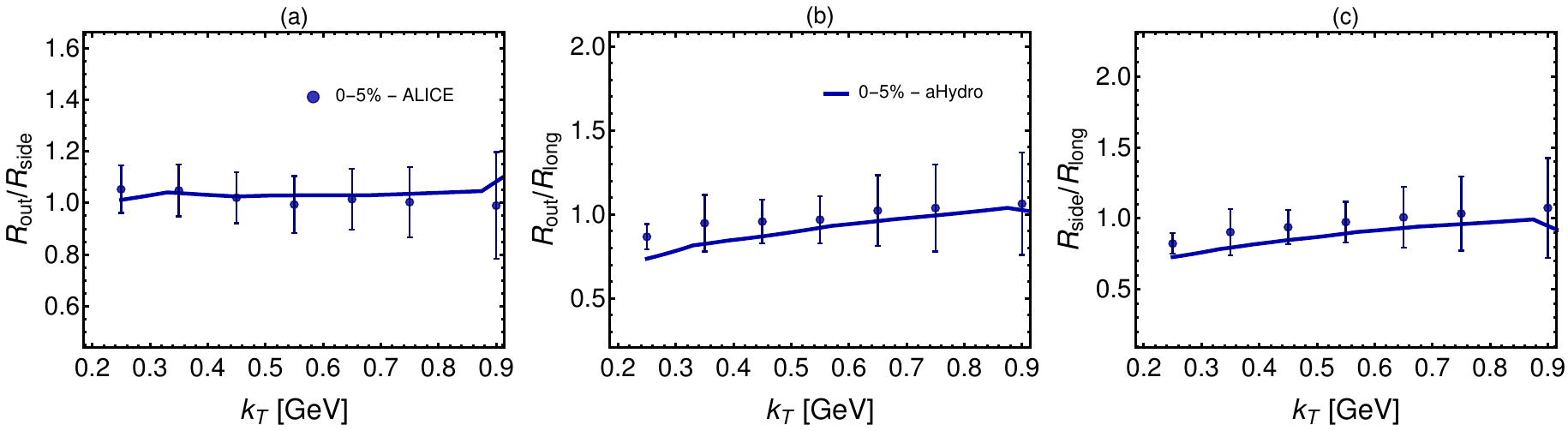}}
\centerline{
\hspace{-1.5mm}
\includegraphics[width=1\linewidth]{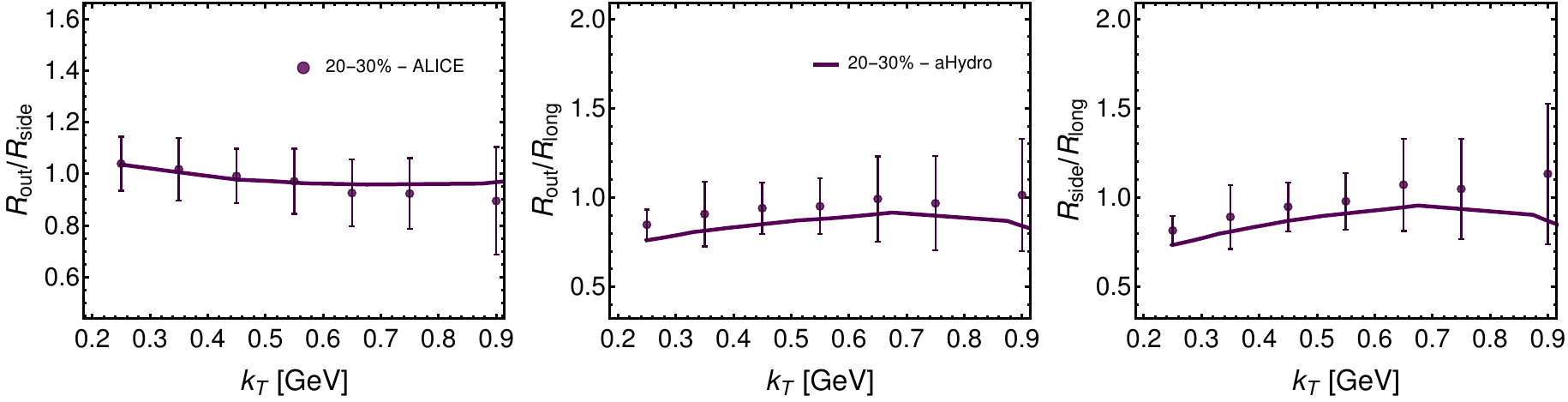}}
\caption{The HBT radii ratios at $\sqrt{s_{NN}} = 2.76$ TeV are shown as a function of $(k_T)$ for $\pi^+ \pi^+ $ in 0-5\% and 20-30\% centrality classes in the top row and bottom row respectively. The solid lines are the aHydroQP predictions and the experimental data are from the ALICE Collaboration~\cite{Graczykowski:2014hoa}.}
\label{fig:HBTratios}
\end{figure}

\subsubsection{2.76 TeV Pb-Pb collisions}

In Ref.~\cite{Alqahtani:2017tnq}, 2.76 TeV Pb-Pb collisions were considered.  I summarize the results found here.  In Fig.~\ref{fig:spectra2.76}, I show the spectra of pions, kaons, and protons as a function of the transverse momentum $p_T$ in four centrality classes 0-5\%, 5-10\%, 10-20\%, and 20-30\%. As can be seen from these comparisons, the aHydroQP model shows good agreement with the experimental data over the entire $p_T$ range shown, with some discrepancies at high $p_T$ in relatively higher centrality classes as shown in panel (d). From the spectra, one can compute the average transverse momentum for identified hadrons $\langle p_T \rangle_i$ and the total charged particle multiplicity $dN/d\eta$. In the left panel of Fig.~\ref{fig:ptavg}, I show the multiplicity as a function of pseudorapidity.  In the right panel, I show the average transverse momentum as a function of centrality for pions, kaons, and protons. On the left, I show the multiplicity in five centrality classes  (0-5\%, 5-10\%, 10-20\%, 20-30\%, and 30-40\%, from top to bottom), which demonstrates that aHydroQP describes the multiplicity quite well compared to available experimental data. In the right panel, I show $\langle p_T \rangle_i$ as a function of centrality.  Again, we see that aHydroQP agrees with the data quite well.

In Fig.~\ref{fig:v2}, the resulting identified elliptic flow is displayed as a function of $p_T$ in two centrality classes, 20-30\% and 30-40\%. As observed in this figure, aHydroQP describes the data well up to $p_T \sim 2$ GeV. However, at higher $p_T$, aHydroQP predictions deviate from the data, which is a common occurrence in hydrodynamic calculations since at high transverse momentum hard physics starts to dominate.   Next, comparisons of HBT radii ratios $R_{\rm out}/R_{\rm side}$, $R_{\rm out}/R_{\rm long}$, and $R_{\rm side}/R_{\rm long}$ are presented as a function of the pair relative momentum, $k_T$. The top row of Fig.~\ref{fig:HBTratios} shows the HBT ratios in the 0-5\% centrality class, while the bottom row displays the ratios in the 20-30\% centrality class. In both centrality classes, aHydroQP predictions agree well with experimental data.  In Ref.~\cite{Alqahtani:2017tnq}, we provide HBT ratios in more centrality classes, along with results for $R_{\rm side}$, $R_{\rm long}$, and $R_{\rm out}$, and more comparisons to experimental data.

\subsubsection{5.02 TeV Pb-Pb collisions}

\begin{figure}[t]
\centerline{
\includegraphics[width=0.95\linewidth]{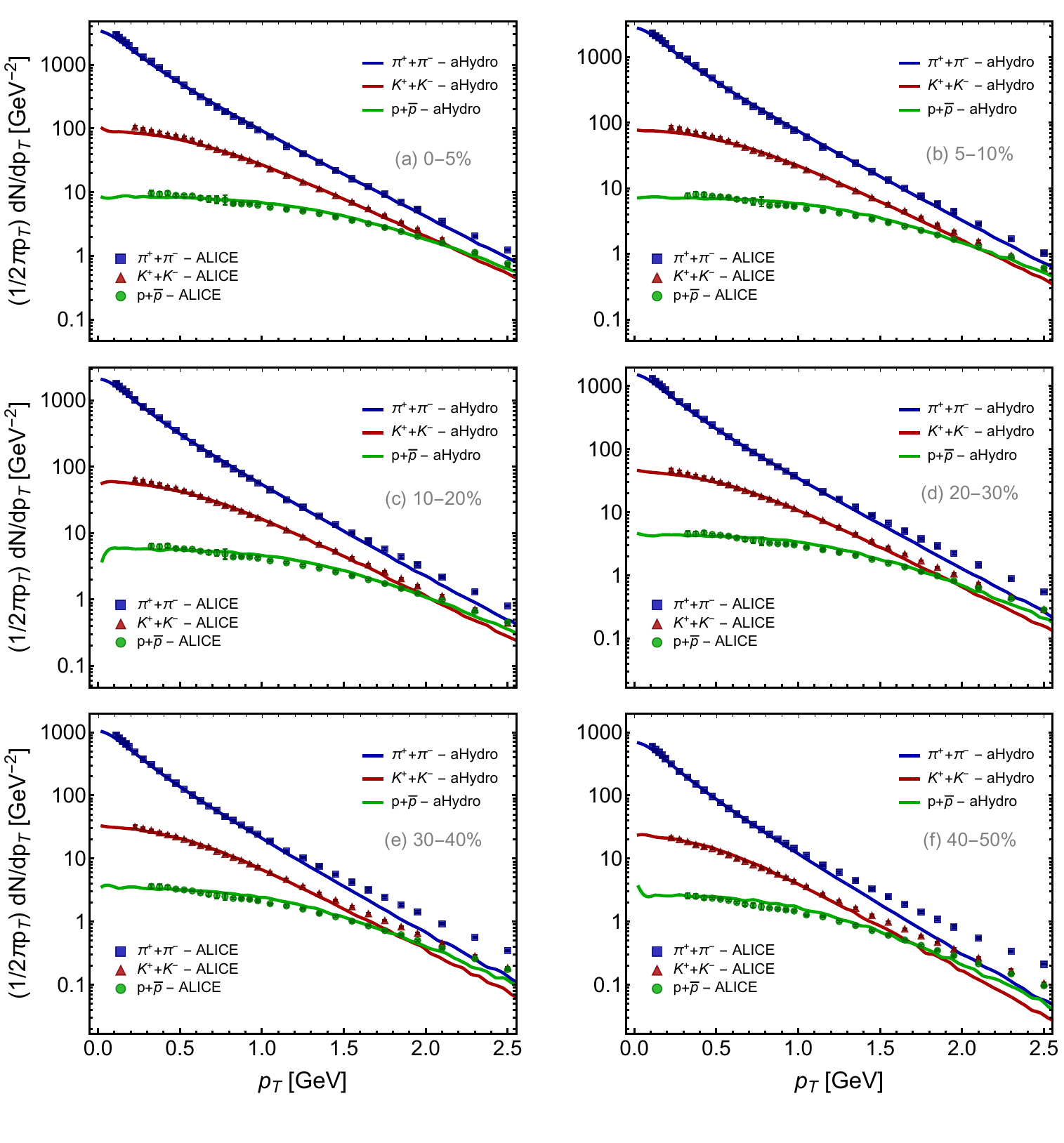}}
\caption{Combined transverse momentum spectra of pions, kaons, and protons for \mbox{5.02 TeV} Pb-Pb collisions in different centrality classes. The solid lines are the predictions of 3+1D aHydroQP and the points are experimental results from the ALICE Collaboration~\cite{ALICE:2019hno}. }
\label{fig:spectra}
\end{figure}

\begin{figure}[t]
\centerline{
\includegraphics[width=0.95\linewidth]{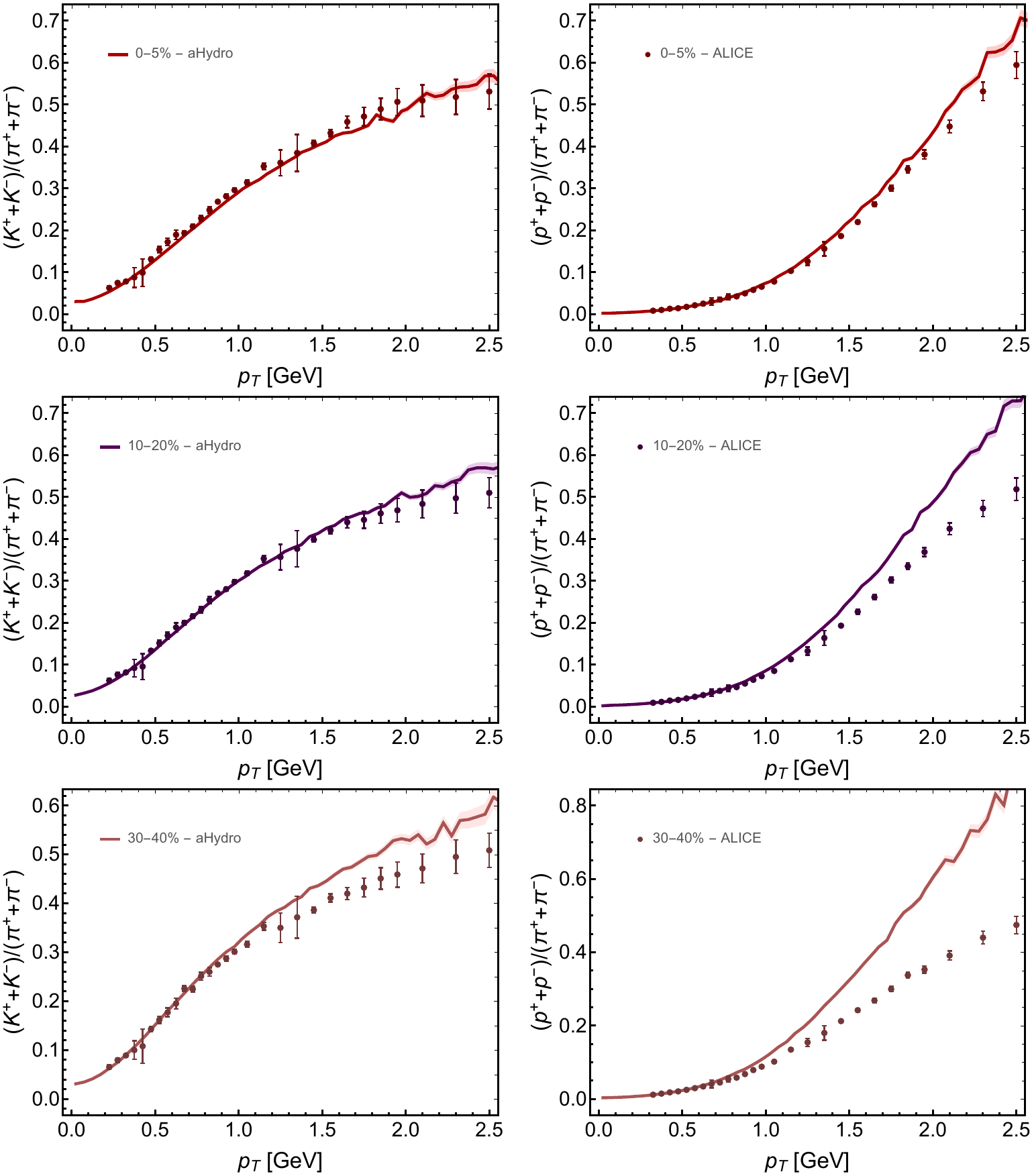}}
\caption{ The $K/\pi$ (left) and $p/\pi$ (right) ratios as a function of $p_T$ measured in Pb-Pb collisions at 5.02 TeV in different centrality classes. Solid lines are predictions of aHydroQP model where symbols with error bars are experimental data from Ref.~\cite{ALICE:2019hno}.}
\label{fig:ratios}
\end{figure}

\begin{figure}[t]
\centerline{
\includegraphics[width=0.95\linewidth]{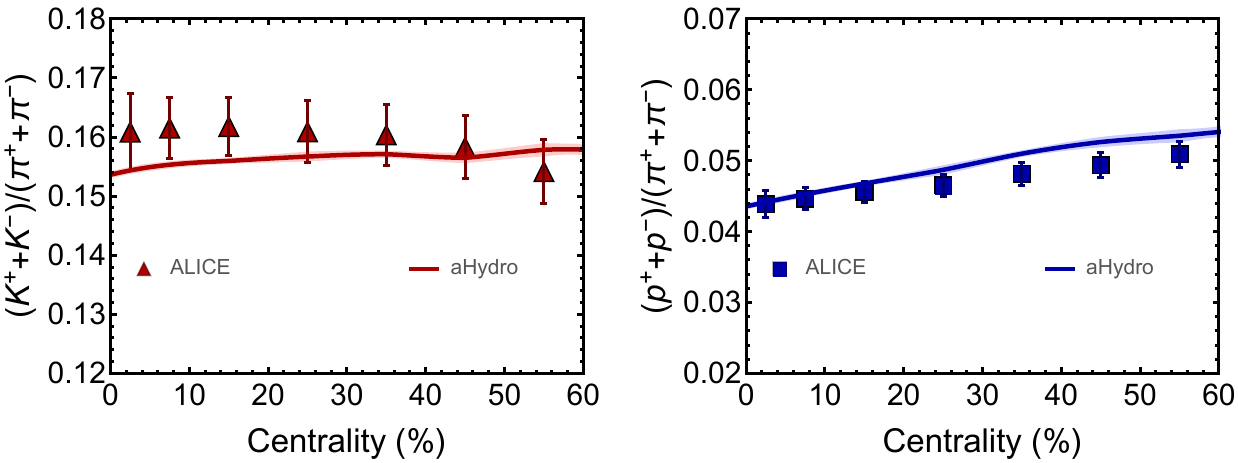}}
\caption{Transverse-momentum integrated $K/\pi$ (left) and $p/\pi$ (right) ratios as a function of centrality at 5.02 TeV. Solid lines are predictions of aHydroQP model.  Symbols with error bars are experimental data from the ALICE Collaboration~\cite{ALICE:2019hno}.   }
\label{fig:ratios-cent}
\end{figure}

In this section, I present comparisons between aHydroQP predictions and 5.02 TeV Pb-Pb collision data gathered by the ALICE collaboration.  This work appeared originally in Ref.~\cite{Alqahtani:2020paa}. I focus on two key free parameters: the initial central temperature \(T_0\) and the specific shear viscosity \(\eta/s\).  These parameters are determined by fitting to the transverse momentum spectra of pions, kaons, and protons in the 0-5\% and 30-40\% centrality classes. The obtained parameters from the spectra fit were: \(T_0 = 630\) MeV and \(\eta/s = 0.159\).  I note that the initial temperature derived is only slightly higher than that found at 2.76 TeV, which was \(T_0^\text{2.76 TeV} = 600\) MeV~\cite{Alqahtani:2017tnq}. The best fit value for \(\eta/s\) matches that found at 2.76 TeV \cite{Alqahtani:2017tnq}.

I begin by presenting aHydroQP's predictions for the transverse momentum distribution of identified hadrons in 5.02 TeV Pb-Pb collisions. These  predictions are compared with experimental data from the ALICE collaboration \cite{Jacazio:2017dvy,ALICE:2019hno}. In Fig.~\ref{fig:spectra}, I present the combined spectra of pions, kaons, and protons as a function of transverse momentum across six distinct centrality classes. At low centralities, aHydroQP demonstrates remarkable consistency with the data, as depicted in Fig.~\ref{fig:spectra}(a). Conversely, for more peripheral collisions, satisfactory agreement is observed only for \(p_T \lesssim 1\) GeV.

In Fig.~\ref{fig:ratios}, the \(K/\pi\) ratio (left column) and \(p/\pi\) ratio (right column) are presented as a function of \(p_T\) in three distinct centrality classes and compared to experimental data. Notably, the agreement between aHydroQP and the data at \(p_T \lesssim 1\) GeV is very good across all centrality bins. Specifically, in the 0-5\% centrality class, the agreement between aHydroQP and the data for the \(K/\pi\) ratio persists up to \(p_T \sim 2.5\) GeV, while for \(p/\pi\), it extends up to \(p_T \sim 1.5\) GeV. The aHydroQP predictions for the integrated \(K/\pi\) and \(p/\pi\) ratios as a function of centrality are shown in the left and right panels of Fig.~\ref{fig:ratios-cent}, respectively. In both panels, we see that aHydroQP describes the centrality dependence of the integrated multiplicity ratios quite well across a wide range of centralities.

\subsubsection{200 GeV Au-Au collisions}

\begin{figure*}[t!]
\centerline{
\includegraphics[width=0.95\linewidth]{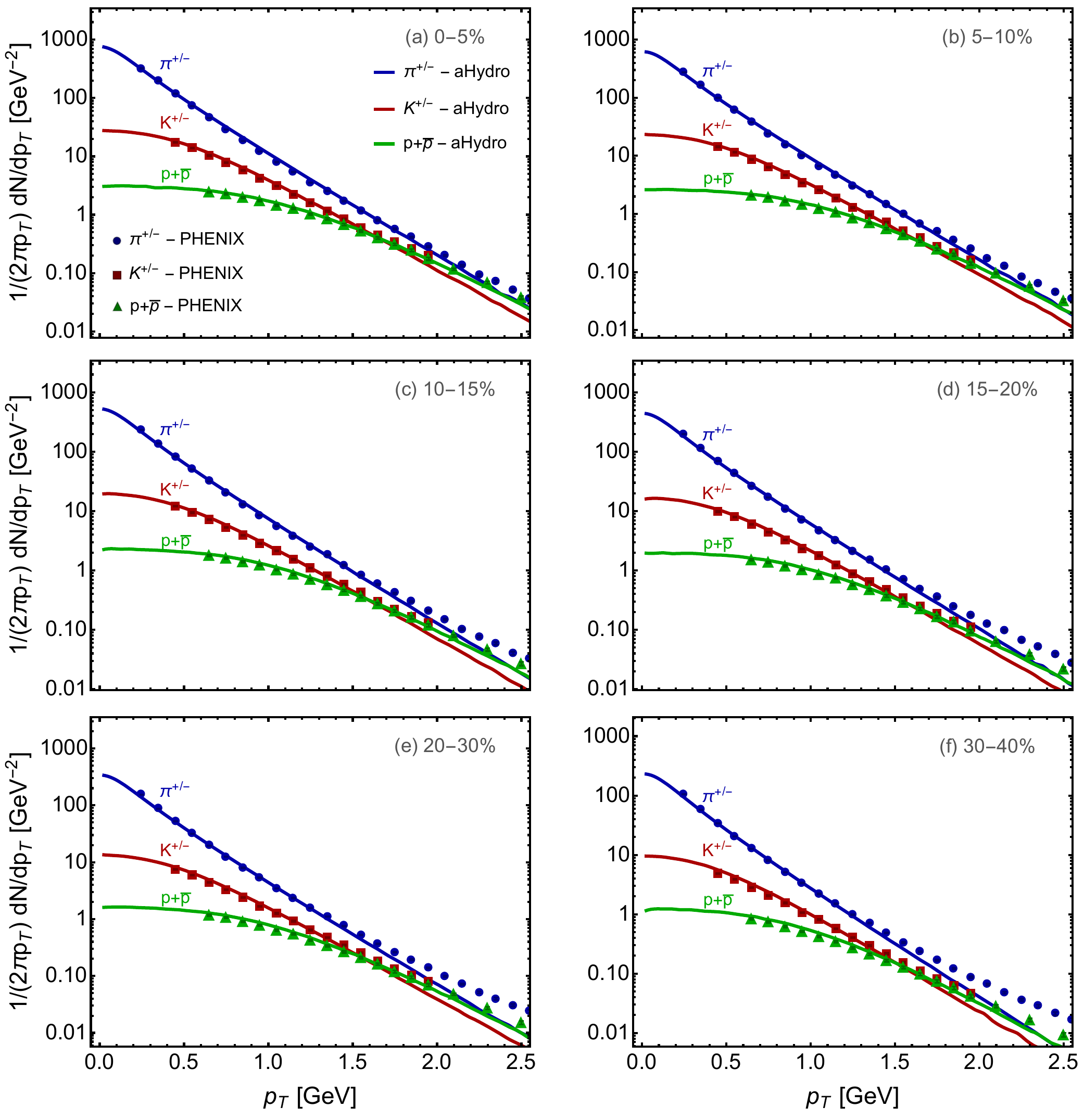}
}
\caption{Pion, kaon, and proton spectra for 200 GeV Au-Au collisions compared to experimental observations by the PHENIX Collaboration \cite{PHENIX:2003iij}.  The panels show the centrality classes (a) 0-5\%, (b) 5-10\%, (c) 10-15\%, (d) 15-20\%, (e) 20-30\%, and (f) 30-40\%. 
}
\label{fig:spectra-all}
\end{figure*}

\begin{figure*}[t!]
\centerline{
\includegraphics[width=0.75\linewidth]{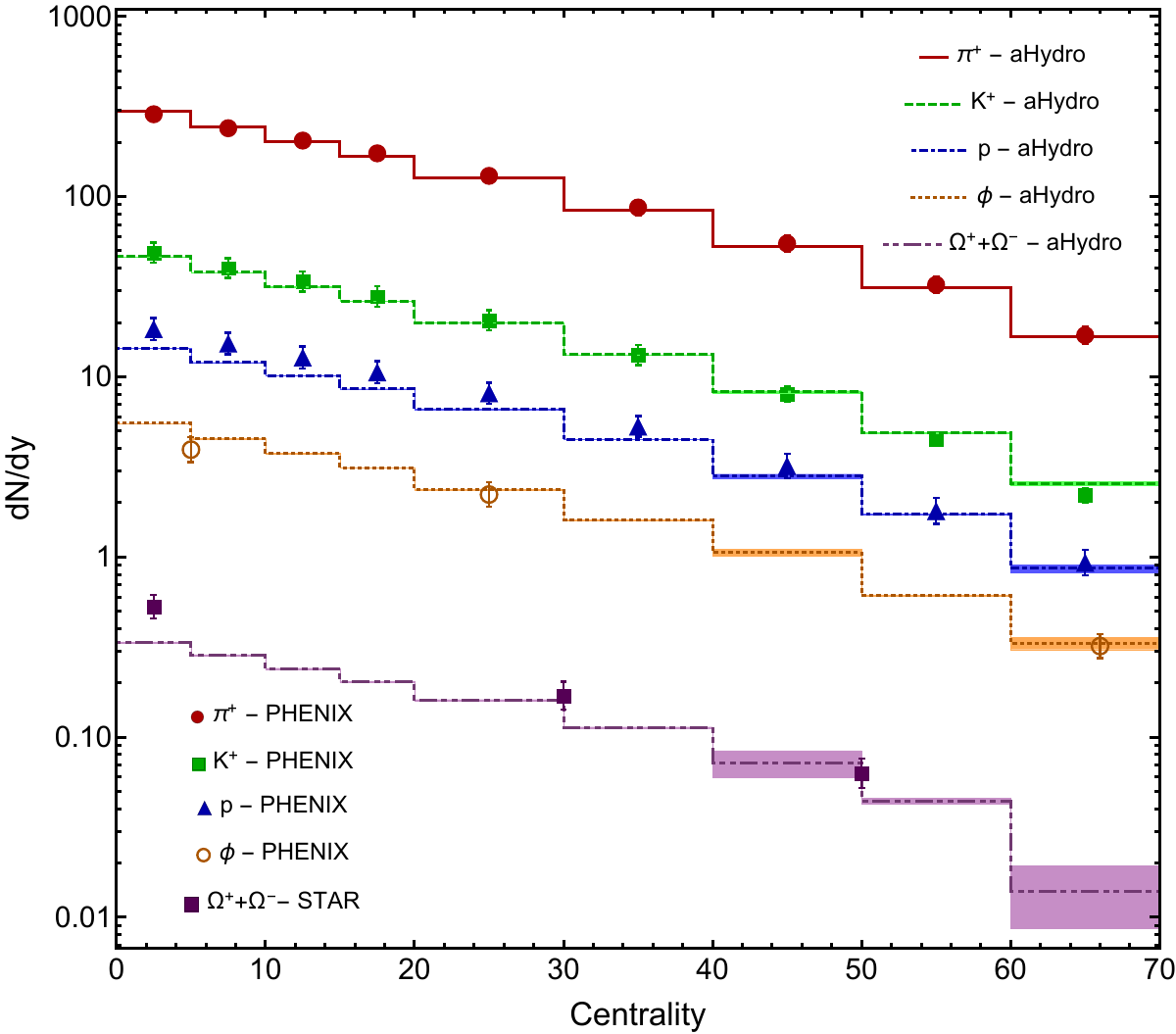}
}
\caption{Identified particle multiplicities as a function of centrality for 200 GeV Au-Au collisions.  From top to bottom, the particles shown are $\pi^+$, $K^+$, $p$, $\phi$, and $\Omega^+ + \Omega^-$.  The data for $\pi^+$, $K^+$, and $p$ are from the PHENIX Collaboration \cite{Adler:2003cb}.  Data for the $\phi$ meson production are also from the PHENIX Collaboration \cite{Adams:2004ux}.  The data for $\Omega^+ + \Omega^-$ comes from the STAR Collaboration \cite{Adams:2006ke}.  The aHydroQP theory results are binned using the centrality bins used by PHENIX Collaboration for $\pi^+$, $K^+$, and $p$.
}
\label{fig:identified-particle-multiplicity}
\end{figure*}

Next, I review results obtained using aHydroQP in 200 GeV Au-Au collisions \cite{Almaalol:2018gjh}.  Drawing from our prior investigation of 2.76 TeV collisions at LHC \cite{Alqahtani:2017tnq}, we adopted a fixed switching (freeze-out) temperature of $T_{\rm FO} = 130$ MeV. This choice leaves only the shear viscosity to entropy density ratio $\eta/s$ and the initial central temperature $T_0$ (the center of the system for a $b=0$ collision) as independent parameters. Similarly to before, we assume that $\bar\eta$ is independent of the temperature. To determine $T_0$ and $\bar\eta$, we conducted comparisons between model predictions and observed pion, proton, and kaon spectra in the 0-5\% and 30-40\% centrality classes. These comparisons yielded $T_0 = 455$ MeV at $\tau_0 = 0.25$ fm/c and $\bar\eta = 0.179$. The resulting fits to the pion, kaon, and proton spectra are depicted in Fig.~\ref{fig:spectra-all} and compared with experimental data from the PHENIX Collaboration \cite{PHENIX:2003iij}. As demonstrated by this figure, the model effectively describes the identified particle spectra with this parameter set. However, in high centrality classes, we observe that the model underestimates hadron production at large transverse momenta, $p_T \gtrsim 1.5$ GeV.

In Fig.~\ref{fig:identified-particle-multiplicity}, I show our findings for the identified particle multiplicities as a function of centrality. From top to bottom, the particles depicted are $\pi^+$, $K^+$, $p$, $\phi$, and $\Omega^+ + \Omega^-$. The data for $\pi^+$, $K^+$, and $p$ are from the PHENIX Collaboration Ref.~\cite{Adler:2003cb}, while the data for the $\phi$ meson are from the PHENIX Collaboration Ref.~\cite{Adams:2004ux}. The $\Omega^+ + \Omega^-$ data are from the STAR collaboration \cite{Adams:2006ke}. The aHydroQP model results are binned according to the centrality bins used by the PHENIX Collaboration for $\pi^+$, $K^+$, and $p$. As illustrated by this figure, aHydroQP, coupled with our customized version of Therminator 2, effectively replicates the centrality dependence of observed identified particle multiplicities. This is particularly noteworthy given that we utilized a single iso-thermal switching (freeze-out) temperature, which is relatively low ($T_{\rm FO} = 130$ MeV), yet are able to reasonably reproduce observed identified particle multiplicities not only for central collisions but across many centrality classes.

\begin{figure*}[t!]
\centerline{
\includegraphics[width=1\linewidth]{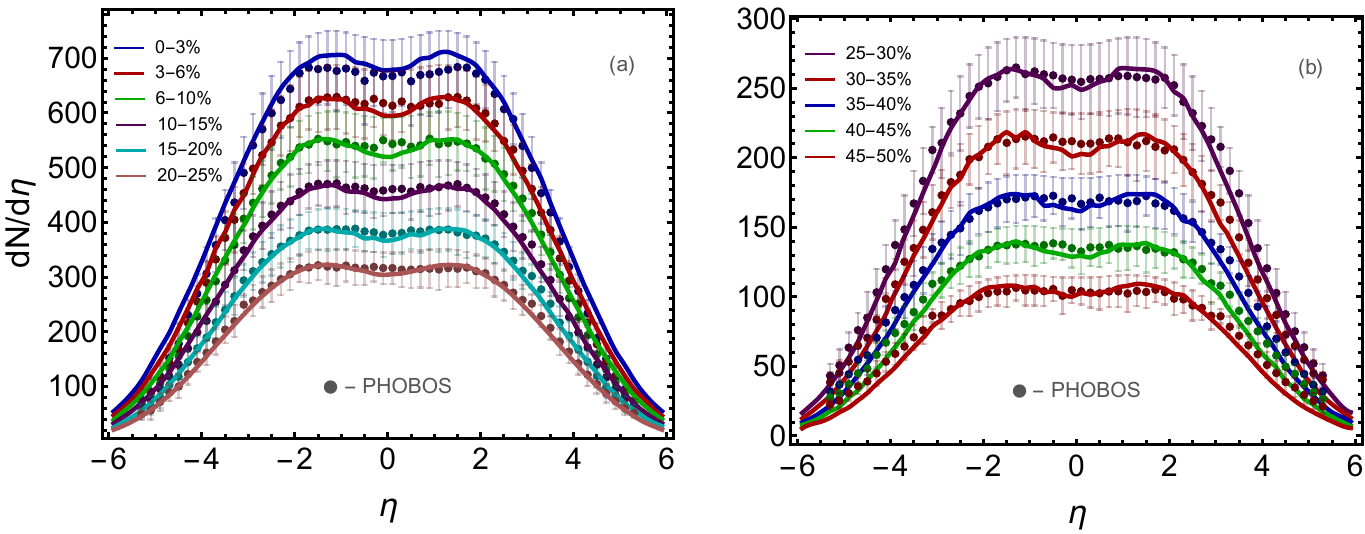}
}
\caption{aHydroQP results for charged particle multiplicity in different centrality classes (solid lines) for 200 GeV Au-Au collisions compared to experimental data from the PHOBOS collaboration \cite{Alver:2010ck}.  Panel (a) shows centrality classes in the range 0-25\% and panel (b) shows centrality classes in the range 25-50\%.}
\label{fig:nPlotall}
\end{figure*}

Finally, in Fig.~\ref{fig:nPlotall} I present the aHydroQP model results for the charged particle multiplicity as a function of pseudorapidity in 200 GeV Au-Au collisions.  In this figure, I compare to the data reported by the PHOBOS Collaboration \cite{Alver:2010ck}.  In the figure, panel (a) shows centrality classes in the range 0-25\% and panel (b) shows centrality classes in the range 25-50\%.  As can be seen from this figure, once the initial central temperature and shear viscosity to entropy density ratio are fit to the identified spectra, aHydroQP is able to describe the centrality and pseudorapidity dependence of the charged particle multiplicity quite well.

\subsubsection{5.02 and 8.16 TeV $p$-Pb collisions}

\begin{figure}[t]
\begin{center}
\includegraphics[width=0.6\linewidth]{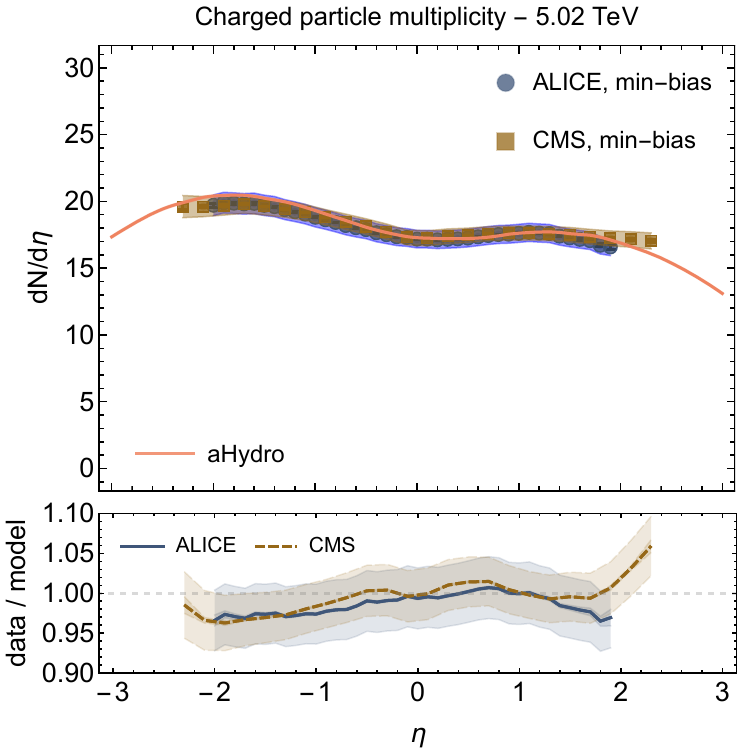}
\end{center}
\vspace{-5mm}
\caption{Min-bias charged particle multiplicity $dN/d\eta$ for $\sqrt{s_{NN}} = $ 5.02 TeV $p$-Pb collisions.
ALICE and CMS collaboration data are from Refs.~\cite{ALICE:2012xs} and \cite{CMS:2017shj}, respectively.  The top panel compares the results of the aHydroQP model to the experimental data while the bottom panel shows the relative error.  In both panels, the shaded regions indicate the reported experimental uncertainty.}
\label{fig:dndeta-5TeV}
\end{figure}

\begin{figure}[t]
\begin{center}
\includegraphics[width=1\linewidth]{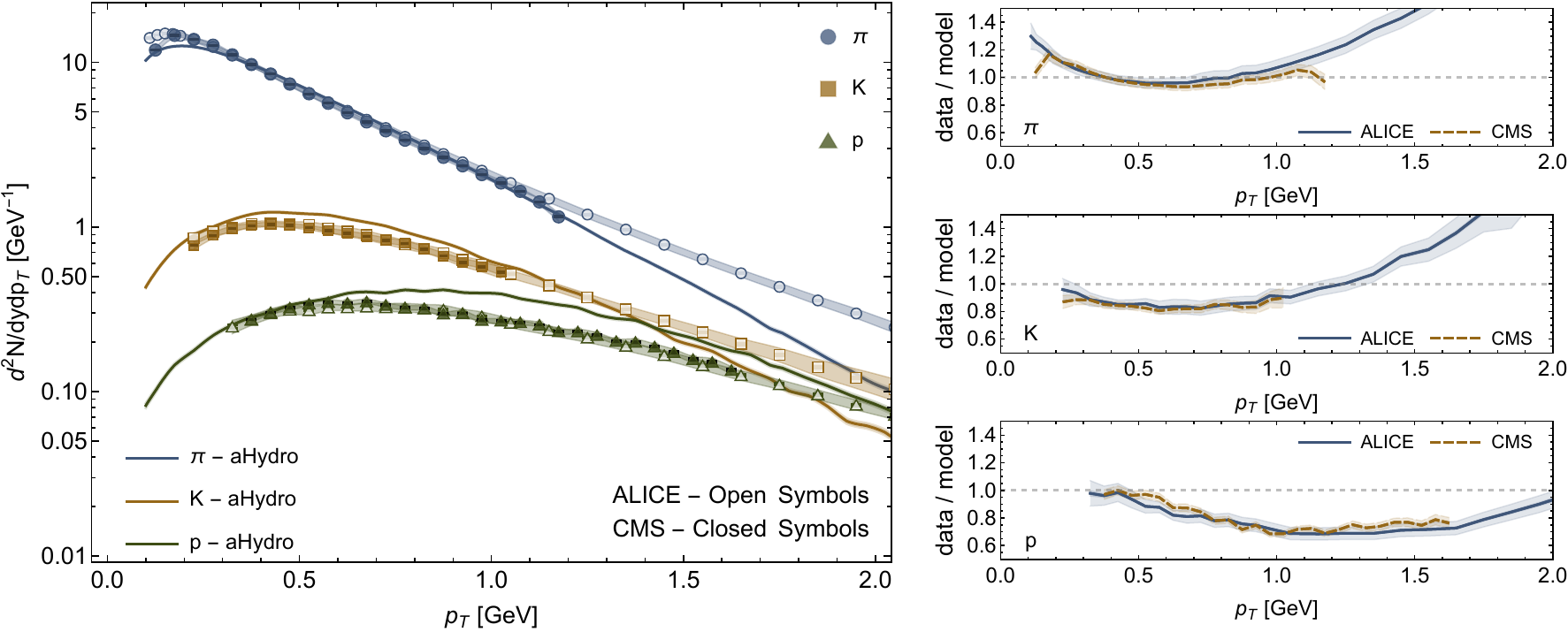}
\end{center}
\vspace{-5mm}
\caption{Min-bias identified particle spectra of pions, kaons, and protons in $p$-Pb collisions at $\sqrt{s_{NN}} = $ 5.02 TeV compared to aHydroQP predictions.  Experimental data from the ALICE and CMS collaborations are from Refs.~\cite{ALICE:2013wgn} and \cite{CMS:2013pdl}, respectively.  The left panel shows model comparisons to the data and the three right panels show the relative error.  In both panels, the shaded regions indicate the reported experimental uncertainty.} 
\label{fig:spectra-5TeV}
\end{figure}

Next, I discuss recent results obtained for $p$-Pb collisions at 5.02 and \mbox{8.16 TeV} in Ref.~\cite{Strickland:2024oat}, which used the same setup as the RHIC and LHC results presented above.  Here, I focus on 5.02 TeV and results for 8.16 TeV can be found in Ref.~\cite{Strickland:2024oat}. In this case we, again, use a tilted profile and fit the central width $\Delta\varsigma$ to the pseudorapidity distribution of charged hadron production.  This gives \mbox{$\Delta\varsigma = 1.8$} for $\sqrt{s_{NN}} =$ 5.02 TeV collisions and \mbox{$\Delta\varsigma = 2.1$} for \mbox{$\sqrt{s_{NN}} =$ 8.16 TeV} collisions.  The width of the Gaussian tails, $\sigma_{\varsigma}$, is largely unconstrained based on available $p$-Pb data.  As a result, we have used $\sigma_{\varsigma} = 1.6$, which was used previously in Pb-Pb collisions at both $\sqrt{s_{NN}} =$ 5.02 TeV and 2.76 TeV~\cite{Alqahtani:2017jwl,Alqahtani:2017tnq,Alqahtani:2020paa}.

The resulting initial energy density at a given transverse position ${\vec x}_\perp$ and spatial rapidity $\varsigma$ was computed using~\cite{Bozek:2013uha} 
\be
\varepsilon({\vec x}_\perp,\varsigma) \propto (1-\chi) \rho(\varsigma) \Big[ W_p({\vec x}_\perp) g(\varsigma) + W_A({\vec x}_\perp) g(-\varsigma)\Big] %
+ \; \chi \rho(\varsigma) C({\vec x}_\perp) \, ,
\ee
where $W_A({\vec x}_\perp)$ is the wounded-nucleon density for nucleus $A$ \cite{Florkowski2010-tl}, $C({\vec x}_\perp)$ is the binary collision density~\cite{Florkowski2010-tl}, and $g(\varsigma)$ is the tilt function introduced earlier.  To compute $W_p$, we parameterize the proton overlap function as~\cite{Skands:2014pea,dEnterria:2020dwq}
\be
T_p({\vec b})= \frac{n}{2\pi r^2_p \,\Gamma (2/n)} \exp{[-(b/r_p)^n]}\,,
\label{eq:overlap}
\ee
with $n = 1.85$ and $r_p = 0.975$ fm.

To evaluate the tuning of aHydroQP in $p$-Pb collisions, Figs.~\ref{fig:dndeta-5TeV} and \ref{fig:spectra-5TeV} present comparisons of our aHydroQP $p$-Pb results with standard soft hadron observables measured at LHC energies. Fig.~\ref{fig:dndeta-5TeV} presents our aHydroQP results for the charged particle multiplicity as a function of pseudorapidity, $\eta \equiv \tfrac{1}{2} \ln((p+p_z)/(p-p_z))$, at $\sqrt{s_{NN}} = $ 5.02 and \mbox{$\sqrt{s_{NN}} = $ 8.16 TeV}. In this figure, comparisons are made with data from the ALICE \cite{ALICE:2012xs,ALICE:2013wgn} and CMS \cite{CMS:2013pdl} collaborations. The top panel displays the model results alongside ALICE and CMS data, while the bottom panel shows the relative error (data/model) of the aHydroQP results. As evident from  Fig.~\ref{fig:dndeta-5TeV}, the aHydroQP model can reproduce the charged particle multiplicity measurements of the ALICE and CMS collaborations to within approximately 5\% in the reported pseudorapidity range. This level of agreement is comparable to or better than all model comparisons presented in Refs.~\cite{ALICE:2012xs,ALICE:2013wgn,CMS:2013pdl}.

In Fig.~\ref{fig:spectra-5TeV}, the computed transverse momentum spectra in $p$-Pb collisions at $\sqrt{s_{NN}} = $ 5.02 TeV for pions, kaons, and protons are compared to data from the ALICE \cite{ALICE:2013wgn} and CMS collaborations \cite{CMS:2013pdl}. In the left panel, comparisons of the identified spectra obtained using aHydroQP with experimental data are shown, while the three right panels show the relative error (data/model). As shown in this figure, aHydroQP is capable of reproducing the experimental observations for $p_T \lesssim 1.5$ GeV reasonably well. The remaining discrepancies are comparable to those encountered with other models presented in Refs.~\cite{ALICE:2013wgn,CMS:2013pdl}, thereby instilling confidence in the ability of aHydroQP to provide a reasonable description of min-bias $p$-Pb collisions.  For the details concerning the $p$-Pb results presented in this subsection, see Ref.~\cite{Strickland:2024oat}.

\section{Phenomenological results from viscous anisotropic hydrodynamics}

Finally, I would like to highlight recent phenomenological results that make use of viscous anisotropic hydrodynamics (vaHydro or VAH).  VAH is a model that  goes beyond leading-order aHydro by including non-spheroidal corrections in a linearized manner akin to second order viscous hydrodynamics~\cite{Bazow:2013ifa,Bazow:2015cha,Bazow:2016yra,McNelis:2018jho,McNelis:2021zji,Liyanage:2023nds}.  In a recent paper \cite{Liyanage:2023nds} a comprehensive Bayesian analysis using the JETSCAPE framework was performed, replacing the free streaming and hydrodynamical model components by the VAH model.  Prior to showing the main results of this study, I quickly sketch the VAH framework.  For details, I refer the reader to Refs.~\cite{Bazow:2013ifa,Bazow:2015cha,Bazow:2016yra,McNelis:2018jho,McNelis:2021zji,Liyanage:2023nds}.

To setup VAH, one starts by decomposing the spatial projector $\Delta^{\mu\nu}$ into projectors along the beam direction ($z^\mu$) and a transverse projector $\Xi^{\mu\nu}$ such that $\Delta^{\mu\nu}=\Xi^{\mu\nu}-z^\mu z^\nu$. Multiplying these projectors by distinct longitudinal and transverse pressures, the energy-momentum tensor can then decomposed as \cite{Molnar:2016vvu}
\be
\label{eq:vah}
T^{\mu\nu} = \varepsilon u^\mu u^\nu + P_L z^\mu z^\nu - P_T  \Xi^{\mu\nu} + 2 W_{T z}^{(\mu} z^{\nu)} + \pi_T^{\mu \nu}.
\ee
This decomposition allows VAH to evolve the longitudinal and transverse pressures $P_L$ and $P_T$ separately, treating them at the same level as the thermal pressure in standard viscous hydrodynamics.  Importantly, it is not by assumed that their differences from the thermal pressure and from one another are small.

The evolution equations for the energy density and the flow velocity are obtained from the conservation laws for energy and momentum
\be
\partial_\mu T^{\mu \nu} = 0 \, .
\ee
The equilibrium pressure $P(\varepsilon)$ is taken from lattice QCD calculations conducted by the HotQCD collaboration \cite{Bazavov:2017dsy}. The dynamical evolution equations for the non-equilibrium flows $P_L$, $P_T$, $W_{T z}^{(\mu} z^{\nu)}$, and $\pi_T^{\mu \nu}$ are derived under the assumption that the fluid's microscopic physics can be described by the relativistic Boltzmann equation with a medium-dependent mass \cite{Alqahtani:2016rth, Tinti:2016bav, McNelis:2018jho}, as was the case with aHydroQP. The relaxation times for $P_L$ and $P_T$ are expressed in terms of those for the bulk and shear viscous pressures \cite{McNelis:2018jho}, which are parameterized as
\ba
\tau_\pi &=& \frac{\eta}{s\beta_\pi} \, , \\
\tau_\Pi &=& \frac{\zeta}{s\beta_\Pi} \, .
\ea
The functions $\beta_\pi$, $\beta_\Pi$, along with all necessary anisotropic transport coefficients, are computed within the quasiparticle kinetic theory model in Ref.~\cite{McNelis:2018jho}.

For the temperature-dependent specific shear and bulk viscosities, $\bigl(\eta/s\bigr)(T)$ and $\bigl(\zeta/s\bigr)(T)$, the authors of Ref.~ \cite{Liyanage:2023nds} used the same parameterizations as the JETSCAPE Collaboration \cite{JETSCAPE:2020mzn}, namely
\be
 \Bigl(\frac{\eta}{s}\Bigr)(T) = \max\left[\left.\frac{\eta}{s}\right\vert_{\rm lin}\!\!\!\!(T), \; 0\right] ,
\ee
with
\be
\left.\frac{\eta}{s}\right\vert_{\rm lin}\!\!(T) = a_{\rm low}\, (T{-}T_{\eta})\, \Theta(T_{\eta}{-}T)+ (\eta/s)_{\rm kink} + a_{\rm high}\, (T{-}T_{\eta})\, \Theta(T{-}T_{\eta}) \, ,
\ee
and
\be
 \Bigl(\frac{\zeta}{s}\Bigr)(T) = \frac{(\zeta/s )_{\max}\Lambda^2}{\Lambda^2+ \left( T-T_\zeta\right)^2} \, ,
\ee
with
\be
\Lambda = w_{\zeta} \Bigl(1 + \lambda_{\zeta} \sign \left(T{-}T_\zeta\right) \Bigr) .
\ee

In addition to the eight parameters related to the viscosities that are included in the above parameterizations, there was one additional parameter inferred from the experimental data, namely the initial pressure ratio $R=(P_L/P_T)_0$ at the time $\tau_0$ that VAH was initialized
\ba
    P_{T,0} &=& \frac{3}{2{+}R}\, P_0 \, , \\
    P_{L,0} &=& \frac{3R}{2+R}\, P_0 \, .
\label{eq:R}
\ea
Above $P_0{\,\equiv\,}P(\varepsilon_0)$ is the equilibrium pressure at the initial proper time. The authors assumed that the initial bulk viscous pressure was zero. The initial flow profile was assumed to be given by Bjorken flow in the longitudinal direction, with zero initial transverse velocities, and the initial residual shear stresses $W_{T z}^\mu$ and $\pi_T^{\mu \nu}$ were taken to be zero.  At freeze-out, as done in aHydroQP, the authors sampled from an anisotropic momentum-space distribution that is non-negative for all momentum.

\begin{figure}[t]
\centering
\includegraphics[width=\linewidth]{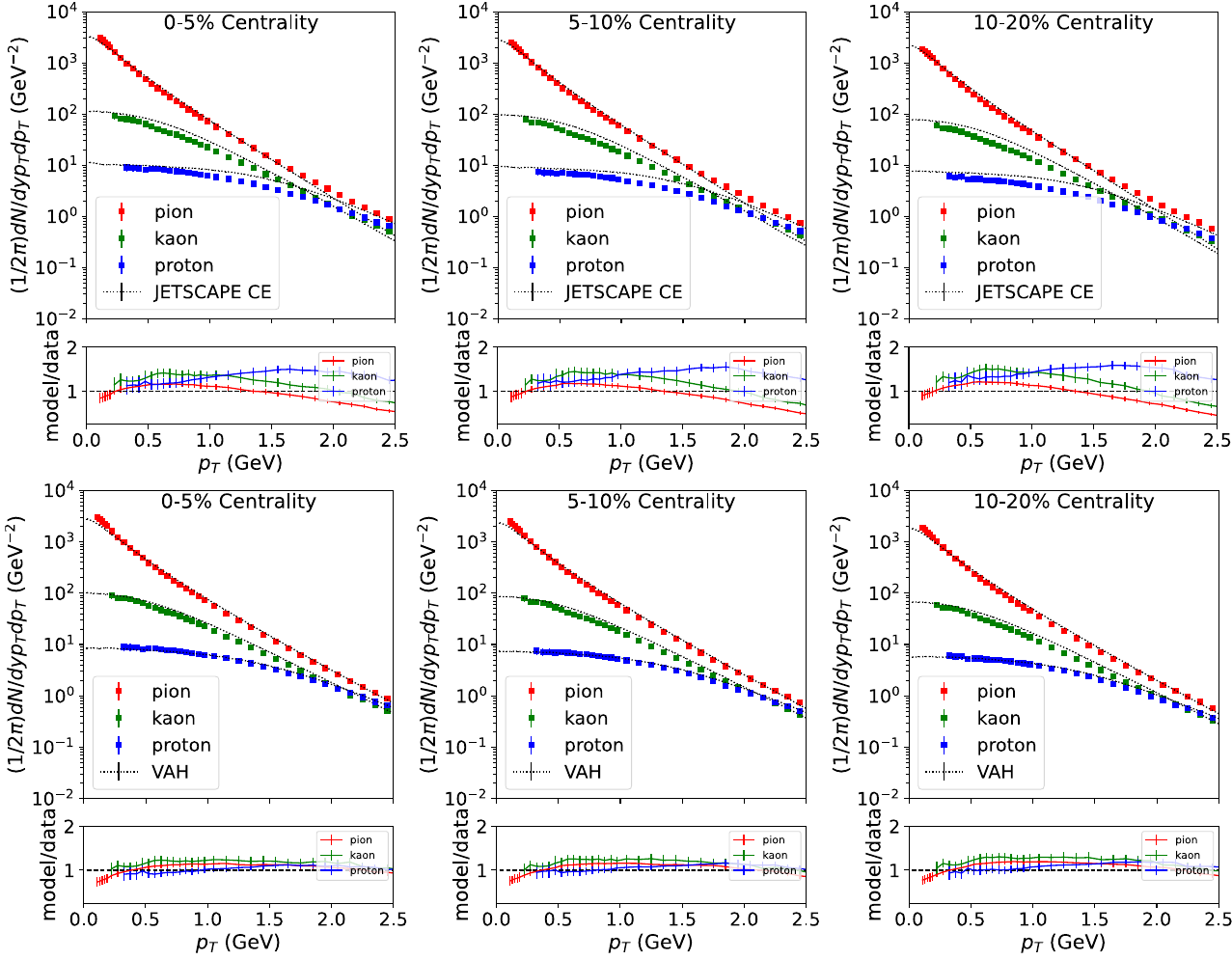}
\caption{The $p_T$-spectra of identified hadrons, $\pi$ (red), $K$ (green), $p$ (blue), predicted by the best-fit parameters of the calibrated JETSCAPE SIMS model with Chapman-Enskog (CE) particlization \cite{JETSCAPE:2020shq,JETSCAPE:2020mzn} (top row) and the calibrated VAH model \cite{Liyanage:2023nds} (bottom row), for 2.76 TeV Pb-Pb collisions at three collision centralities (from left to right: 0-5\%, 5-10\%, 10-20\% \cite{ALICE:2015dtd}). Figure from Ref.~\cite{Heinz:2023kzr}.}
\label{fig:VAH-2}       
\end{figure}

\begin{figure}[t]
\centering
\includegraphics[width=\linewidth]{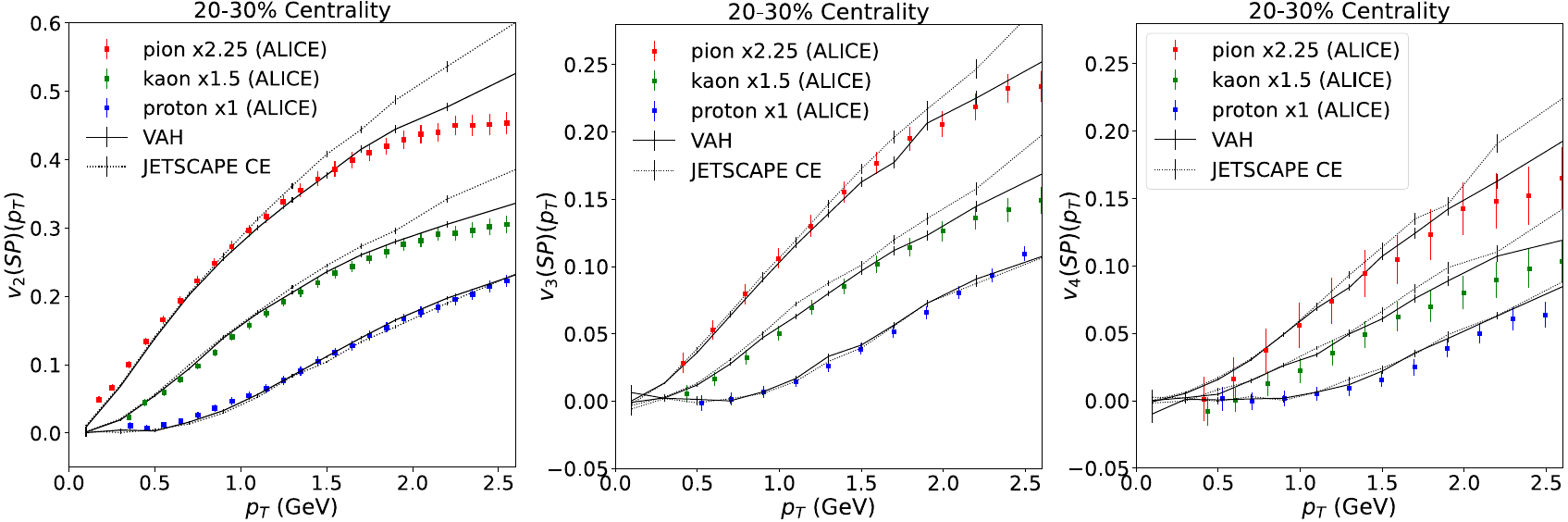}
\caption{The $p_T$-differential elliptic $v_2\{\mathrm{SP}\}$ (left), triangular $v_3\{\mathrm{SP}\}$ (middle), and quadrangular $v_4\{\mathrm{SP}\}$ (right) flows for identified pions, kaons, and protons.  Results predicted by the best-fit parameters of the calibrated JETSCAPE SIMS model with Chapman-Enskog (CE) particlization \cite{JETSCAPE:2020shq,JETSCAPE:2020mzn} are indicated by dotted lines and the calibrated VAH model \cite{Liyanage:2023nds} by solid lines.  Results are for 2.76 TeV Pb-Pb collisions at 20-30\% centrality.  The experimental data are from the ALICE Collaboration \cite{Noferini:2012ps,ALICE:2011ab}. Figure from Ref.~\cite{Heinz:2023kzr}. }  
\label{fig:VAH-3}       
\end{figure}

\begin{figure}[t]
\centering
\includegraphics[width=0.9\textwidth,clip]{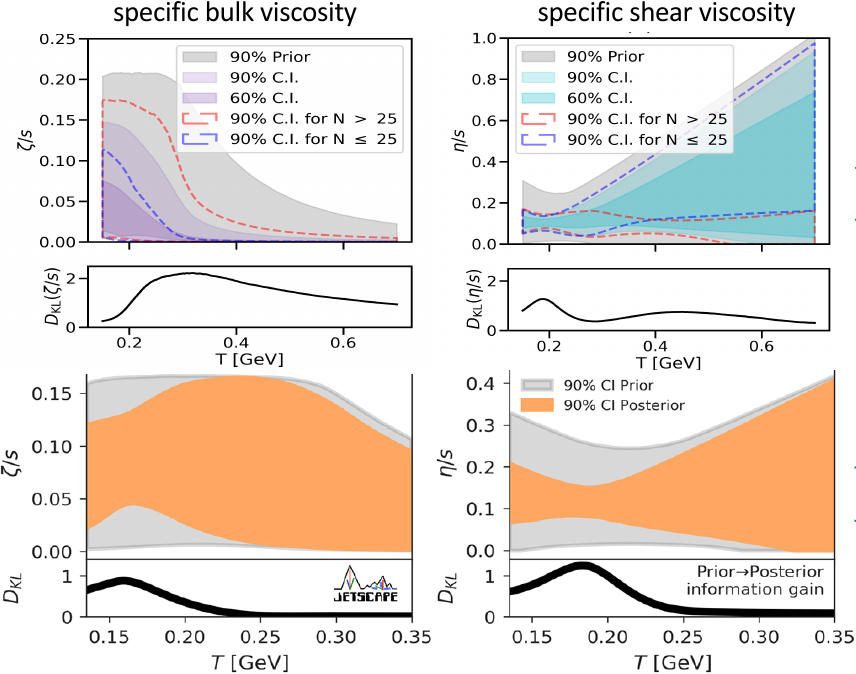}
\hspace*{1mm}
\caption{The prior (gray) and posterior (colored) credibility intervals for the temperature-dependent specific bulk (left) and shear (right) viscosities inferred from the VAH model \cite{Liyanage:2023nds} (top) and from the JETSCAPE SIMS model \cite{JETSCAPE:2020shq} (bottom). The lower subpanels illustrate the Kullback-Leibler relative entropy $D_\mathrm{KL}$, quantifying the information gain provided by the experimental data utilized for model calibration. Note that the temperature ranges differ between the top and bottom rows.  Figure from Ref.~\cite{Heinz:2023kzr}. }
\label{fig:VAH-1}       
\end{figure}

The VAH model replaces the free-streaming and relativistic viscous hydrodynamics stages by a single viscous anisotropic hydrodynamic module, coupled to its own anisotropic freeze-out/particlization routine. This eliminates the free-streaming time parameter (and an associated discontinuity in the equation of state \cite{JETSCAPE:2020mzn}) and extends the sensitivity of the model to the QGP viscosities into the far-from-equilibrium stage.  This allows VAH to be applied at much higher temperatures than in other available hydrodynamical evolution models.  

In Fig.~\ref{fig:VAH-2}, I show results from Refs.~\cite{Heinz:2023kzr,Liyanage:2023nds} for the $p_T$-spectra of identified hadrons, $\pi$ (red), $K$ (green), $p$ (blue), predicted by the best-fit parameters of the calibrated JETSCAPE SIMS model with Chapman-Enskog (CE) particlization \cite{JETSCAPE:2020shq,JETSCAPE:2020mzn} (top row) and the calibrated VAH model \cite{Liyanage:2023nds} (bottom row), for 2.76 TeV Pb-Pb collisions at three collision centralities (from left to right: 0-5\%, 5-10\%, 10-20\% \cite{ALICE:2015dtd}).  As can be seen from this figure the VAH best fit results agree better with the experimental data than the calibrated JETSCAPE SIMS result.  This is quantified by the model/data ratios, which are closer to unity when using the calibrated VAH model than when using the calibrated JETSCAPE SIMS result.

In Fig.~\ref{fig:VAH-3}, I show the VAH results \cite{Heinz:2023kzr,Liyanage:2023nds} for the $p_T$-differential elliptic $v_2\{\mathrm{SP}\}$ (left), triangular $v_3\{\mathrm{SP}\}$ (middle), and quadrangular $v_4\{\mathrm{SP}\}$ (right) flows for identified pions, kaons, and protons.  The results predicted by the best-fit parameters of the calibrated JETSCAPE SIMS model with Chapman-Enskog (CE) particlization \cite{JETSCAPE:2020shq,JETSCAPE:2020mzn} are indicated by dotted lines and the calibrated VAH model \cite{Liyanage:2023nds} by solid lines.  Results are for 2.76 TeV Pb-Pb collisions at 20-30\% centrality.  The experimental data are from the ALICE Collaboration \cite{Noferini:2012ps,ALICE:2011ab}.  As this figure demonstrates, the VAH predictions exhibit better agreement with the data compared to the JETSCAPE SIMS predictions, across all three particlization models examined in Refs.~\cite{JETSCAPE:2020shq,JETSCAPE:2020mzn} and at all available collision centralities. 

Finally, from the Bayesian analysis performed in Refs.~\cite{Liyanage:2023nds} it is possible to extract confidence intervals on the extracted transport coefficients.   In Fig.~\ref{fig:VAH-1}, I show the VAH results  \cite{Heinz:2023kzr,Liyanage:2023nds} for the prior (gray) and posterior (colored) credibility intervals for the temperature-dependent specific bulk (left) and shear (right) viscosities inferred from the VAH model \cite{Liyanage:2023nds} (top) and from the JETSCAPE SIMS model \cite{JETSCAPE:2020shq} (bottom).   In the subpanels, the authors plotted the Kullback-Leibler divergence $D_\mathrm{KL}$, which quantifies the information gain provided by the experimental data.  From this figure, we see that $D_\mathrm{KL}$ is non-zero up to temperatures of around 700\,MeV in the VAH model and, at low temperatures, the VAH model provides tighter constraints than JETSCAPE SIMS.  I note, importantly, that the temperature dependence of the bulk viscosity to entropy density ratio extracted using VAH (see the 60\% confidence interval) is consistent with the result obtained using leading-order non-linearized aHydroQP, in which it was found that the bulk viscosity to entropy density ratio was below 5\% at all temperatures (see Fig.~\ref{fig:zeta}).

\section{Conclusions and outlook}

In this chapter, I have made a connection between our understanding of non-equilibrium attractors and the phenomenological predictions of anisotropic hydrodynamics.  The motivation for aHydro comes from the understanding that a high degree of momentum-space anisotropy exists at early times in the QGP's lifetime and also near the transverse edges.   This can be easily seen from the attractor since small $w = \tau T$ corresponds either to small $\tau$ (early times) or low temperature regions (transverse and longitudinal edges).  

I demonstrated that aHydro provides the most accurate reproduction of exact solutions to the Boltzmann equation for 0+1D systems and the associated attractor for the one-particle distribution function.  Although not discussed in detail here, this also holds true for non-trivial conformal flows like Gubser flow \cite{Nopoush:2014qba,Martinez:2017ibh} and non-conformal systems \cite{Nopoush:2014pfa}.  I also discussed tests of different freeze-out formulations using QCD EKT simulations in which it was found that, like in RTA, an attractor for the full one-particle distribution function exists and its form is better described by the anisotropic aHydro ansatz that linearized viscous hydrodynamics ansatze.  This is particularly important for peripheral collisions and small systems, where the lifetime of the QGP may be short and the system may not have time enough to approach a near-equilibrium configuration for which linearized approaches are applicable.

Taking this together, I then reviewed various phenomenological applications of aHydroQP, including to Pb-Pb collisions at 2.76 and 5.02 TeV, Au-Au collisions at 200 GeV, and finally $p$-Pb collisions at 5.02 and \mbox{8.16 TeV}.  In all cases, one finds good agreement between the aHydroQP results and experimental data for spectra, multiplicities, elliptic flow, HBT radii, etc.  I closed by highlighting some very recent results using the viscous anisotropic hydrodynamics (VAH) framework, which relies on single anisotropy parameter (spheroidal deformation) plus linearized corrections in the spirit of second order viscous hydrodynamics.  The results shown indicate that the calibrated VAH did a better job at reproducing experimental observations than the calibrated JETSCAPE SIMS framework.  In addition, the VAH model allows one to dispense with the free streaming phase and, therefore, access early times when the temperature is large.  This allows VAH to, in principle, constrain QGP transport coefficients at much higher temperatures that standard simulation chains.

\section*{Acknowledgments}

I thank my collaborators for their contributions to the work reported herein.  This work was supported by the U.S.\ Department of Energy, Office of Science, Office of Nuclear Physics (Nuclear Theory) under contract number DE-SC0013470.

\bibliographystyle{ws-rv-van}
\bibliography{strickland}

\begin{thebibliography}{185}
\providecommand{\natexlab}[1]{#1}
\providecommand{\url}[1]{\texttt{#1}}
\expandafter\ifx\csname urlstyle\endcsname\relax
  \providecommand{\doi}[1]{doi: #1}\else
  \providecommand{\doi}{doi: \begingroup \urlstyle{rm}\Url}\fi

\bibitem{Heinz:2013th}
U.~Heinz and R.~Snellings, {Collective flow and viscosity in relativistic
  heavy-ion collisions}, \emph{Ann. Rev. Nucl. Part. Sci.} {\bf 63}, \penalty0
  123--151  (2013).
\newblock \doi{10.1146/annurev-nucl-102212-170540}.

\bibitem{Chapman_Enskog}
S.~Chapman and T.~G. Cowling, \emph{Cambridge mathematical library: The
  mathematical theory of non-uniform gases: An account of the kinetic theory of
  viscosity, thermal conduction and diffusion in gases}. Cambridge University
  Press, Cambridge, England  (January, 1991).

\bibitem{Heller:2013fn}
M.~P. Heller, R.~A. Janik, and P.~Witaszczyk, {Hydrodynamic Gradient Expansion
  in Gauge Theory Plasmas}, \emph{Phys. Rev. Lett.} {\bf 110}\penalty0 (21),
  \penalty0 211602  (2013).
\newblock \doi{10.1103/PhysRevLett.110.211602}.

\bibitem{Buchel:2016cbj}
A.~Buchel, M.~P. Heller, and J.~Noronha, {Entropy Production, Hydrodynamics,
  and Resurgence in the Primordial Quark-Gluon Plasma from Holography},
  \emph{Phys. Rev. D}. {\bf 94}\penalty0 (10), \penalty0 106011  (2016).
\newblock \doi{10.1103/PhysRevD.94.106011}.

\bibitem{Heller:2016rtz}
M.~P. Heller, A.~Kurkela, M.~Spali\'nski, and V.~Svensson, {Hydrodynamization
  in kinetic theory: Transient modes and the gradient expansion}, \emph{Phys.
  Rev. D}. {\bf 97}\penalty0 (9), \penalty0 091503  (2018).
\newblock \doi{10.1103/PhysRevD.97.091503}.

\bibitem{Denicol:2016bjh}
G.~S. Denicol and J.~Noronha, {Divergence of the Chapman-Enskog expansion in
  relativistic kinetic theory}  (8, 2016).

\bibitem{Heinz:2004qz}
U.~W. Heinz.
\newblock {Concepts of heavy ion physics}.
\newblock In \emph{{2nd CERN-CLAF School of High Energy Physics}}, pp. 165--238
   (7, 2004).

\bibitem{Chesler:2008hg}
P.~M. Chesler and L.~G. Yaffe, {Horizon formation and far-from-equilibrium
  isotropization in supersymmetric Yang-Mills plasma}, \emph{Phys. Rev. Lett.}
  {\bf 102}, \penalty0 211601  (2009).
\newblock \doi{10.1103/PhysRevLett.102.211601}.

\bibitem{Beuf:2009cx}
G.~Beuf, M.~P. Heller, R.~A. Janik, and R.~Peschanski, {Boost-invariant early
  time dynamics from AdS/CFT}, \emph{JHEP}. {\bf 10}, \penalty0 043  (2009).
\newblock \doi{10.1088/1126-6708/2009/10/043}.

\bibitem{Chesler:2009cy}
P.~M. Chesler and L.~G. Yaffe, {Boost invariant flow, black hole formation, and
  far-from-equilibrium dynamics in N = 4 supersymmetric Yang-Mills theory},
  \emph{Phys. Rev. D}. {\bf 82}, \penalty0 026006  (2010).
\newblock \doi{10.1103/PhysRevD.82.026006}.

\bibitem{Heller:2011ju}
M.~P. Heller, R.~A. Janik, and P.~Witaszczyk, {The characteristics of
  thermalization of boost-invariant plasma from holography}, \emph{Phys. Rev.
  Lett.} {\bf 108}, \penalty0 201602  (2012).
\newblock \doi{10.1103/PhysRevLett.108.201602}.

\bibitem{Heller:2012je}
M.~P. Heller, R.~A. Janik, and P.~Witaszczyk, {A numerical relativity approach
  to the initial value problem in asymptotically Anti-de Sitter spacetime for
  plasma thermalization - an ADM formulation}, \emph{Phys. Rev. D}. {\bf 85},
  \penalty0 126002  (2012).
\newblock \doi{10.1103/PhysRevD.85.126002}.

\bibitem{Heller:2012km}
M.~P. Heller, D.~Mateos, W.~van~der Schee, and D.~Trancanelli, {Strong Coupling
  Isotropization of Non-Abelian Plasmas Simplified}, \emph{Phys. Rev. Lett.}
  {\bf 108}, \penalty0 191601  (2012).
\newblock \doi{10.1103/PhysRevLett.108.191601}.

\bibitem{vanderSchee:2012qj}
W.~van~der Schee, {Holographic thermalization with radial flow}, \emph{Phys.
  Rev. D}. {\bf 87}\penalty0 (6), \penalty0 061901  (2013).
\newblock \doi{10.1103/PhysRevD.87.061901}.

\bibitem{Casalderrey-Solana:2013aba}
J.~Casalderrey-Solana, M.~P. Heller, D.~Mateos, and W.~van~der Schee, {From
  full stopping to transparency in a holographic model of heavy ion
  collisions}, \emph{Phys. Rev. Lett.} {\bf 111}, \penalty0 181601  (2013).
\newblock \doi{10.1103/PhysRevLett.111.181601}.

\bibitem{vanderSchee:2013pia}
W.~van~der Schee, P.~Romatschke, and S.~Pratt, {Fully Dynamical Simulation of
  Central Nuclear Collisions}, \emph{Phys. Rev. Lett.} {\bf 111}\penalty0 (22),
  \penalty0 222302  (2013).
\newblock \doi{10.1103/PhysRevLett.111.222302}.

\bibitem{Heller:2013oxa}
M.~P. Heller, D.~Mateos, W.~van~der Schee, and M.~Triana, {Holographic
  isotropization linearized}, \emph{JHEP}. {\bf 09}, \penalty0 026  (2013).
\newblock \doi{10.1007/JHEP09(2013)026}.

\bibitem{Keegan:2015avk}
L.~Keegan, A.~Kurkela, P.~Romatschke, W.~van~der Schee, and Y.~Zhu, {Weak and
  strong coupling equilibration in nonabelian gauge theories}, \emph{JHEP}.
  {\bf 04}, \penalty0 031  (2016).
\newblock \doi{10.1007/JHEP04(2016)031}.

\bibitem{Chesler:2015bba}
P.~M. Chesler, {Colliding shock waves and hydrodynamics in small systems},
  \emph{Phys. Rev. Lett.} {\bf 115}\penalty0 (24), \penalty0 241602  (2015).
\newblock \doi{10.1103/PhysRevLett.115.241602}.

\bibitem{Kurkela:2015qoa}
A.~Kurkela and Y.~Zhu, {Isotropization and hydrodynamization in weakly coupled
  heavy-ion collisions}, \emph{Phys. Rev. Lett.} {\bf 115}\penalty0 (18),
  \penalty0 182301  (2015).
\newblock \doi{10.1103/PhysRevLett.115.182301}.

\bibitem{Chesler:2016ceu}
P.~M. Chesler, {How big are the smallest drops of quark-gluon plasma?},
  \emph{JHEP}. {\bf 03}, \penalty0 146  (2016).
\newblock \doi{10.1007/JHEP03(2016)146}.

\bibitem{Attems:2016ugt}
M.~Attems, J.~Casalderrey-Solana, D.~Mateos, I.~Papadimitriou,
  D.~Santos-Oliv\'an, C.~F. Sopuerta, M.~Triana, and M.~Zilh\~ao,
  {Thermodynamics, transport and relaxation in non-conformal theories},
  \emph{JHEP}. {\bf 10}, \penalty0 155  (2016).
\newblock \doi{10.1007/JHEP10(2016)155}.

\bibitem{Attems:2016tby}
M.~Attems, J.~Casalderrey-Solana, D.~Mateos, D.~Santos-Oliv\'an, C.~F.
  Sopuerta, M.~Triana, and M.~Zilh\~ao, {Holographic Collisions in
  Non-conformal Theories}, \emph{JHEP}. {\bf 01}, \penalty0 026  (2017).
\newblock \doi{10.1007/JHEP01(2017)026}.

\bibitem{Attems:2017zam}
M.~Attems, J.~Casalderrey-Solana, D.~Mateos, D.~Santos-Oliv\'an, C.~F.
  Sopuerta, M.~Triana, and M.~Zilh\~ao, {Paths to equilibrium in non-conformal
  collisions}, \emph{JHEP}. {\bf 06}, \penalty0 154  (2017).
\newblock \doi{10.1007/JHEP06(2017)154}.

\bibitem{Florkowski:2017olj}
W.~Florkowski, M.~P. Heller, and M.~Spalinski, {New theories of relativistic
  hydrodynamics in the LHC era}, \emph{Rept. Prog. Phys.} {\bf 81}\penalty0
  (4), \penalty0 046001  (2018).
\newblock \doi{10.1088/1361-6633/aaa091}.

\bibitem{Florkowski:2010cf}
W.~Florkowski and R.~Ryblewski, {Highly-anisotropic and strongly-dissipative
  hydrodynamics for early stages of relativistic heavy-ion collisions},
  \emph{Phys. Rev. C}. {\bf 83}, \penalty0 034907  (2011).
\newblock \doi{10.1103/PhysRevC.83.034907}.

\bibitem{Martinez:2010sc}
M.~Martinez and M.~Strickland, {Dissipative Dynamics of Highly Anisotropic
  Systems}, \emph{Nucl. Phys. A}. {\bf 848}, \penalty0 183--197  (2010).
\newblock \doi{10.1016/j.nuclphysa.2010.08.011}.

\bibitem{Ryblewski:2010ch}
R.~Ryblewski and W.~Florkowski, {Highly anisotropic hydrodynamics -- discussion
  of the model assumptions and forms of the initial conditions}, \emph{Acta
  Phys. Polon. B}. {\bf 42}, \penalty0 115--138  (2011).
\newblock \doi{10.5506/APhysPolB.42.115}.

\bibitem{Martinez:2012tu}
M.~Martinez, R.~Ryblewski, and M.~Strickland, {Boost-Invariant
  (2+1)-dimensional Anisotropic Hydrodynamics}, \emph{Phys. Rev. C}. {\bf 85},
  \penalty0 064913  (2012).
\newblock \doi{10.1103/PhysRevC.85.064913}.

\bibitem{Ryblewski:2012rr}
R.~Ryblewski and W.~Florkowski, {Highly-anisotropic hydrodynamics in 3+1
  space-time dimensions}, \emph{Phys. Rev. C}. {\bf 85}, \penalty0 064901
  (2012).
\newblock \doi{10.1103/PhysRevC.85.064901}.

\bibitem{Bazow:2013ifa}
D.~Bazow, U.~W. Heinz, and M.~Strickland, {Second-order (2+1)-dimensional
  anisotropic hydrodynamics}, \emph{Phys. Rev. C}. {\bf 90}\penalty0 (5),
  \penalty0 054910  (2014).
\newblock \doi{10.1103/PhysRevC.90.054910}.

\bibitem{Tinti:2013vba}
L.~Tinti and W.~Florkowski, {Projection method and new formulation of
  leading-order anisotropic hydrodynamics}, \emph{Phys. Rev. C}. {\bf
  89}\penalty0 (3), \penalty0 034907  (2014).
\newblock \doi{10.1103/PhysRevC.89.034907}.

\bibitem{Nopoush:2014pfa}
M.~Nopoush, R.~Ryblewski, and M.~Strickland, {Bulk viscous evolution within
  anisotropic hydrodynamics}, \emph{Phys. Rev. C}. {\bf 90}\penalty0 (1),
  \penalty0 014908  (2014).
\newblock \doi{10.1103/PhysRevC.90.014908}.

\bibitem{Tinti:2015xwa}
L.~Tinti, {Anisotropic matching principle for the hydrodynamic expansion},
  \emph{Phys. Rev. C}. {\bf 94}\penalty0 (4), \penalty0 044902  (2016).
\newblock \doi{10.1103/PhysRevC.94.044902}.

\bibitem{Bazow:2015cha}
D.~Bazow, U.~W. Heinz, and M.~Martinez, {Nonconformal viscous anisotropic
  hydrodynamics}, \emph{Phys. Rev. C}. {\bf 91}\penalty0 (6), \penalty0 064903
  (2015).
\newblock \doi{10.1103/PhysRevC.91.064903}.

\bibitem{Strickland:2015utc}
M.~Strickland, M.~Nopoush, and R.~Ryblewski, {Anisotropic hydrodynamics for
  conformal Gubser flow}, \emph{Nucl. Phys. A}. {\bf 956}, \penalty0 268--271
  (2016).
\newblock \doi{10.1016/j.nuclphysa.2016.02.014}.

\bibitem{Alqahtani:2015qja}
M.~Alqahtani, M.~Nopoush, and M.~Strickland, {Quasiparticle equation of state
  for anisotropic hydrodynamics}, \emph{Phys. Rev. C}. {\bf 92}\penalty0 (5),
  \penalty0 054910  (2015).
\newblock \doi{10.1103/PhysRevC.92.054910}.

\bibitem{Molnar:2016vvu}
E.~Molnar, H.~Niemi, and D.~H. Rischke, {Derivation of anisotropic dissipative
  fluid dynamics from the Boltzmann equation}, \emph{Phys. Rev. D}. {\bf
  93}\penalty0 (11), \penalty0 114025  (2016).
\newblock \doi{10.1103/PhysRevD.93.114025}.

\bibitem{Molnar:2016gwq}
E.~Moln\'ar, H.~Niemi, and D.~H. Rischke, {Closing the equations of motion of
  anisotropic fluid dynamics by a judicious choice of a moment of the Boltzmann
  equation}, \emph{Phys. Rev. D}. {\bf 94}\penalty0 (12), \penalty0 125003
  (2016).
\newblock \doi{10.1103/PhysRevD.94.125003}.

\bibitem{Alqahtani:2016rth}
M.~Alqahtani, M.~Nopoush, and M.~Strickland, {Quasiparticle anisotropic
  hydrodynamics for central collisions}, \emph{Phys. Rev. C}. {\bf 95}\penalty0
  (3), \penalty0 034906  (2017).
\newblock \doi{10.1103/PhysRevC.95.034906}.

\bibitem{Bluhm:2015raa}
M.~Bluhm and T.~Sch\"afer, {Dissipative fluid dynamics for the dilute Fermi gas
  at unitarity: Anisotropic fluid dynamics}, \emph{Phys. Rev. A}. {\bf
  92}\penalty0 (4), \penalty0 043602  (2015).
\newblock \doi{10.1103/PhysRevA.92.043602}.

\bibitem{Bluhm:2015bzi}
M.~Bluhm and T.~Schaefer, {Model-independent determination of the shear
  viscosity of a trapped unitary Fermi gas: Application to high temperature
  data}, \emph{Phys. Rev. Lett.} {\bf 116}\penalty0 (11), \penalty0 115301
  (2016).
\newblock \doi{10.1103/PhysRevLett.116.115301}.

\bibitem{Alqahtani:2017jwl}
M.~Alqahtani, M.~Nopoush, R.~Ryblewski, and M.~Strickland, {(3+1)D
  Quasiparticle Anisotropic Hydrodynamics for Ultrarelativistic Heavy-Ion
  Collisions}, \emph{Phys. Rev. Lett.} {\bf 119}\penalty0 (4), \penalty0 042301
   (2017).
\newblock \doi{10.1103/PhysRevLett.119.042301}.

\bibitem{Alqahtani:2017tnq}
M.~Alqahtani, M.~Nopoush, R.~Ryblewski, and M.~Strickland, {Anisotropic
  hydrodynamic modeling of 2.76 TeV Pb-Pb collisions}, \emph{Phys. Rev. C}.
  {\bf 96}\penalty0 (4), \penalty0 044910  (2017).
\newblock \doi{10.1103/PhysRevC.96.044910}.

\bibitem{Heller:2015dha}
M.~P. Heller and M.~Spalinski, {Hydrodynamics Beyond the Gradient Expansion:
  Resurgence and Resummation}, \emph{Phys. Rev. Lett.} {\bf 115}\penalty0 (7),
  \penalty0 072501  (2015).
\newblock \doi{10.1103/PhysRevLett.115.072501}.

\bibitem{Romatschke:2017vte}
P.~Romatschke, {Relativistic Fluid Dynamics Far From Local Equilibrium},
  \emph{Phys. Rev. Lett.} {\bf 120}\penalty0 (1), \penalty0 012301  (2018).
\newblock \doi{10.1103/PhysRevLett.120.012301}.

\bibitem{Bemfica:2017wps}
F.~S. Bemfica, M.~M. Disconzi, and J.~Noronha, {Causality and existence of
  solutions of relativistic viscous fluid dynamics with gravity}, \emph{Phys.
  Rev. D}. {\bf 98}\penalty0 (10), \penalty0 104064  (2018).
\newblock \doi{10.1103/PhysRevD.98.104064}.

\bibitem{Spalinski:2017mel}
M.~Spali\'nski, {On the hydrodynamic attractor of Yang\textendash{}Mills
  plasma}, \emph{Phys. Lett. B}. {\bf 776}, \penalty0 468--472  (2018).
\newblock \doi{10.1016/j.physletb.2017.11.059}.

\bibitem{Strickland:2013uga}
M.~Strickland, {Thermalization and isotropization in heavy-ion collisions},
  \emph{Pramana}. {\bf 84}\penalty0 (5), \penalty0 671--684  (2015).
\newblock \doi{10.1007/s12043-015-0972-1}.

\bibitem{Berges:2004ce}
J.~Berges, S.~Borsanyi, and C.~Wetterich, {Prethermalization}, \emph{Phys. Rev.
  Lett.} {\bf 93}, \penalty0 142002  (2004).
\newblock \doi{10.1103/PhysRevLett.93.142002}.

\bibitem{Bjorken:1982qr}
J.~D. Bjorken, {Highly Relativistic Nucleus-Nucleus Collisions: The Central
  Rapidity Region}, \emph{Phys. Rev. D}. {\bf 27}, \penalty0 140--151  (1983).
\newblock \doi{10.1103/PhysRevD.27.140}.

\bibitem{Kovtun:2005ev}
P.~K. Kovtun and A.~O. Starinets, {Quasinormal modes and holography},
  \emph{Phys. Rev. D}. {\bf 72}, \penalty0 086009  (2005).
\newblock \doi{10.1103/PhysRevD.72.086009}.

\bibitem{Romatschke:2017ejr}
P.~Romatschke and U.~Romatschke, \emph{{Relativistic Fluid Dynamics In and Out
  of Equilibrium}}. Cambridge Monographs on Mathematical Physics, Cambridge
  University Press  (5, 2019).
\newblock ISBN 978-1-108-48368-1, 978-1-108-75002-8.
\newblock \doi{10.1017/9781108651998}.

\bibitem{Berges:2020fwq}
J.~Berges, M.~P. Heller, A.~Mazeliauskas, and R.~Venugopalan, {QCD
  thermalization: Ab initio approaches and interdisciplinary connections},
  \emph{Rev. Mod. Phys.} {\bf 93}\penalty0 (3), \penalty0 035003  (2021).
\newblock \doi{10.1103/RevModPhys.93.035003}.

\bibitem{Chesler:2010bi}
P.~M. Chesler and L.~G. Yaffe, {Holography and colliding gravitational shock
  waves in asymptotically AdS$_{5}$ spacetime}, \emph{Phys. Rev. Lett.} {\bf
  106}, \penalty0 021601  (2011).
\newblock \doi{10.1103/PhysRevLett.106.021601}.

\bibitem{Chesler:2013lia}
P.~M. Chesler and L.~G. Yaffe, {Numerical solution of gravitational dynamics in
  asymptotically anti-de Sitter spacetimes}, \emph{JHEP}. {\bf 07}, \penalty0
  086  (2014).
\newblock \doi{10.1007/JHEP07(2014)086}.

\bibitem{Chesler:2015wra}
P.~M. Chesler and L.~G. Yaffe, {Holography and off-center collisions of
  localized shock waves}, \emph{JHEP}. {\bf 10}, \penalty0 070  (2015).
\newblock \doi{10.1007/JHEP10(2015)070}.

\bibitem{Denicol:2014xca}
G.~S. Denicol, U.~W. Heinz, M.~Martinez, J.~Noronha, and M.~Strickland, {New
  Exact Solution of the Relativistic Boltzmann Equation and its Hydrodynamic
  Limit}, \emph{Phys. Rev. Lett.} {\bf 113}\penalty0 (20), \penalty0 202301
  (2014).
\newblock \doi{10.1103/PhysRevLett.113.202301}.

\bibitem{Denicol:2014tha}
G.~S. Denicol, U.~W. Heinz, M.~Martinez, J.~Noronha, and M.~Strickland,
  {Studying the validity of relativistic hydrodynamics with a new exact
  solution of the Boltzmann equation}, \emph{Phys. Rev. D}. {\bf 90}\penalty0
  (12), \penalty0 125026  (2014).
\newblock \doi{10.1103/PhysRevD.90.125026}.

\bibitem{Bazow:2015dha}
D.~Bazow, G.~S. Denicol, U.~Heinz, M.~Martinez, and J.~Noronha, {Analytic
  solution of the Boltzmann equation in an expanding system}, \emph{Phys. Rev.
  Lett.} {\bf 116}\penalty0 (2), \penalty0 022301  (2016).
\newblock \doi{10.1103/PhysRevLett.116.022301}.

\bibitem{Strickland:2017kux}
M.~Strickland, J.~Noronha, and G.~Denicol, {Anisotropic nonequilibrium
  hydrodynamic attractor}, \emph{Phys. Rev. D}. {\bf 97}\penalty0 (3),
  \penalty0 036020  (2018).
\newblock \doi{10.1103/PhysRevD.97.036020}.

\bibitem{Blaizot:2017ucy}
J.-P. Blaizot and L.~Yan, {Fluid dynamics of out of equilibrium boost invariant
  plasmas}, \emph{Phys. Lett. B}. {\bf 780}, \penalty0 283--286  (2018).
\newblock \doi{10.1016/j.physletb.2018.02.058}.

\bibitem{Strickland:2018ayk}
M.~Strickland, {The non-equilibrium attractor for kinetic theory in relaxation
  time approximation}, \emph{JHEP}. {\bf 12}, \penalty0 128  (2018).
\newblock \doi{10.1007/JHEP12(2018)128}.

\bibitem{Kurkela:2018wud}
A.~Kurkela, A.~Mazeliauskas, J.-F. Paquet, S.~Schlichting, and D.~Teaney,
  {Matching the Nonequilibrium Initial Stage of Heavy Ion Collisions to
  Hydrodynamics with QCD Kinetic Theory}, \emph{Phys. Rev. Lett.} {\bf
  122}\penalty0 (12), \penalty0 122302  (2019).
\newblock \doi{10.1103/PhysRevLett.122.122302}.

\bibitem{Kurkela:2018vqr}
A.~Kurkela, A.~Mazeliauskas, J.-F. Paquet, S.~Schlichting, and D.~Teaney,
  {Effective kinetic description of event-by-event pre-equilibrium dynamics in
  high-energy heavy-ion collisions}, \emph{Phys. Rev. C}. {\bf 99}\penalty0
  (3), \penalty0 034910  (2019).
\newblock \doi{10.1103/PhysRevC.99.034910}.

\bibitem{Kurkela:2019set}
A.~Kurkela, W.~van~der Schee, U.~A. Wiedemann, and B.~Wu, {Early- and Late-Time
  Behavior of Attractors in Heavy-Ion Collisions}, \emph{Phys. Rev. Lett.} {\bf
  124}\penalty0 (10), \penalty0 102301  (2020).
\newblock \doi{10.1103/PhysRevLett.124.102301}.

\bibitem{Strickland:2019hff}
M.~Strickland and U.~Tantary, {Exact solution for the non-equilibrium attractor
  in number-conserving relaxation time approximation}, \emph{JHEP}. {\bf 10},
  \penalty0 069  (2019).
\newblock \doi{10.1007/JHEP10(2019)069}.

\bibitem{Blaizot:2019scw}
J.-P. Blaizot and L.~Yan, {Emergence of hydrodynamical behavior in expanding
  ultra-relativistic plasmas}, \emph{Annals Phys.} {\bf 412}, \penalty0 167993
  (2020).
\newblock \doi{10.1016/j.aop.2019.167993}.

\bibitem{Denicol:2019lio}
G.~S. Denicol and J.~Noronha, {Exact hydrodynamic attractor of an
  ultrarelativistic gas of hard spheres}, \emph{Phys. Rev. Lett.} {\bf
  124}\penalty0 (15), \penalty0 152301  (2020).
\newblock \doi{10.1103/PhysRevLett.124.152301}.

\bibitem{Brewer:2019oha}
J.~Brewer, L.~Yan, and Y.~Yin, {Adiabatic hydrodynamization in
  rapidly-expanding quark\textendash{}gluon plasma}, \emph{Phys. Lett. B}. {\bf
  816}, \penalty0 136189  (2021).
\newblock \doi{10.1016/j.physletb.2021.136189}.

\bibitem{Almaalol:2020rnu}
D.~Almaalol, A.~Kurkela, and M.~Strickland, {Nonequilibrium Attractor in
  High-Temperature QCD Plasmas}, \emph{Phys. Rev. Lett.} {\bf 125}\penalty0
  (12), \penalty0 122302  (2020).
\newblock \doi{10.1103/PhysRevLett.125.122302}.

\bibitem{Ambrus:2021fej}
V.~E. Ambrus, S.~Schlichting, and C.~Werthmann, {Development of transverse flow
  at small and large opacities in conformal kinetic theory}, \emph{Phys. Rev.
  D}. {\bf 105}\penalty0 (1), \penalty0 014031  (2022).
\newblock \doi{10.1103/PhysRevD.105.014031}.

\bibitem{Blaizot:2021cdv}
J.-P. Blaizot and L.~Yan, {Attractor and fixed points in Bjorken flows},
  \emph{Phys. Rev. C}. {\bf 104}\penalty0 (5), \penalty0 055201  (2021).
\newblock \doi{10.1103/PhysRevC.104.055201}.

\bibitem{Jaiswal:2022udf}
S.~Jaiswal, J.-P. Blaizot, R.~S. Bhalerao, Z.~Chen, A.~Jaiswal, and L.~Yan,
  {From moments of the distribution function to hydrodynamics: The nonconformal
  case}, \emph{Phys. Rev. C}. {\bf 106}\penalty0 (4), \penalty0 044912  (2022).
\newblock \doi{10.1103/PhysRevC.106.044912}.

\bibitem{Alalawi:2022pmg}
H.~Alalawi and M.~Strickland, {Far-from-equilibrium attractors for massive
  kinetic theory in the relaxation time approximation}, \emph{JHEP}. {\bf 12},
  \penalty0 143  (2022).
\newblock \doi{10.1007/JHEP12(2022)143}.

\bibitem{Mullins:2022fbx}
N.~Mullins, G.~S. Denicol, and J.~Noronha, {Far-from-equilibrium kinetic
  dynamics of \ensuremath{\lambda}\ensuremath{\phi}4 theory in an expanding
  universe}, \emph{Phys. Rev. D}. {\bf 106}\penalty0 (5), \penalty0 056024
  (2022).
\newblock \doi{10.1103/PhysRevD.106.056024}.

\bibitem{Ambrus:2022qya}
V.~E. Ambrus, S.~Schlichting, and C.~Werthmann, {Establishing the Range of
  Applicability of Hydrodynamics in High-Energy Collisions}, \emph{Phys. Rev.
  Lett.} {\bf 130}\penalty0 (15), \penalty0 152301  (2023).
\newblock \doi{10.1103/PhysRevLett.130.152301}.

\bibitem{Ambrus:2022koq}
V.~E. Ambrus, S.~Schlichting, and C.~Werthmann, {Opacity dependence of
  transverse flow, preequilibrium, and applicability of hydrodynamics in
  heavy-ion collisions}, \emph{Phys. Rev. D}. {\bf 107}\penalty0 (9), \penalty0
  094013  (2023).
\newblock \doi{10.1103/PhysRevD.107.094013}.

\bibitem{Rocha:2022ind}
G.~S. Rocha, G.~S. Denicol, and J.~Noronha, {Perturbative approaches in
  relativistic kinetic theory and the emergence of first-order hydrodynamics},
  \emph{Phys. Rev. D}. {\bf 106}\penalty0 (3), \penalty0 036010  (2022).
\newblock \doi{10.1103/PhysRevD.106.036010}.

\bibitem{Du:2022bel}
X.~Du, M.~P. Heller, S.~Schlichting, and V.~Svensson, {Exponential approach to
  the hydrodynamic attractor in Yang-Mills kinetic theory}, \emph{Phys. Rev.
  D}. {\bf 106}\penalty0 (1), \penalty0 014016  (2022).
\newblock \doi{10.1103/PhysRevD.106.014016}.

\bibitem{Baier:2007ix}
R.~Baier, P.~Romatschke, D.~T. Son, A.~O. Starinets, and M.~A. Stephanov,
  {Relativistic viscous hydrodynamics, conformal invariance, and holography},
  \emph{JHEP}. {\bf 04}, \penalty0 100  (2008).
\newblock \doi{10.1088/1126-6708/2008/04/100}.

\bibitem{Baym:1984np}
G.~Baym, {THERMAL EQUILIBRATION IN ULTRARELATIVISTIC HEAVY ION COLLISIONS},
  \emph{Phys. Lett. B}. {\bf 138}, \penalty0 18--22  (1984).
\newblock \doi{10.1016/0370-2693(84)91863-X}.

\bibitem{Florkowski:2013lza}
W.~Florkowski, R.~Ryblewski, and M.~Strickland, {Anisotropic Hydrodynamics for
  Rapidly Expanding Systems}, \emph{Nucl. Phys. A}. {\bf 916}, \penalty0
  249--259  (2013).
\newblock \doi{10.1016/j.nuclphysa.2013.08.004}.

\bibitem{Florkowski:2013lya}
W.~Florkowski, R.~Ryblewski, and M.~Strickland, {Testing viscous and
  anisotropic hydrodynamics in an exactly solvable case}, \emph{Phys. Rev. C}.
  {\bf 88}, \penalty0 024903  (2013).
\newblock \doi{10.1103/PhysRevC.88.024903}.

\bibitem{Florkowski:2014sfa}
W.~Florkowski, E.~Maksymiuk, R.~Ryblewski, and M.~Strickland, {Exact solution
  of the (0+1)-dimensional Boltzmann equation for a massive gas}, \emph{Phys.
  Rev. C}. {\bf 89}\penalty0 (5), \penalty0 054908  (2014).
\newblock \doi{10.1103/PhysRevC.89.054908}.

\bibitem{Du:2023bwi}
X.~Du, S.~Ochsenfeld, and S.~Schlichting, {Universality of energy-momentum
  response in kinetic theories}, \emph{Phys. Lett. B}. {\bf 845}, \penalty0
  138161  (2023).
\newblock \doi{10.1016/j.physletb.2023.138161}.

\bibitem{Baier:2000sb}
R.~Baier, A.~H. Mueller, D.~Schiff, and D.~T. Son, {'Bottom up' thermalization
  in heavy ion collisions}, \emph{Phys. Lett. B}. {\bf 502}, \penalty0 51--58
  (2001).
\newblock \doi{10.1016/S0370-2693(01)00191-5}.

\bibitem{Chattopadhyay:2021ive}
C.~Chattopadhyay, S.~Jaiswal, L.~Du, U.~Heinz, and S.~Pal, {Non-conformal
  attractor in boost-invariant plasmas}, \emph{Phys. Lett. B}. {\bf 824},
  \penalty0 136820  (2022).
\newblock \doi{10.1016/j.physletb.2021.136820}.

\bibitem{Jaiswal:2021uvv}
S.~Jaiswal, C.~Chattopadhyay, L.~Du, U.~Heinz, and S.~Pal, {Nonconformal
  kinetic theory and hydrodynamics for Bjorken flow}, \emph{Phys. Rev. C}. {\bf
  105}\penalty0 (2), \penalty0 024911  (2022).
\newblock \doi{10.1103/PhysRevC.105.024911}.

\bibitem{anderson1974relativistic}
J.~Anderson and H.~Witting, Relativistic quantum transport coefficients,
  \emph{Physica}. {\bf 74}\penalty0 (3), \penalty0 489--495  (1974).

\bibitem{Denicol:2010xn}
G.~S. Denicol, T.~Koide, and D.~H. Rischke, {Dissipative relativistic fluid
  dynamics: a new way to derive the equations of motion from kinetic theory},
  \emph{Phys. Rev. Lett.} {\bf 105}, \penalty0 162501  (2010).
\newblock \doi{10.1103/PhysRevLett.105.162501}.

\bibitem{Denicol:2011fa}
G.~S. Denicol, J.~Noronha, H.~Niemi, and D.~H. Rischke, {Origin of the
  Relaxation Time in Dissipative Fluid Dynamics}, \emph{Phys. Rev. D}. {\bf
  83}, \penalty0 074019  (2011).
\newblock \doi{10.1103/PhysRevD.83.074019}.

\bibitem{Muller:1967zza}
I.~Muller, {Zum Paradoxon der Warmeleitungstheorie}, \emph{Z. Phys.} {\bf 198},
  \penalty0 329--344  (1967).
\newblock \doi{10.1007/BF01326412}.

\bibitem{Israel:1976tn}
W.~Israel, {Nonstationary irreversible thermodynamics: A Causal relativistic
  theory}, \emph{Annals Phys.} {\bf 100}, \penalty0 310--331  (1976).
\newblock \doi{10.1016/0003-4916(76)90064-6}.

\bibitem{Israel:1979wp}
W.~Israel and J.~M. Stewart, {Transient relativistic thermodynamics and kinetic
  theory}, \emph{Annals Phys.} {\bf 118}, \penalty0 341--372  (1979).
\newblock \doi{10.1016/0003-4916(79)90130-1}.

\bibitem{Denicol:2012cn}
G.~S. Denicol, H.~Niemi, E.~Molnar, and D.~H. Rischke, {Derivation of transient
  relativistic fluid dynamics from the Boltzmann equation}, \emph{Phys. Rev.
  D}. {\bf 85}, \penalty0 114047  (2012).
\newblock \doi{10.1103/PhysRevD.85.114047}.
\newblock [Erratum: Phys.Rev.D 91, 039902 (2015)].

\bibitem{Denicol:2014loa}
G.~S. Denicol, {Kinetic foundations of relativistic dissipative fluid
  dynamics}, \emph{J. Phys. G}. {\bf 41}\penalty0 (12), \penalty0 124004
  (2014).
\newblock \doi{10.1088/0954-3899/41/12/124004}.

\bibitem{Jaiswal:2013vta}
A.~Jaiswal, {Relativistic third-order dissipative fluid dynamics from kinetic
  theory}, \emph{Phys. Rev. C}. {\bf 88}, \penalty0 021903  (2013).
\newblock \doi{10.1103/PhysRevC.88.021903}.

\bibitem{Jaiswal:2013npa}
A.~Jaiswal, {Relativistic dissipative hydrodynamics from kinetic theory with
  relaxation time approximation}, \emph{Phys. Rev. C}. {\bf 87}\penalty0 (5),
  \penalty0 051901  (2013).
\newblock \doi{10.1103/PhysRevC.87.051901}.

\bibitem{Muronga:2003ta}
A.~Muronga, {Causal theories of dissipative relativistic fluid dynamics for
  nuclear collisions}, \emph{Phys. Rev. C}. {\bf 69}, \penalty0 034903  (2004).
\newblock \doi{10.1103/PhysRevC.69.034903}.

\bibitem{Romatschke:2003ms}
P.~Romatschke and M.~Strickland, {Collective modes of an anisotropic quark
  gluon plasma}, \emph{Phys. Rev. D}. {\bf 68}, \penalty0 036004  (2003).
\newblock \doi{10.1103/PhysRevD.68.036004}.

\bibitem{Romatschke:2004jh}
P.~Romatschke and M.~Strickland, {Collective modes of an anisotropic
  quark-gluon plasma II}, \emph{Phys. Rev. D}. {\bf 70}, \penalty0 116006
  (2004).
\newblock \doi{10.1103/PhysRevD.70.116006}.

\bibitem{Rebhan:2008uj}
A.~Rebhan, M.~Strickland, and M.~Attems, {Instabilities of an anisotropically
  expanding non-Abelian plasma: 1D+3V discretized hard-loop simulations},
  \emph{Phys. Rev. D}. {\bf 78}, \penalty0 045023  (2008).
\newblock \doi{10.1103/PhysRevD.78.045023}.

\bibitem{Janik:2005zt}
R.~A. Janik and R.~B. Peschanski, {Asymptotic perfect fluid dynamics as a
  consequence of Ads/CFT}, \emph{Phys. Rev. D}. {\bf 73}, \penalty0 045013
  (2006).
\newblock \doi{10.1103/PhysRevD.73.045013}.

\bibitem{Liddle:1994dx}
A.~R. Liddle, P.~Parsons, and J.~D. Barrow, {Formalizing the slow roll
  approximation in inflation}, \emph{Phys. Rev. D}. {\bf 50}, \penalty0
  7222--7232  (1994).
\newblock \doi{10.1103/PhysRevD.50.7222}.

\bibitem{Florkowski:2016zsi}
W.~Florkowski, R.~Ryblewski, and M.~Spali\'nski, {Gradient expansion for
  anisotropic hydrodynamics}, \emph{Phys. Rev. D}. {\bf 94}\penalty0 (11),
  \penalty0 114025  (2016).
\newblock \doi{10.1103/PhysRevD.94.114025}.

\bibitem{Aniceto:2024pyc}
I.~Aniceto, J.~Noronha, and M.~Spali\'nski, {An analytic approach to the RTA
  Boltzmann attractor}  (1, 2024).

\bibitem{Florkowski:2014sda}
W.~Florkowski and E.~Maksymiuk, {Exact solution of the (0+1)-dimensional
  Boltzmann equation for massive Bose-Einstein and Fermi-Dirac gases}, \emph{J.
  Phys. G}. {\bf 42}\penalty0 (4), \penalty0 045106  (2015).
\newblock \doi{10.1088/0954-3899/42/4/045106}.

\bibitem{Czyz:1986mr}
W.~Czyz and W.~Florkowski, {Kinetic Coefficients for Quark - Anti-quark
  Plasma}, \emph{Acta Phys. Polon. B}. {\bf 17}, \penalty0 819--837  (1986).

\bibitem{Bialas:1984wv}
A.~Bialas and W.~Czyz, {Boost Invariant Boltzmann-vlasov Equations for
  Relativistic Quark - Anti-quark Plasma}, \emph{Phys. Rev. D}. {\bf 30},
  \penalty0 2371  (1984).
\newblock \doi{10.1103/PhysRevD.30.2371}.

\bibitem{Bialas:1987en}
A.~Bialas, W.~Czyz, A.~Dyrek, and W.~Florkowski, {Oscillations of Quark - Gluon
  Plasma Generated in Strong Color Fields}, \emph{Nucl. Phys. B}. {\bf 296},
  \penalty0 611--624  (1988).
\newblock \doi{10.1016/0550-3213(88)90035-1}.

\bibitem{Baym:1985tna}
G.~Baym, {ENTROPY PRODUCTION AND THE EVOLUTION OF ULTRARELATIVISTIC HEAVY ION
  COLLISIONS}, \emph{Nucl. Phys. A}. {\bf 418}, \penalty0 525C--537C  (1984).
\newblock \doi{10.1016/0375-9474(84)90573-6}.

\bibitem{Heiselberg:1995sh}
H.~Heiselberg and X.-N. Wang, {Expansion, thermalization and entropy production
  in high-energy nuclear collisions}, \emph{Phys. Rev. C}. {\bf 53}, \penalty0
  1892--1902  (1996).
\newblock \doi{10.1103/PhysRevC.53.1892}.

\bibitem{Wong:1996va}
S.~M.~H. Wong, {Thermal and chemical equilibration in relativistic heavy ion
  collisions}, \emph{Phys. Rev. C}. {\bf 54}, \penalty0 2588--2599  (1996).
\newblock \doi{10.1103/PhysRevC.54.2588}.

\bibitem{Averbeck:2015jja}
R.~Averbeck, J.~W. Harris, and B.~Schenke, \emph{{Heavy-Ion Physics at the
  LHC}}, In ed. T.~Sch\"orner-Sadenius, \emph{{The Large Hadron Collider}:
  {Harvest of Run 1}}, pp. 355--420.
\newblock  (2015).
\newblock \doi{10.1007/978-3-319-15001-7_9}.

\bibitem{Jeon:2016uym}
S.~Jeon and U.~Heinz, \emph{{Introduction to Hydrodynamics}}, In ed. X.-N.
  Wang, \emph{{Quark-Gluon Plasma 5}}, pp. 131--187.
\newblock  (2016).
\newblock \doi{10.1142/9789814663717_0003}.

\bibitem{Teaney:2003kp}
D.~Teaney, {The Effects of viscosity on spectra, elliptic flow, and HBT radii},
  \emph{Phys. Rev. C}. {\bf 68}, \penalty0 034913  (2003).
\newblock \doi{10.1103/PhysRevC.68.034913}.

\bibitem{Arnold:2002zm}
P.~B. Arnold, G.~D. Moore, and L.~G. Yaffe, {Effective kinetic theory for high
  temperature gauge theories}, \emph{JHEP}. {\bf 01}, \penalty0 030  (2003).
\newblock \doi{10.1088/1126-6708/2003/01/030}.

\bibitem{York:2014wja}
M.~C. Abraao~York, A.~Kurkela, E.~Lu, and G.~D. Moore, {UV cascade in classical
  Yang-Mills theory via kinetic theory}, \emph{Phys. Rev. D}. {\bf 89}\penalty0
  (7), \penalty0 074036  (2014).
\newblock \doi{10.1103/PhysRevD.89.074036}.

\bibitem{Mueller:1999pi}
A.~H. Mueller, {The Boltzmann equation for gluons at early times after a heavy
  ion collision}, \emph{Phys. Lett. B}. {\bf 475}, \penalty0 220--224  (2000).
\newblock \doi{10.1016/S0370-2693(00)00084-8}.

\bibitem{Arnold:2003zc}
P.~B. Arnold, G.~D. Moore, and L.~G. Yaffe, {Transport coefficients in high
  temperature gauge theories. 2. Beyond leading log}, \emph{JHEP}. {\bf 05},
  \penalty0 051  (2003).
\newblock \doi{10.1088/1126-6708/2003/05/051}.

\bibitem{Keegan:2016cpi}
L.~Keegan, A.~Kurkela, A.~Mazeliauskas, and D.~Teaney, {Initial conditions for
  hydrodynamics from weakly coupled pre-equilibrium evolution}, \emph{JHEP}.
  {\bf 08}, \penalty0 171  (2016).
\newblock \doi{10.1007/JHEP08(2016)171}.

\bibitem{Kurkela:2018oqw}
A.~Kurkela and A.~Mazeliauskas, {Chemical equilibration in weakly coupled QCD},
  \emph{Phys. Rev. D}. {\bf 99}\penalty0 (5), \penalty0 054018  (2019).
\newblock \doi{10.1103/PhysRevD.99.054018}.

\bibitem{Kurkela:2018xxd}
A.~Kurkela and A.~Mazeliauskas, {Chemical Equilibration in Hadronic
  Collisions}, \emph{Phys. Rev. Lett.} {\bf 122}, \penalty0 142301  (2019).
\newblock \doi{10.1103/PhysRevLett.122.142301}.

\bibitem{Mueller:1999fp}
A.~H. Mueller, {Toward equilibration in the early stages after a high-energy
  heavy ion collision}, \emph{Nucl. Phys. B}. {\bf 572}, \penalty0 227--240
  (2000).
\newblock \doi{10.1016/S0550-3213(99)00502-7}.

\bibitem{Kovchegov:2000hz}
Y.~V. Kovchegov, {Classical initial conditions for ultrarelativistic heavy ion
  collisions}, \emph{Nucl. Phys. A}. {\bf 692}, \penalty0 557--582  (2001).
\newblock \doi{10.1016/S0375-9474(01)00652-2}.

\bibitem{Lappi:2011ju}
T.~Lappi, {Gluon spectrum in the glasma from JIMWLK evolution}, \emph{Phys.
  Lett. B}. {\bf 703}, \penalty0 325--330  (2011).
\newblock \doi{10.1016/j.physletb.2011.08.011}.

\bibitem{Borsanyi:2010cj}
S.~Borsanyi, G.~Endrodi, Z.~Fodor, A.~Jakovac, S.~D. Katz, S.~Krieg, C.~Ratti,
  and K.~K. Szabo, {The QCD equation of state with dynamical quarks},
  \emph{JHEP}. {\bf 11}, \penalty0 077  (2010).
\newblock \doi{10.1007/JHEP11(2010)077}.

\bibitem{HotQCD:2014kol}
A.~Bazavov et~al., {Equation of state in ( 2+1 )-flavor QCD}, \emph{Phys. Rev.
  D}. {\bf 90}, \penalty0 094503  (2014).
\newblock \doi{10.1103/PhysRevD.90.094503}.

\bibitem{Heller:2018qvh}
M.~P. Heller and V.~Svensson, {How does relativistic kinetic theory remember
  about initial conditions?}, \emph{Phys. Rev. D}. {\bf 98}\penalty0 (5),
  \penalty0 054016  (2018).
\newblock \doi{10.1103/PhysRevD.98.054016}.

\bibitem{Giacalone:2019ldn}
G.~Giacalone, A.~Mazeliauskas, and S.~Schlichting, {Hydrodynamic attractors,
  initial state energy and particle production in relativistic nuclear
  collisions}, \emph{Phys. Rev. Lett.} {\bf 123}\penalty0 (26), \penalty0
  262301  (2019).
\newblock \doi{10.1103/PhysRevLett.123.262301}.

\bibitem{Gelis:2010nm}
F.~Gelis, E.~Iancu, J.~Jalilian-Marian, and R.~Venugopalan, {The Color Glass
  Condensate}, \emph{Ann. Rev. Nucl. Part. Sci.} {\bf 60}, \penalty0 463--489
  (2010).
\newblock \doi{10.1146/annurev.nucl.010909.083629}.

\bibitem{Teaney:2009qa}
D.~A. Teaney, \emph{{Viscous Hydrodynamics and the Quark Gluon Plasma}}, In
  eds. R.~C. Hwa and X.-N. Wang, \emph{{Quark-gluon plasma 4}}, pp. 207--266.
\newblock  (2010).
\newblock \doi{10.1142/9789814293297_0004}.

\bibitem{Dusling:2009df}
K.~Dusling, G.~D. Moore, and D.~Teaney, {Radiative energy loss and v(2) spectra
  for viscous hydrodynamics}, \emph{Phys. Rev. C}. {\bf 81}, \penalty0 034907
  (2010).
\newblock \doi{10.1103/PhysRevC.81.034907}.

\bibitem{Bozek:2013uha}
P.~Bozek and W.~Broniowski, {Collective dynamics in high-energy proton-nucleus
  collisions}, \emph{Phys. Rev. C}. {\bf 88}\penalty0 (1), \penalty0 014903
  (2013).
\newblock \doi{10.1103/PhysRevC.88.014903}.

\bibitem{Shen:2016zpp}
C.~Shen, J.-F. Paquet, G.~S. Denicol, S.~Jeon, and C.~Gale, {Collectivity and
  electromagnetic radiation in small systems}, \emph{Phys. Rev. C}. {\bf
  95}\penalty0 (1), \penalty0 014906  (2017).
\newblock \doi{10.1103/PhysRevC.95.014906}.

\bibitem{Weller:2017tsr}
R.~D. Weller and P.~Romatschke, {One fluid to rule them all: viscous
  hydrodynamic description of event-by-event central p+p, p+Pb and Pb+Pb
  collisions at $\sqrt{s}=5.02$ TeV}, \emph{Phys. Lett. B}. {\bf 774},
  \penalty0 351--356  (2017).
\newblock \doi{10.1016/j.physletb.2017.09.077}.

\bibitem{Mantysaari:2017cni}
H.~M\"antysaari, B.~Schenke, C.~Shen, and P.~Tribedy, {Imprints of fluctuating
  proton shapes on flow in proton-lead collisions at the LHC}, \emph{Phys.
  Lett. B}. {\bf 772}, \penalty0 681--686  (2017).
\newblock \doi{10.1016/j.physletb.2017.07.038}.

\bibitem{Strickland:2018exs}
M.~Strickland, {Small system studies: A theory overview}, \emph{Nucl. Phys. A}.
  {\bf 982}, \penalty0 92--98  (2019).
\newblock \doi{10.1016/j.nuclphysa.2018.09.071}.

\bibitem{Alqahtani:2017mhy}
M.~Alqahtani, M.~Nopoush, and M.~Strickland, {Relativistic anisotropic
  hydrodynamics}, \emph{Prog. Part. Nucl. Phys.} {\bf 101}, \penalty0 204--248
  (2018).
\newblock \doi{10.1016/j.ppnp.2018.05.004}.

\bibitem{Alalawi:2021jwn}
H.~Alalawi, M.~Alqahtani, and M.~Strickland, {Resummed Relativistic Dissipative
  Hydrodynamics}, \emph{Symmetry}. {\bf 14}\penalty0 (2), \penalty0 329
  (2022).
\newblock \doi{10.3390/sym14020329}.

\bibitem{Nopoush:2019vqc}
M.~Nopoush and M.~Strickland, {Including off-diagonal anisotropies in
  anisotropic hydrodynamics}, \emph{Phys. Rev. C}. {\bf 100}\penalty0 (1),
  \penalty0 014904  (2019).
\newblock \doi{10.1103/PhysRevC.100.014904}.

\bibitem{Ryblewski:2017ybw}
R.~Ryblewski, {Thermodynamically consistent formulation of quasiparticle
  viscous hydrodynamics}, \emph{Acta Phys. Polon. Supp.} {\bf 10}, \penalty0
  1073  (2017).
\newblock \doi{10.5506/APhysPolBSupp.10.1073}.

\bibitem{Ryu:2015vwa}
S.~Ryu, J.~F. Paquet, C.~Shen, G.~S. Denicol, B.~Schenke, S.~Jeon, and C.~Gale,
  {Importance of the Bulk Viscosity of QCD in Ultrarelativistic Heavy-Ion
  Collisions}, \emph{Phys. Rev. Lett.} {\bf 115}\penalty0 (13), \penalty0
  132301  (2015).
\newblock \doi{10.1103/PhysRevLett.115.132301}.

\bibitem{Bass:2017zyn}
S.~A. Bass, J.~E. Bernhard, and J.~S. Moreland, {Determination of
  Quark-Gluon-Plasma Parameters from a Global Bayesian Analysis}, \emph{Nucl.
  Phys. A}. {\bf 967}, \penalty0 67--73  (2017).
\newblock \doi{10.1016/j.nuclphysa.2017.05.052}.

\bibitem{Rougemont:2017tlu}
R.~Rougemont, R.~Critelli, J.~Noronha-Hostler, J.~Noronha, and C.~Ratti,
  {Dynamical versus equilibrium properties of the QCD phase transition: A
  holographic perspective}, \emph{Phys. Rev. D}. {\bf 96}\penalty0 (1),
  \penalty0 014032  (2017).
\newblock \doi{10.1103/PhysRevD.96.014032}.

\bibitem{Bozek:2010bi}
P.~Bozek and I.~Wyskiel, {Directed flow in ultrarelativistic heavy-ion
  collisions}, \emph{Phys. Rev. C}. {\bf 81}, \penalty0 054902  (2010).
\newblock \doi{10.1103/PhysRevC.81.054902}.

\bibitem{Alqahtani:2020paa}
M.~Alqahtani and M.~Strickland, {Bulk observables at 5.02~TeV using
  quasiparticle anisotropic hydrodynamics}, \emph{Eur. Phys. J. C}. {\bf
  81}\penalty0 (11), \penalty0 1022  (2021).
\newblock \doi{10.1140/epjc/s10052-021-09832-z}.

\bibitem{Almaalol:2018gjh}
D.~Almaalol, M.~Alqahtani, and M.~Strickland, {Anisotropic hydrodynamic
  modeling of 200 GeV Au-Au collisions}, \emph{Phys. Rev. C}. {\bf 99}\penalty0
  (4), \penalty0 044902  (2019).
\newblock \doi{10.1103/PhysRevC.99.044902}.

\bibitem{Chojnacki:2011hb}
M.~Chojnacki, A.~Kisiel, W.~Florkowski, and W.~Broniowski, {THERMINATOR 2:
  THERMal heavy IoN generATOR 2}, \emph{Comput. Phys. Commun.} {\bf 183},
  \penalty0 746--773  (2012).
\newblock \doi{10.1016/j.cpc.2011.11.018}.

\bibitem{kent-code-library}
{M. Strickland et al}.
\newblock {Kent Code Library}.
\newblock \url{http://personal.kent.edu/~mstrick6/code}  (2022).

\bibitem{Abelev:2013vea}
B.~Abelev et~al., {Centrality dependence of $\pi$, K, p production in Pb-Pb
  collisions at $\sqrt{s_{NN}}$ = 2.76 TeV}, \emph{Phys. Rev. C}. {\bf 88},
  \penalty0 044910  (2013).
\newblock \doi{10.1103/PhysRevC.88.044910}.

\bibitem{Abbas:2013bpa}
E.~Abbas et~al., {Centrality dependence of the pseudorapidity density
  distribution for charged particles in Pb-Pb collisions at $\sqrt{s_{\rm NN}}$
  = 2.76 TeV}, \emph{Phys. Lett. B}. {\bf 726}, \penalty0 610--622  (2013).
\newblock \doi{10.1016/j.physletb.2013.09.022}.

\bibitem{Adam:2015kda}
J.~Adam et~al., {Centrality evolution of the charged\textendash{}particle
  pseudorapidity density over a broad pseudorapidity range in Pb\textendash{}Pb
  collisions at $\sqrt{s_{\rm NN}}=2.76$ TeV}, \emph{Phys. Lett. B}. {\bf 754},
  \penalty0 373--385  (2016).
\newblock \doi{10.1016/j.physletb.2015.12.082}.

\bibitem{Abelev:2014pua}
B.~B. Abelev et~al., {Elliptic flow of identified hadrons in Pb-Pb collisions
  at $ \sqrt{s_{\mathrm{NN}}}=2.76 $ TeV}, \emph{JHEP}. {\bf 06}, \penalty0 190
   (2015).
\newblock \doi{10.1007/JHEP06(2015)190}.

\bibitem{Graczykowski:2014hoa}
L.~K. Graczykowski, {Pion femtoscopy measurements in ALICE at the LHC},
  \emph{EPJ Web Conf.} {\bf 71}, \penalty0 00051  (2014).
\newblock \doi{10.1051/epjconf/20147100051}.

\bibitem{ALICE:2019hno}
S.~Acharya et~al., {Production of charged pions, kaons, and (anti-)protons in
  Pb-Pb and inelastic $pp$ collisions at $\sqrt {s_{NN}}$ = 5.02 TeV},
  \emph{Phys. Rev. C}. {\bf 101}\penalty0 (4), \penalty0 044907  (2020).
\newblock \doi{10.1103/PhysRevC.101.044907}.

\bibitem{Jacazio:2017dvy}
N.~Jacazio, {Production of identified charged hadrons in Pb\textendash{}Pb
  collisions at $\sqrt{{s}_{NN}}=$ 5.02 TeV}, \emph{Nucl. Phys. A}. {\bf 967},
  \penalty0 421--424  (2017).
\newblock \doi{10.1016/j.nuclphysa.2017.05.023}.

\bibitem{PHENIX:2003iij}
S.~S. Adler et~al., {Identified charged particle spectra and yields in Au+Au
  collisions at S(NN)**1/2 = 200-GeV}, \emph{Phys. Rev. C}. {\bf 69}, \penalty0
  034909  (2004).
\newblock \doi{10.1103/PhysRevC.69.034909}.

\bibitem{Adler:2003cb}
S.~S. Adler et~al., {Identified charged particle spectra and yields in Au+Au
  collisions at S(NN)**1/2 = 200-GeV}, \emph{Phys. Rev. C}. {\bf 69}, \penalty0
  034909  (2004).
\newblock \doi{10.1103/PhysRevC.69.034909}.

\bibitem{Adams:2004ux}
J.~Adams et~al., {phi meson production in Au + Au and p+p collisions at
  s(NN)**(1/2) = 200-GeV}, \emph{Phys. Lett. B}. {\bf 612}, \penalty0 181--189
  (2005).
\newblock \doi{10.1016/j.physletb.2004.12.082}.

\bibitem{Adams:2006ke}
J.~Adams et~al., {Scaling Properties of Hyperon Production in Au+Au Collisions
  at s**(1/2) = 200-GeV}, \emph{Phys. Rev. Lett.} {\bf 98}, \penalty0 062301
  (2007).
\newblock \doi{10.1103/PhysRevLett.98.062301}.

\bibitem{Alver:2010ck}
B.~Alver et~al., {Phobos results on charged particle multiplicity and
  pseudorapidity distributions in Au+Au, Cu+Cu, d+Au, and p+p collisions at
  ultra-relativistic energies}, \emph{Phys. Rev. C}. {\bf 83}, \penalty0 024913
   (2011).
\newblock \doi{10.1103/PhysRevC.83.024913}.

\bibitem{ALICE:2012xs}
B.~Abelev et~al., {Pseudorapidity density of charged particles in $p$ + Pb
  collisions at $\sqrt{s_{NN}}=5.02$ TeV}, \emph{Phys. Rev. Lett.} {\bf
  110}\penalty0 (3), \penalty0 032301  (2013).
\newblock \doi{10.1103/PhysRevLett.110.032301}.

\bibitem{CMS:2017shj}
A.~M. Sirunyan et~al., {Pseudorapidity distributions of charged hadrons in
  proton-lead collisions at $\sqrt{s_{_\mathrm{NN}}} =$ 5.02 and 8.16 TeV},
  \emph{JHEP}. {\bf 01}, \penalty0 045  (2018).
\newblock \doi{10.1007/JHEP01(2018)045}.

\bibitem{ALICE:2013wgn}
B.~B. Abelev et~al., {Multiplicity Dependence of Pion, Kaon, Proton and Lambda
  Production in p-Pb Collisions at $\sqrt{s_{NN}}$ = 5.02 TeV}, \emph{Phys.
  Lett. B}. {\bf 728}, \penalty0 25--38  (2014).
\newblock \doi{10.1016/j.physletb.2013.11.020}.

\bibitem{CMS:2013pdl}
S.~Chatrchyan et~al., {Study of the Production of Charged Pions, Kaons, and
  Protons in pPb Collisions at $\sqrt{s_{NN}} =\ $ 5.02 $\,\text {TeV}$},
  \emph{Eur. Phys. J. C}. {\bf 74}\penalty0 (6), \penalty0 2847  (2014).
\newblock \doi{10.1140/epjc/s10052-014-2847-x}.

\bibitem{Strickland:2024oat}
M.~Strickland, S.~Thapa, and R.~Vogt, {Bottomonium suppression in 5.02 and 8.16
  TeV p-Pb collisions}  (1, 2024).

\bibitem{Florkowski2010-tl}
W.~Florkowski, \emph{Phenomenology of ultra-relativistic heavy-ion collisions}.
  World Scientific Publishing, Singapore, Singapore  (Mar., 2010).

\bibitem{Skands:2014pea}
P.~Skands, S.~Carrazza, and J.~Rojo, {Tuning PYTHIA 8.1: the Monash 2013 Tune},
  \emph{Eur. Phys. J. C}. {\bf 74}\penalty0 (8), \penalty0 3024  (2014).
\newblock \doi{10.1140/epjc/s10052-014-3024-y}.

\bibitem{dEnterria:2020dwq}
D.~d'Enterria and C.~Loizides, {Progress in the Glauber Model at Collider
  Energies}, \emph{Ann. Rev. Nucl. Part. Sci.} {\bf 71}, \penalty0 315--344
  (2021).
\newblock \doi{10.1146/annurev-nucl-102419-060007}.

\bibitem{Bazow:2016yra}
D.~Bazow, U.~W. Heinz, and M.~Strickland, {Massively parallel simulations of
  relativistic fluid dynamics on graphics processing units with CUDA},
  \emph{Comput. Phys. Commun.} {\bf 225}, \penalty0 92--113  (2018).
\newblock \doi{10.1016/j.cpc.2017.01.015}.

\bibitem{McNelis:2018jho}
M.~McNelis, D.~Bazow, and U.~Heinz, {(3+1)-dimensional anisotropic fluid
  dynamics with a lattice QCD equation of state}, \emph{Phys. Rev. C}. {\bf
  97}\penalty0 (5), \penalty0 054912  (2018).
\newblock \doi{10.1103/PhysRevC.97.054912}.

\bibitem{McNelis:2021zji}
M.~McNelis, D.~Bazow, and U.~Heinz, {Anisotropic fluid dynamical simulations of
  heavy-ion collisions}, \emph{Comput. Phys. Commun.} {\bf 267}, \penalty0
  108077  (2021).
\newblock \doi{10.1016/j.cpc.2021.108077}.

\bibitem{Liyanage:2023nds}
D.~Liyanage, O.~S\"urer, M.~Plumlee, S.~M. Wild, and U.~Heinz, {Bayesian
  calibration of viscous anisotropic hydrodynamic simulations of heavy-ion
  collisions}, \emph{Phys. Rev. C}. {\bf 108}\penalty0 (5), \penalty0 054905
  (2023).
\newblock \doi{10.1103/PhysRevC.108.054905}.

\bibitem{Bazavov:2017dsy}
A.~Bazavov, P.~Petreczky, and J.~H. Weber, {Equation of State in 2+1 Flavor QCD
  at High Temperatures}, \emph{Phys. Rev. D}. {\bf 97}\penalty0 (1), \penalty0
  014510  (2018).
\newblock \doi{10.1103/PhysRevD.97.014510}.

\bibitem{Tinti:2016bav}
L.~Tinti, A.~Jaiswal, and R.~Ryblewski, {Quasiparticle second-order viscous
  hydrodynamics from kinetic theory}, \emph{Phys. Rev. D}. {\bf 95}\penalty0
  (5), \penalty0 054007  (2017).
\newblock \doi{10.1103/PhysRevD.95.054007}.

\bibitem{JETSCAPE:2020mzn}
D.~Everett et~al., {Multisystem Bayesian constraints on the transport
  coefficients of QCD matter}, \emph{Phys. Rev. C}. {\bf 103}\penalty0 (5),
  \penalty0 054904  (2021).
\newblock \doi{10.1103/PhysRevC.103.054904}.

\bibitem{JETSCAPE:2020shq}
D.~Everett et~al., {Phenomenological constraints on the transport properties of
  QCD matter with data-driven model averaging}, \emph{Phys. Rev. Lett.} {\bf
  126}\penalty0 (24), \penalty0 242301  (2021).
\newblock \doi{10.1103/PhysRevLett.126.242301}.

\bibitem{ALICE:2015dtd}
J.~Adam et~al., {Centrality dependence of the nuclear modification factor of
  charged pions, kaons, and protons in Pb-Pb collisions at $\sqrt{s_{\rm
  NN}}=2.76$ TeV}, \emph{Phys. Rev. C}. {\bf 93}\penalty0 (3), \penalty0 034913
   (2016).
\newblock \doi{10.1103/PhysRevC.93.034913}.

\bibitem{Heinz:2023kzr}
U.~Heinz, D.~Liyanage, and C.~Gantenberg.
\newblock {Bayesian calibration of viscous anisotropic hydrodynamic (VAH)
  simulations of heavy-ion collisions}.
\newblock In \emph{{30th International Conference on Ultrarelativstic
  Nucleus-Nucleus Collisions}}  (11, 2023).

\bibitem{Noferini:2012ps}
F.~Noferini, {Anisotropic flow of identified particles in Pb-Pb collisions at
  $\sqrt{s_{NN}} = 2.76$ TeV measured with ALICE at the LHC}, \emph{Nucl. Phys.
  A}. {\bf 904-905}, \penalty0 483c--486c  (2013).
\newblock \doi{10.1016/j.nuclphysa.2013.02.058}.

\bibitem{ALICE:2011ab}
K.~Aamodt et~al., {Higher harmonic anisotropic flow measurements of charged
  particles in Pb-Pb collisions at $\sqrt{s_{NN}}$=2.76 TeV}, \emph{Phys. Rev.
  Lett.} {\bf 107}, \penalty0 032301  (2011).
\newblock \doi{10.1103/PhysRevLett.107.032301}.

\bibitem{Nopoush:2014qba}
M.~Nopoush, R.~Ryblewski, and M.~Strickland, {Anisotropic hydrodynamics for
  conformal Gubser flow}, \emph{Phys. Rev. D}. {\bf 91}\penalty0 (4), \penalty0
  045007  (2015).
\newblock \doi{10.1103/PhysRevD.91.045007}.

\bibitem{Martinez:2017ibh}
M.~Martinez, M.~McNelis, and U.~Heinz, {Anisotropic fluid dynamics for Gubser
  flow}, \emph{Phys. Rev. C}. {\bf 95}\penalty0 (5), \penalty0 054907  (2017).
\newblock \doi{10.1103/PhysRevC.95.054907}.

\end{thebibliography}


\end{document}